\documentclass{aa}  

\usepackage{graphicx}
\usepackage{txfonts}
\usepackage{upgreek}   
\usepackage{amsmath}
\usepackage[breaklinks=true,hidelinks]{hyperref}

\newcommand{\lsun}{\mbox{L}_\odot}
\newcommand{\rsun}{\mbox{R}_\odot}

\newcommand{\lstar}{L_\star}

\newcommand{\rstar}{R_\star}
\newcommand{\teff}{T_{\rm eff}}

\usepackage[dvipsnames]{xcolor}
\definecolor{green}{rgb}{0.0, 0.5, 0.0}

\newcommand{\GG}[1]{}

\graphicspath{{./}{fig/}}

\begin{document}

   \title{Short- and long-term variations of the high mass accretion rate classical T~Tauri star DR~Tau}
   \titlerunning{Short- and long-term variations of DR~Tau}

   \author{Gabriella Zsidi
           \inst{1,2,3,4}
           \and
           Ágnes Kóspál\inst{1,2,5,6}
           \and
           Péter Ábrahám\inst{1,2,5,7}
           \and
           Evelyne Alecian\inst{8}
           \and
           Silvia H. P. Alencar\inst{9}
           \and
           J\'{e}r\^{o}me Bouvier\inst{8}
           \and
           Gaitee A. J. Hussain\inst{10}
           \and
           Carlo F. Manara\inst{11}
           \and
           Michal Siwak\inst{1,2, 12}
           \and
           R\'obert Szab\'o\inst{1,2,5}
           \and
           Zs\'ofia Bora\inst{1,2}
           \and
           Borb\'ala Cseh\inst{1,2,15}
           \and
           Csilla Kalup\inst{1,2}
           \and
           Csaba Kiss\inst{1,2}
           \and
           Levente Kriskovics\inst{1,2}
           \and
           M\'aria Kun\inst{1,2}
           \and
           Andr\'as P\'al\inst{1,2}
           \and
           \'Ad\'am S\'odor\inst{1,2}
           \and
           Kriszti\'an S\'arneczky\inst{1,2}
           \and
           R\'obert Szak\'ats\inst{1,2}
           \and
           Kriszti\'an Vida\inst{1,2}
           \and
           J\'ozsef Vink\'o\inst{1,2}
           \and
           Zsófia M. Szab\'o \inst{1,2,13,14}
          }

    \institute{Konkoly Observatory, HUN-REN Research Centre for Astronomy and Earth Sciences, Konkoly-Thege Miklós út 15-17, 1121 Budapest, Hungary.            
    \email{g.zsidi@herts.ac.uk}
    \and
    CSFK, MTA Centre of Excellence, Budapest, Konkoly Thege Mikl\'os \'ut 15-17, 1121, Hungary
    \and
    School of Physics and Astronomy, Sir William Henry Bragg Building, Woodhouse Ln.,
    University of Leeds, Leeds LS2 9JT, UK
    \and
    Centre for Astrophysics Research, University of Hertfordshire, College Lane, Hatfield, Hertfordshire, AL10 9AB, UK
    \and
    Institute of Physics and Astronomy, ELTE, Institute of Physics, P\'azm\'any P\'eter s\'et\'any 1/A, 1117 Budapest, Hungary
    \and
    Max Planck Institute for Astronomy, K\"onigstuhl 17, 69117 Heidelberg, Germany
    \and
     Institute for Astronomy (IfA), University of Vienna, T\"urkenschanzstrasse 17, A-1180 Vienna 
    \and
    Univ. Grenoble Alpes, CNRS, IPAG, 38000, Grenoble, France
    \and
    Departamento de Física, Universidade Federal de Minas Gerais, Belo Horizonte, MG, 31270-901, Brazil
    \and
    Science Division, Directorate of Science, European Space Research and Technology Centre (ESA/ESTEC), Keplerlaan 1, 2201 AZ, Noordwijk, The Netherlands
    \and
    European Southern Observatory, Karl-Schwarzschild-Strasse 2, 85748 Garching bei München, Germany
    \and
    Mt. Suhora Astronomical Observatory, University of the National Education Commission, ul. Podchor\k{a}\.zych 2, 30-084 Krak{\'o}w, Poland
    \and
    Max-Planck-Institut für Radioastronomie, Auf dem Hügel 69, 53121, Bonn, Germany
    \and
    Scottish Universities Physics Alliance (SUPA), School of Physics and Astronomy, University of St Andrews, North Haugh, St Andrews, KY16 9SS, UK
    \and
    MTA-ELTE Lend{\"u}let "Momentum" Milky Way Research Group, Hungary}

   \date{Received \today}


  \abstract
   {Classical T~Tauri stars are newly formed, low mass stars which may display both periodic and random variations in their brightness. These systems are surrounded by a circumstellar disk, from which material is falling onto the stellar surface. The interaction between the star and the circumstellar disk is time-dependent, leading to short or long-term physical changes in the physical environment, and hence variability of the system.}
   {DR Tau is a highly variable young star. By compiling a large dataset with high-cadence photometric, and high-resolution spectroscopic observations, we aim to examine the short- and long-term variability of the system, and identify the underlying physical mechanisms.   
   }
   {We combine multifilter ground-based optical, near-infrared, and space-based mid-infrared (Spitzer Space Telescope) monitoring observations from 2009, 2017 and 2021 with high-cadence optical Kepler K2 and TESS light curves.  
   We complement our photometric dataset with spectropolarimetric monitoring observations obtained with the CFHT/ESPaDOnS instrument in 2016, which provided high-resolution data at optical wavelengths. 
   }
   {Our results reveal that DR Tau exhibits stochastic photometric variability not only on daily, but also on hourly timescale with peak-to-peak amplitude of 1.4 mag probably originating from accretion related variations. Our ground-based multifilter photometry shows that the shape of the light curves are similar at all wavelengths, although the amplitude of the variability decreases with increasing wavelength. This trend towards the infrared wavelengths suggests that part of the disk may be optically thick and invariable. In addition to this, the high-cadence Kepler and TESS data allowed us to carry out a detailed period analysis. The spectroscopic analysis showed that the H$\alpha$ line presents the most complex line profile with several components but the significance of the components changes over time. This suggests the presence and variation of both accretion flow and wind. Broad and narrow components can be clearly distinguished in the He~I and the Ca~II lines, suggesting contribution from both the accretion flow and the post-shock region. The CFHT/ESPaDOnS data suggest that the strength of the longitudinal magnetic field varies between 400 and 1800\,G.}
   {DR~Tau exhibits high level of photometric and spectroscopic variability on both short- and long-timescales, which is caused by the combination of accretion, wind, stellar activity, and obscuration by circumstellar matter; and the significance of the physical mechanisms causing the observed variability changes over time.}
   

   \maketitle

%

\section{Introduction}
Classical T~Tauri stars are newly formed, low mass stars which display both periodic and stochastic brightness variations. 
These systems are surrounded by a circumstellar disk, from which material is falling onto the stellar surface. 
The interaction between the star and the circumstellar disk can be explained by the magnetospheric accretion model, in which the stellar magnetic field truncates the inner disk at a distance of a few stellar radii and redirects the disk material to the star \citep{hartmann2016}. 
The existence of the stellar magnetic field required by the model has been measured by several studies in classical T~Tauri systems based on the Zeeman broadening measurements of the photospheric lines \citep[e.g.,][]{johns-krull2007, johnstone2014, donati2011, donati2019}.

The photospheric spectrum of T~Tauri stars is contaminated by strong emission lines of low-excitation species \citep{joy1945, herczeg2014}. 
The emission lines are often associated with the accretion process and can show strong variability due to changes in the accretion rate \citep{zsidi2022_cr, zsidi2022_vw, fischer2023}, but they can also come from chromospheric activity \citep{manara2013}.
However, other physical processes, such as stellar wind, turbulence or magnetic activity, can also contribute to variability in the spectral line profiles \citep{kurosawa2011}. 
The photospheric absorption lines often appear to be weak as a result of an additional continuum and line emission that arises from the accretion shock; this phenomenon is the veiling \citep{stempels2003, dodin2012, rei2018}.

The classical T~Tauri star DR~Tau is located in the Taurus star forming region, at a distance of 
$187$\,pc according to Gaia DR3 measurements \citep{gaia2023}.
Based on long-term optical photometric measurements, \cite{chavarria1979} reported a brightness increase by 3 magnitudes between 1970 and 1980. 
During the brightening, the spectrum of the object has changed as well, but it preserved the features characteristic of T~Tauri stars. 
Since this brightening occurred, DR~Tau has been repeatedly observed to examine the system from various aspects.
These studies revealed that the system displays irregular light variations on multiple timescales: quasi-periodic signals in the range of $2-10$\,days were found by, e.g., \cite{bouvier1993, bouvier1995, hessman1997, alencar2001}.
Spectral variability was also detected along with the brightness variations. Strong emission lines, such as the hydrogen Balmer lines, Ca\,II lines, and He\,I lines, show significant variations \citep{alencar2001}.
\cite{banzatti2014} studied how the extreme variability of DR~Tau in the UV-to-mid-infrared wavelength range affects the water vapor in the planet forming region of the disk, and found that water might be unaffected by accretion outbursts of the central objects, which are less long-lived and weaker than typical of the EXor outbursts.
The estimated stellar parameters are $\teff=4100$\,K, $\log g = 3.7$, $\lstar=0.52\,\lsun$, $\rstar=1.46\,\rsun$, and $v \sin i = 5$\,km\,s$^{-1}$ \citep{petrov2011}, and DR~Tau has a spectral type of K5 to M0. 
A recent study with ALMA observations reports $i = 5.^{\circ}4^{+2.1}_{-2.6}$ \citep{long2019} for the angle between the disk's plane and the plane of the sky, i.e., the disk is almost face-on.
\cite{kenyon1994} claimed that the observed periodicity of DR~Tau is due to a hot spot on a rotating star with a magnetic axis inclined to the stellar rotation axis.

\cite{alencar2001} examined a series of spectra of DR Tau obtained over the time span of more than a decade. 
They analyzed the line profile variations with the aim of exploring the emission-line region, and to test the general predictions of the magnetospheric accretion model. 
By analyzing the aforementioned strong emission lines, they found variability on multiple timescales due to both infall and outflow. 
Based on their analysis, they suggest that the system is seen nearly pole-on when considering the rotational pole, and \cite{kenyon1994} estimated $i + \beta = 95^{\circ}$, which implies a misalignment between the rotational and the magnetic axis  by $65^{\circ}$ when considering $i<30^{\circ}$.
\cite{petrov2011} analyzed a series of spectra of DR~Tau and found that the veiling varies by more than 10 times with respect to the stellar continuum intensity and is caused by both a non-photospheric continuum and chromospheric line emission filling in the photospheric absorption lines. 

Here, we show a new detailed study of DR Tau. We present new high cadence Kepler K2 and TESS observations, which allow us to examine the short timescale photometric behavior. With our spectropolarimetric observations, we analyze the line profile variations and we determine the stellar magnetic field characteristics. We complemented our data with unpublished Spitzer infrared observations in order to study the relationship between the stellar and disk variability, and we acquired additional $BVR_CI_C$ band observations in order to examine the color information of the variability seen in the K2 light curve. Furthermore, the TESS monitoring was also complemented with contemporaneous multi-filter optical photometry in the $BgVri$ filters.

\section{Observations and data reduction} \label{sec:observations}
\subsection{Photometric observations} 
The Taurus star forming region was observed by the Kepler Space Telescope in the framework of the K2 extended mission \citep{howell2014},  during its Campaign 13. 
DR~Tau was observed as part of the Guest Observing programme associated to this campaign (proposal ID: GO13005, PI: \'A. K\'osp\'al).
An 80-days-long optical measurement sequence without any photometric filter was obtained between 2017 March 8 and May 27 with both 30-minute and 1-minute cadences (long cadence and short cadence, respectively).
The data reduction and the light curve were obtained following the methods of \cite{kospal2018}.

To acquire color information, we observed DR~Tau contemporaneously with the first few weeks of the K2 observing period with the 60/90/180\,cm Schmidt Telescope of Konkoly Observatory (Hungary) using the 4k$\times$4k FLI camera, providing a field-of-view of nearly $70^\prime\times70^\prime$. 
We performed $BVR_CI_C$ band measurements between 2017 March 8 and April 10, always obtaining $3-4$ frames in each band on each clear night. 
We followed the standard steps of CCD data reduction and we performed aperture photometry.
Finally, we transformed the instrumental magnitudes to the standard system using 12 comparison stars from the URAT catalogue taking also into account the color of the stars in the transformation process. 
The typical photometric errors in the standard system were in the range of $2-3\,\%$.

We complemented the optical photometry with mid-infrared observations at 3.6 and 4.5\,$\upmu$m using the Spitzer Space Telescope (proposal ID: 13159, PI: P. \'Abrah\'am). 
The visibility window of Spitzer allowed us to monitor DR~Tau during the final 11 days of the K2 campaign, between 2017 May 17 and May 28, with an average cadence of 20\,hr. 
We used the IRAC instrument in sub-array mode with 0.1\,s frame time.
The S19.12.0 pipeline at the Spitzer Science Centre produced the corrected basic calibrated data (CBCD).
We carried out aperture photometry on the CBCD frames using an aperture of 3 pixels radius ($3.^{\prime\prime}6$), a sky annulus between 3 and 7 pixels, and aperture correction factors of $1.125$ and $1.120$ at 3.6 and 4.5\,$\upmu$m, respectively (IRAC Handbook\footnote{\url{https://irsa.ipac.caltech.edu/data/SPITZER/docs/irac/iracinstrumenthandbook/27/}}).
As a final step, we computed the average and the rms of the flux densities extracted from the five CBCD images corresponding to the individual dithers to obtain the final photometry and its uncertainty.

We complemented our dataset with observations from 2009. 
Optical $BVR_CI_C$ band monitoring observations were carried out between 2009 October 24 and November 7 using the IAC80 telescope of the Instituto de Astrof\'{i}sica de Canarias with a primary mirror of diameter 82\,cm. 
These data were reduced with the same procedure as the 2017 data from Konkoly Observatory.
$JHK_s$ band measurements were carried out by the Telescopio Carlos S\'{a}nchez at the Teide Observatory during the same period. The data were reduced following a similar procedure as described in \cite{kun2011}, i. e., we derived the magnitudes by aperture photometry, and transformed them into the standard system by comparing them the Two Micron All Sky Survey (2MASS)
magnitudes \citep{cutri2003}.
We also analyzed mid-infrared observations from 2009 October$-$November at 3.6 and 4.5\,$\upmu$m using the Spitzer Space Telescope (proposal ID: 60167, PI: P. \'Abrah\'am). We followed the same data reduction procedure for the 2009 Spitzer data, as we did for the 2017 measurements.

\begin{figure*}[ht!]
   \centering
   \includegraphics[height=0.9\textwidth, angle=90]{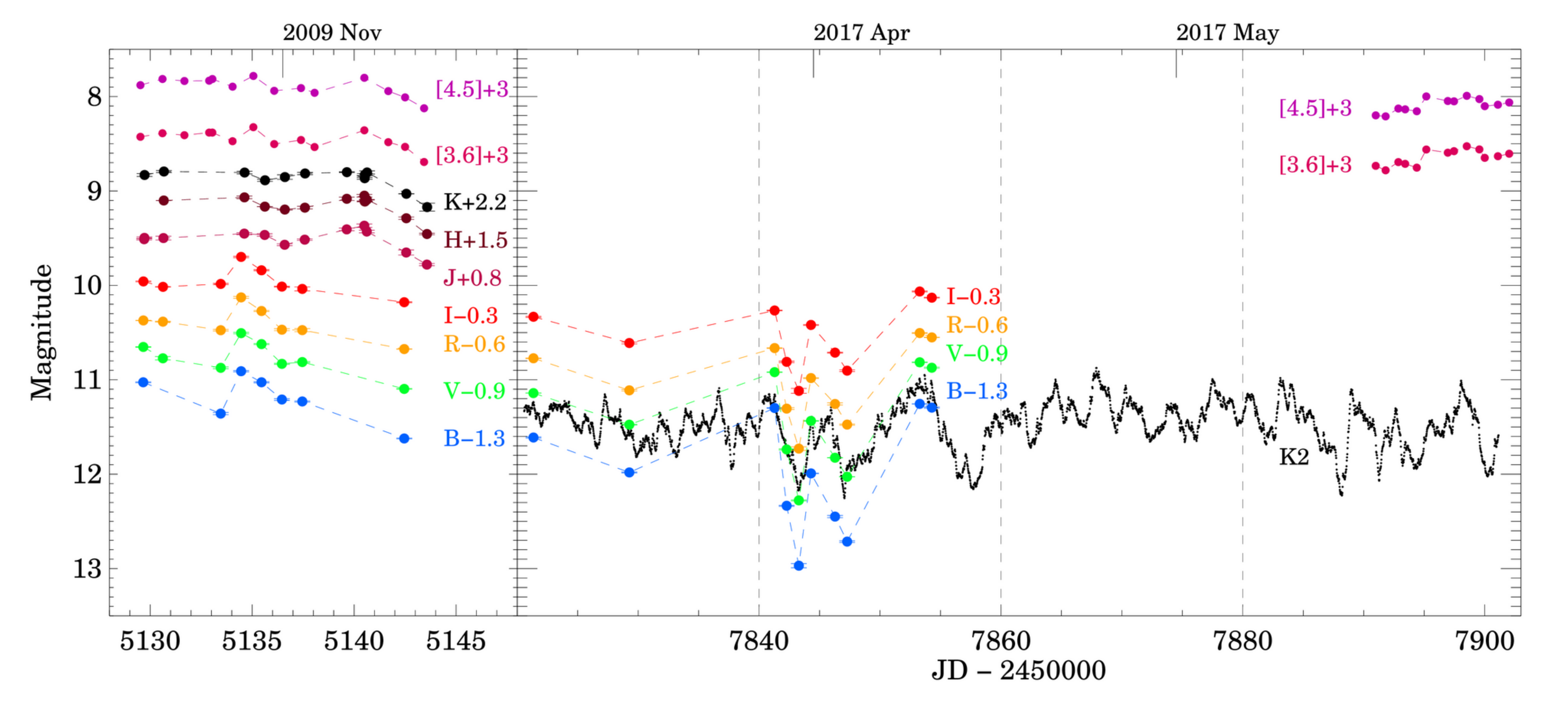}
      \caption{Optical and mid-infrared light curves of DR Tau. The small black dots are the Kepler K2 data with 30 minute cadence, the large filled circles are multiband optical photometry from Konkoly Observatory, and the pink and purple circles represent the Spitzer observations. The different light curves, except from the Kepler K2, are arbitrarily shifted as indicated. The vertical dashed lines divide the sections that we describe in \ref{sect:drtau_lightcurves}. The long tick marks on the top axis indicate the first day of each month.}
         \label{fig:drtau_light_curves}
   \end{figure*}

\begin{figure*}[t!]
   \centering
   \includegraphics[height=0.9\textwidth, angle=90]{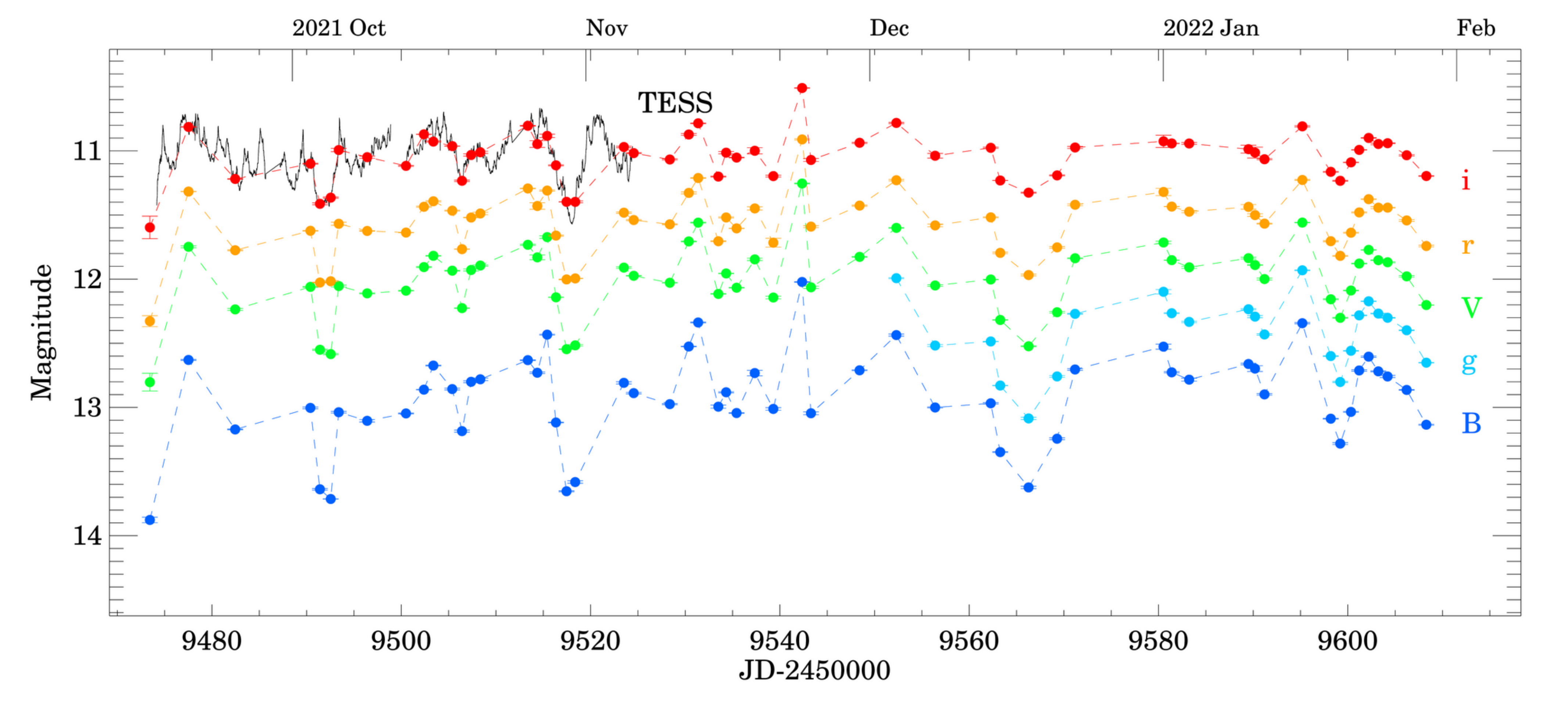}
      \caption{Optical light curves of DR~Tau from 2021-2022. The small black dots are the TESS data obtained with 10 minute cadence, the large filled circles are multifilter optical photometry from Konkoly Observatory obtained with the RC80 telescope in $BgVri$-bands. The long tick marks on the top axis indicate the first day of each month.}
         \label{fig:drtau_light_curves2021}
   \end{figure*}

Furthermore, we analyzed the Transiting Exoplanet Survey Satellite (TESS) observations as well.
DR~Tau was covered by Sectors 43 and 44 of TESS, with 10-minutes cadence observations between 2021 September 16 and November 6.
Imaging data were processed with the same method as described in \cite{pal2020, plachy2021}.
Contemporaneous ground-based multi-filter observations were obtained at the Piszkéstető Mountain Station of Konkoly Observatory with the 80\,cm Ritchey-Chrétien (RC80) telescope in Johnson $BV$ and Sloan $gri$ filters. 
These observations were carried out between 2021 September~15 and 2022 January~28 in the $BVri$-bands, the $g$ filter was added starting from 2021 December~3.
We reduced these data with the same method as the 2017 and 2009 data, however, as the field of view of the RC80 telescope ($18.8^\prime\times18.8^\prime$) is smaller than that of the Schmidt telescope, we could use only four reference stars. 
The results from all optical and infrared photometric observations are shown in Figures \ref{fig:drtau_light_curves} and \ref{fig:drtau_light_curves2021}.

\subsection{Spectropolarimetric observations}

DR Tau was observed with the ESPaDOnS device (proposal ID: 16BF007, PI J. Bouvier) for high resolution spectroscopy and polarimetry mounted on the 3.6 meter Canada-France-Hawaii Telescope (CFHT). 
The resolving power of the instrument in spectropolarimetric mode is about 68\,000 and has a wavelength coverage from 370 to 1050\,nm. 
Nine spectra were obtained between 2016 October 11 and 2016 October 20, however, one spectrum was highly contaminated by the Moon, therefore it is not analyzed here.

The spectropolarimetric mode was used during the observations, i.e., each observation consists in four successive sub-exposure of the same length, between which the polarimeter was rotated to adequate angles for the measurement of circularly polarized spectra. Each observation was reduced using the dedicated UPENA tool running automatically the Libre-ESpRIT pipeline \citep{donati1997}. After completing the usual reduction techniques (bias subtraction, blaze correction, and wavelength calibration), Libre-ESpRIT performs a spectrum optimal extraction, and then combines the four individual spectra to compute the Stokes~$V$ (circularly polarized) and Stokes~$I$ (total intensity) spectra. Finally an automatic normalization is performed on Stokes~$I$. We then carried out telluric correction on all of the spectra using the Molecfit tool\footnote{\url{https://www.eso.org/sci/software/pipelines/skytools/molecfit}}\citep{smette2015, kausch2015}, and we analyzed the corrected spectra.

   \begin{figure*}[ht!]
   \centering
   \includegraphics[width=\textwidth]{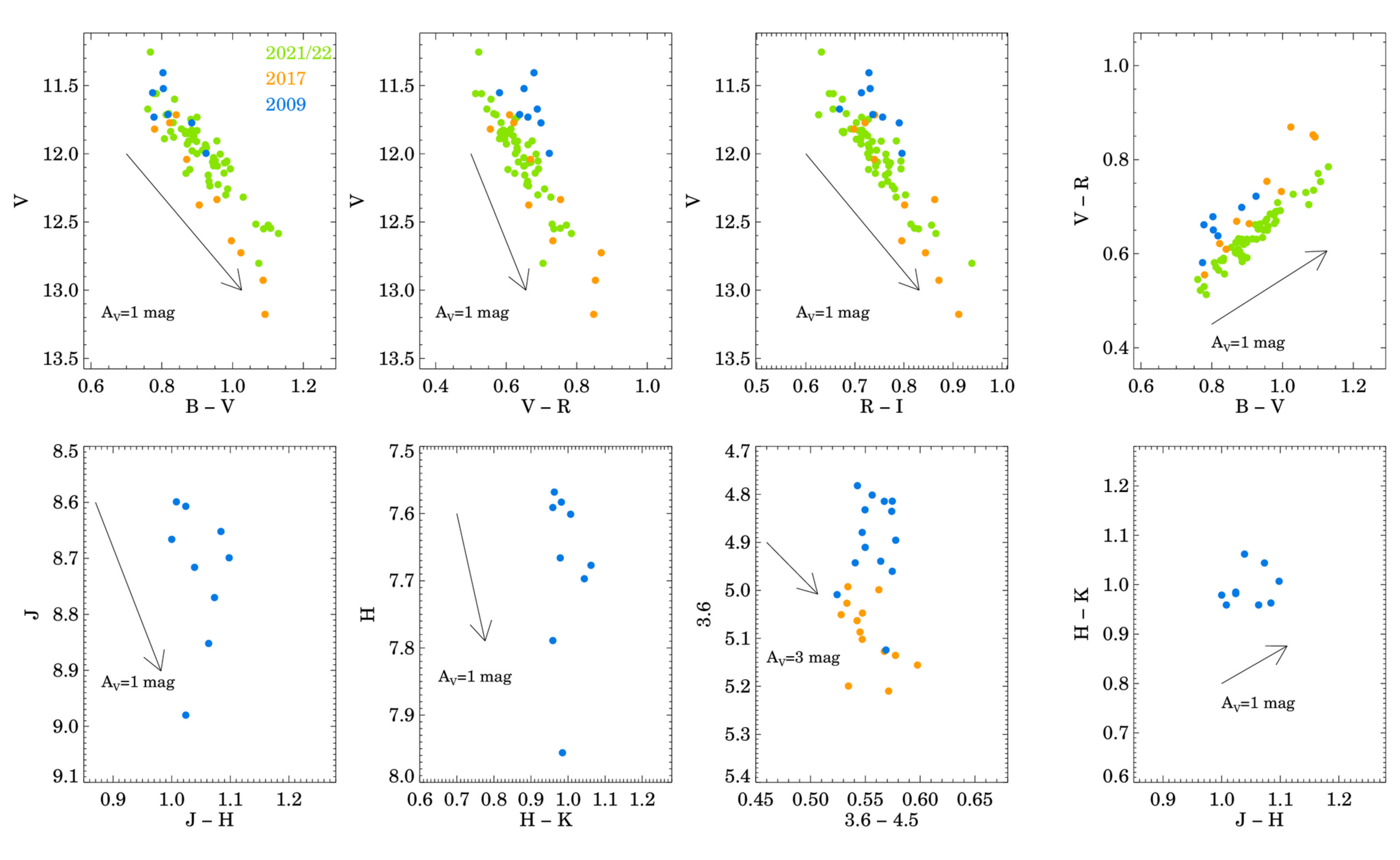}
      \caption{Color-magnitude and color-color diagrams of DR Tau. The filled blue circles show the data obtained in 2009, and the filled orange circles indicate the 2017 data, and the green circles display the 2021/22 data. It must be noted that the 2021/2022 data in the $V-R$ vs. $V$ and the $R-I$ vs. $V$ diagrams were converted from the Sloan $r$ and $i$ filters to the Cousins $R_C$ and $I_C$ filters using the relations by \cite{jordi2006} for better comparability. The black arrows in each panel indicate a change corresponding to $A_V = 1$\,mag except for the $3.6-4.5$ vs. $3.6$ mid-infrared color-magnitude diagram, where it corresponds to $A_V = 3$\,mag.}
         \label{fig:drtau_cmd}
   \end{figure*}

   \begin{figure}[t!]
   \centering
   \includegraphics[width=\columnwidth, angle=0]{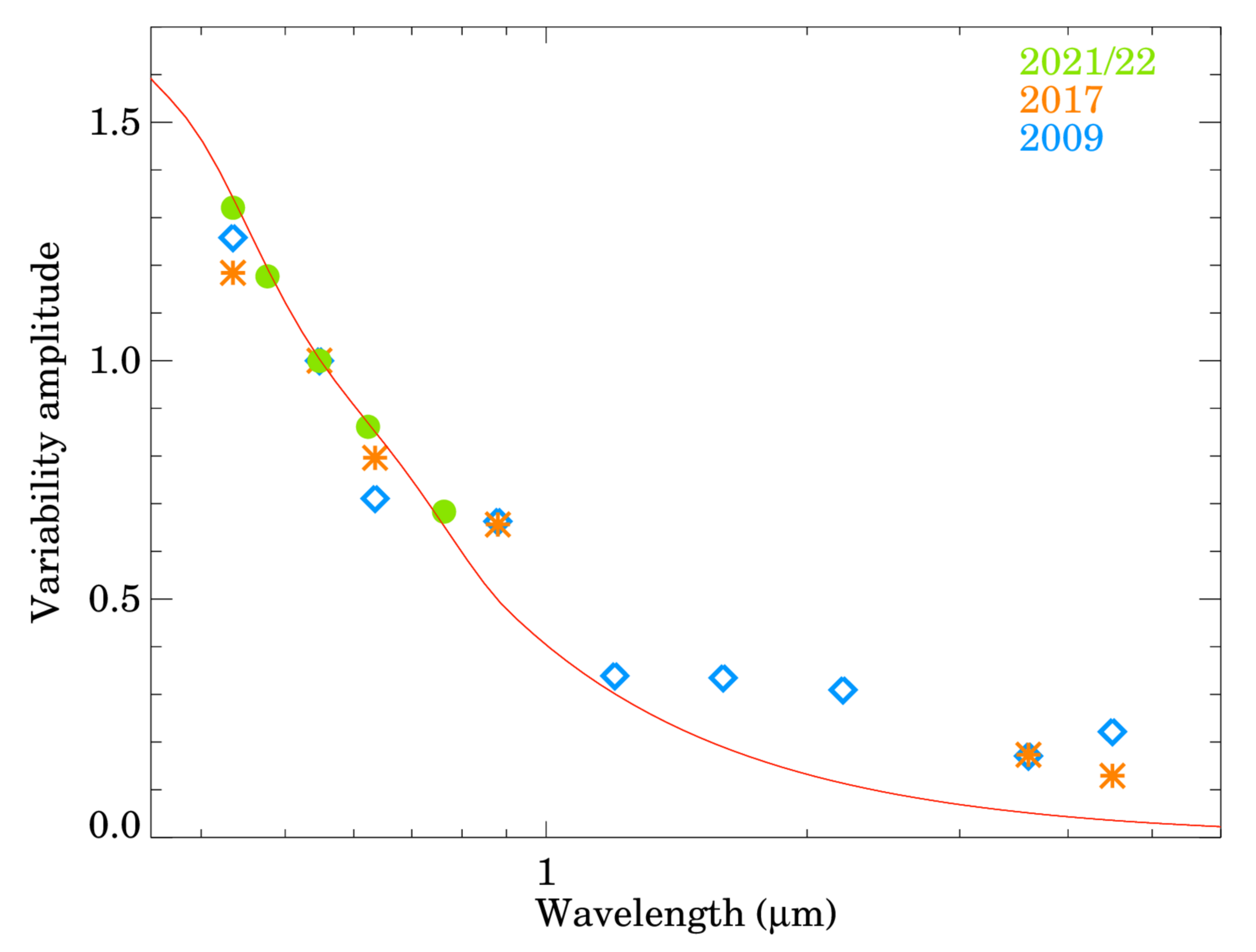}
      \caption{The variability
      amplitudes measured at all wavelengths. The 2021/22 data are shown with green, the 2017 data are plotted with orange, and the 2009 data are indicated with blue symbols. The interstellar reddening law \citep{cardelli1989}, normalized to the $V$-band, is overplotted with a red curve.}
         \label{fig:drtau_var_amp}
   \end{figure}

\section{Results} \label{sect:drtau_results}

In the following, we present the analysis of the photometric and spectroscopic observations.
We examine the variability of the light curves, we look for periodic signals, and study the color variations. 
Moreover, we study the high-resolution spectra, the morphology and the variations of several spectral lines.

\subsection{Light curves} \label{sect:drtau_lightcurves}
 
Figure \ref{fig:drtau_light_curves} shows the light curves of DR~Tau from 2009 and 2017. 
The ground-based observations show variability on daily timescale, whereas the K2 light curve, obtained in 2017, is capable of revealing changes even on a minute timescale.

Considering the temporal characteristics of the data series, the three-month-long K2 observing period can be split into four sections. 
The first section (${\rm JD}-2450000=7820\dots7840$) shows quasi-periodic variations on short timescale with the amplitude of a few tenths of a magnitude, whereas in the second section ($7840\dots7860$), the light curve indicates variations with a peak-to-peak amplitude of 1.4\,mag. 
At the limit of these two sections, 
the flux change occurs rapidly: the brightness decreases by 1.4\,mag in approximately two days. 
The third section of the K2 light curve ($7860\dots7880$) shows again quasi-periodic variations with the amplitude of a few tenths of a magnitude. 
By the end of the K2 observing period ($7880\dots7900$), the brightness variations become again more irregular and have somewhat larger amplitude.

The shapes of the $B$, $V$, $R_C$, and $I_C$ light curves obtained in 2017 are in accordance with the K2 observations, however, the amplitude of the variations is decreasing with increasing wavelength (Fig.~\ref{fig:drtau_light_curves} right panel). 
The comparison between the second section of the K2 light curve and the simultaneous ground based observations shows that the amplitude of the variability is larger in the $B$ and $V$ bands (1.7\,mag and 1.5\,mag, respectively), and smaller in the $R_C$ and the $I_C$ bands (1.2\,mag and 1.1\,mag, respectively) than in the K2 observations. 
The Spitzer observations follow the variations seen at the optical wavelengths but with smaller amplitude relative to the fourth section of the K2 light curve.

The data from 2009 also present variability on daily timescale (Fig.~\ref{fig:drtau_light_curves} left panel). The optical measurements from IAC80 exhibit similar changes, but the amplitude is decreasing with increasing wavelength: the peak-to-peak amplitude in the $B$ band is 0.7\,mag, whereas the maximum amplitude of the brightness changes in the $I$ band is 0.5\,mag. 
The small brightness increase around ${\rm JD} - 2450000=5135$ appears at both optical and infrared wavelengths, but it seems to be slightly delayed on the Spitzer light curve. 
The near-infrared $J$, $H$, $K$ and the Spitzer light curves also have similar shapes and similar peak-to-peak amplitudes ($\sim$0.3\,mag).

The most recent observations of TESS, obtained in 2021, also show high variability. However, the peak-to-peak amplitude of the changes ($\sim$0.86\,mag) is slightly smaller than that of the K2 data (Fig.~\ref{fig:drtau_light_curves2021}).
The stochastic variations are interrupted with a more gradual brightness increase with small amplitude fluctuations around ${\rm JD-2450000}\approx9500$, which lasts for a few days.
Furthermore, a more pronounced dip appears around ${\rm JD-2450000}\approx9518$.
The ground-based multifilter data follow the shape of the TESS light curve and with varying amplitude at different bands. 
This behavior is similar to the one observed in 2017.

The ground-based data from 2021/22 cover a longer period of time (four and a half months) than the TESS observation and the corresponding peak-to-peak amplitudes range from 1.9\,mag in the $B$-band to 1.1\,mag in the $i$-band. The amplitude of the variability is decreasing with increasing wavelength during the entire observing period, and it is more striking when large bursts or dips occur. The light curves are dominated by stochastic changes, with some shorter periods of time when the amplitude of the variability decreases to a few tenth of mag (e.g., from 9570 to 9590 in Fig.\ref{fig:drtau_light_curves2021}).

\subsection{Color variations}
 
The color-magnitude diagrams, including all photometric data from 2009, 2017 and 2021/22, are shown in Fig.~\ref{fig:drtau_cmd}. 
For comparability, we converted the 2021/22 Sloan $r$ and $i$-band data to the Cousins $R_C$ and $I_C$-bands using the empirical relations by \cite{jordi2006}.
The data from 2009 line up with the 2017 measurements in the visible filters, however, the object was slightly brighter during the 2009 observing season, and the 2017 observations cover larger dynamical range.
The 2021/22 observations also line up with the earlier measurements and populate the whole dynamical range, demonstrating the high photometric variability of DR~Tau on monthly timescales. 
When comparing the color changes with the extinction law \citep{cardelli1989} using $R_V = 3.1$, the slope of the 2017  $B-V$ vs. $V$ and the $R-I$ vs. $V$ color-magnitude diagrams differ the most from the slope of the $A_V=1$\,mag extinction arrow: we found steeper slopes in our observations (4.5 and 6.4, respectively) than those predicted by the extinction law.

In contrast, the near-infrared and the mid-infrared color-magnitude diagrams do not show significant color changes. The mid-infrared data (Fig.~\ref{fig:drtau_cmd}, bottom right panel) suggest that the system was observed in a fainter state in 2017 than in 2009, in agreement with the optical data.

The characteristic shape of the light curves are similar at all wavelengths, however, the amplitude of the variability differs. 
The variability amplitude can be determined by plotting the data points measured contemporaneously at different wavelengths as a function of the $V$-band magnitudes.
These plots can be fitted with a line whose slope gives the variability amplitude relative to $V$-band.
We note that for the 2017 data, we used the K2 data as the reference band instead of the $V$-band, since only these are simultaneous with both the ground-based and the Spitzer observations. 
The variability amplitudes calculated this way for the 2017 data were then normalized to the $V$-band for comparability.
The variability amplitudes measured at the different wavelengths are plotted in Fig.~\ref{fig:drtau_var_amp} for all the 2009, 2017, and 2021/22 observing seasons along with the interstellar extinction law using $R_V = 3.1$ \citep{cardelli1989}.

\subsection{Periodogram analysis and rotational period}
 \label{sect:drtau_period}

The photometric data appear to be dominated by stochastic variations.
However, unlike ground-based measurements, the high-cadence observations obtained by the Kepler and the TESS space telescopes are not affected by the alternation of day and night, therefore, they might reveal an underlying periodic variation related to the stellar rotation, which might be hidden for ground-based observations.

\begin{figure}[t!]
   \centering
   \includegraphics[width=\columnwidth]{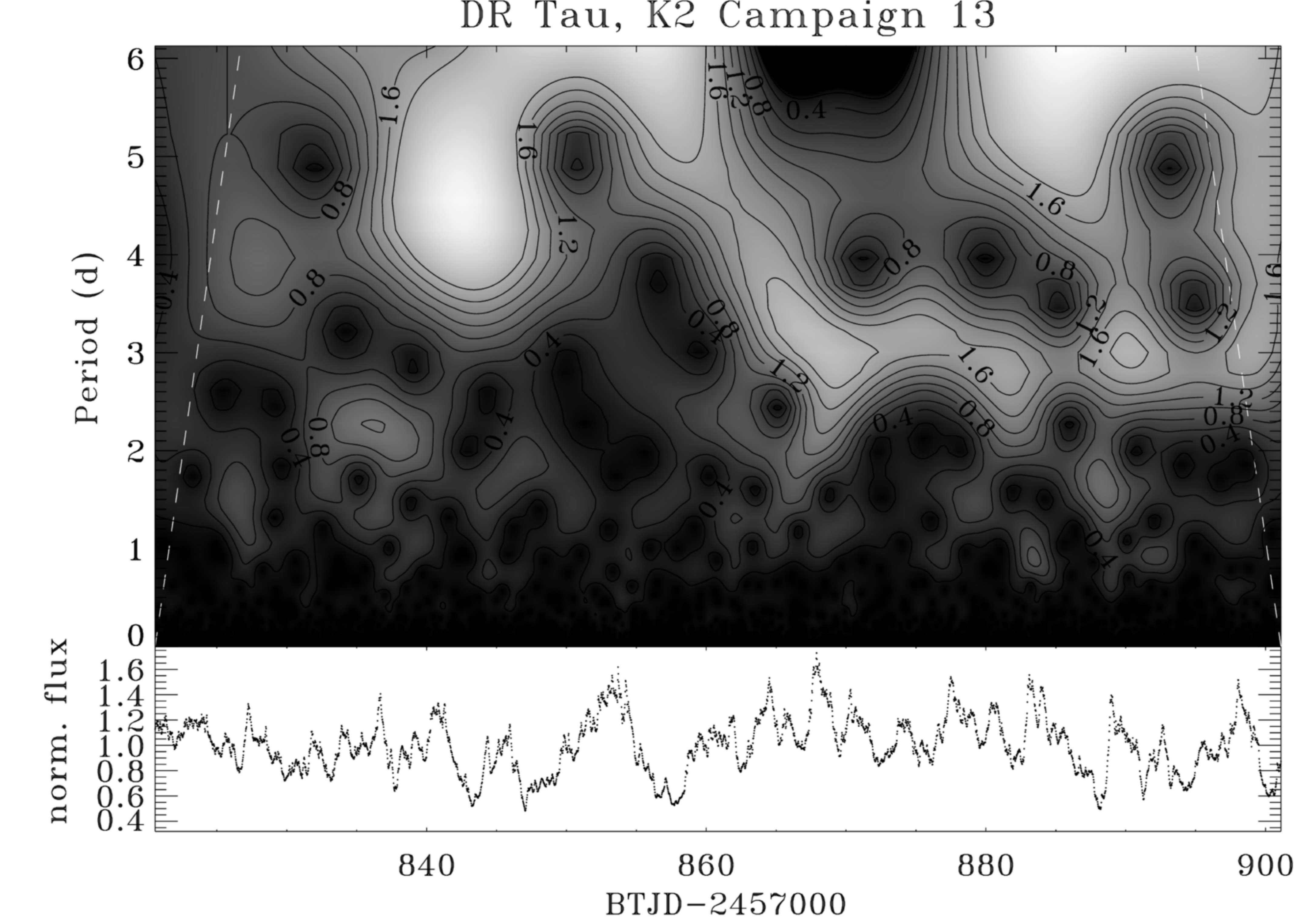}
   \includegraphics[width=\columnwidth]{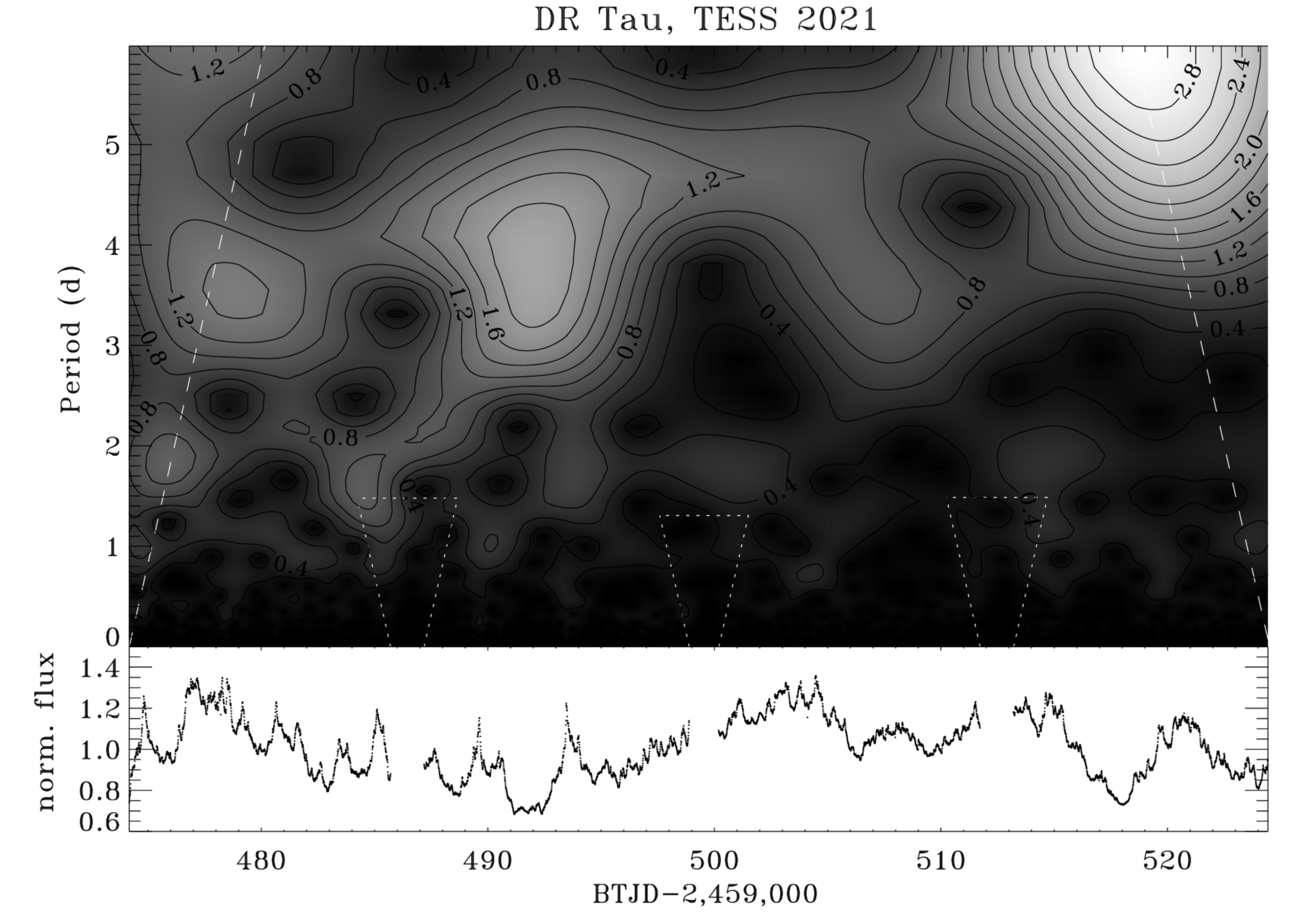}
      \caption{The wavelet analysis for DR Tau. The top panel shows the results from the K2 data, and the bottom panel shows the results from the 2021 TESS data.}
         \label{fig:drtau_wavelet}
   \end{figure}

To find any significant periodicity, we carried out a wavelet analysis first on the K2 long cadence data. 
A wavelet analysis is a period analysis method which yields a two-dimensional spectrum, from which it is possible to study whether a given peak (e.g., the quasi-period or period) is a stable, permanent feature or not.
To obtain a uniform data sampling required for the wavelet analysis, and refill short gaps caused by removal of a few outliers points, we performed a spline interpolation on the 3901 long-cadence data points into a grid of 3943 equally spaced points at 0.5\,h intervals. 
The results obtained with the Morlet-6 wavelet are displayed in the top panel of Fig.~\ref{fig:drtau_wavelet}, which shows the part of the spectrum corresponding to short periods.
The re-sampled light curve is plotted directly below the wavelet spectrum for clarity. 
Significant signals are seen in the trapezoidal region bounded by the two dashed lines.  
The spectrum confirms the primary characteristics of the light curve, the predominantly chaotic noise in its first part, i.e., from the beginning of the observation until $\textrm{BTJD}-2457000=860$.
As time progressed, the light curve itself and the short-periodic part of the wavelet spectrum started to behave in a more regular way: although the light curve and the wavelet spectrum are still affected by higher and lower frequencies, they both started to show fairly constant signal concentrated at almost exactly 3\,days, which could be interpreted as the rotational period of the star. 
An additional pattern appears around $\textrm{BTJD}-2457000=840$ indicating a period around 4-5\,days, however, this pattern is less stable: it firstly increases to 6\,days and later it gradually decreases to the more stable 3\,days period.

We carried out similar analysis for the 2021 TESS data as well, and the resulting wavelet spectrum is shown in the bottom panel of Fig.~\ref{fig:drtau_wavelet}. 
Before computations, the general upward trend seen in the original data was removed by a second order polynomial fit, and then the data were transformed from magnitudes to normalized fluxes.
This operation helps the removal of any long-term trend from the light curve, and it does not affects periods or quasi-periods shorter than the total run length, which is in the interest of this analysis.
The significant results are displayed between the white dashed lines in the Figure, and the dotted white lines delimit the 3 small areas affected by interruptions in data acquisition.
The results confirm the predominantly stochastic behavior of the system, however, the wavelet spectrum hints at a quasi periodic signal at $\sim$3-4\,days in the first half of the observing period.

Earlier studies also found similar periodic signals in photometric and spectroscopic datasets. \cite{bouvier1993} obtained UBVRI photometric data under the COYOTES program and applied the string-length method \citep{dworetsky1983} and periodogram analysis \citep{horne1986}. These methods revealed significant periods at 2.8 days and 7.3 days, however, none of these periods could be unambiguously assigned to the stellar rotation as both could have been consistent with the the measured $v \sin i$ of the star \citep[$v \sin i \leq 10$km$/$s,][]{hartmann1989}. \cite{bouvier1995} obtained additional 40-days-long UBVRI photometric data as part of the COYOTES II program. Using the same methods as in \cite{bouvier1993}, a period of 7.3\,days was reported and a period of 9\,days was also found with larger uncertainty. In contrast, \cite{hessman1997} reported a period around 4.5-5\,days based on the analysis of the equivalent widths of strong emission lines, such as H$\alpha$ and Ca\,II, in their dataset of 121 echelle spectra. \cite{alencar2001} obtained 103 spectra and carried out periodogram analysis on the intensity variations of different line. They analyzed the data on a month-to-month basis and found periods of 4 to 9 days, however, they could not recover a single period that describes the variations of all the lines or the whole data set.

The results from the wavelet analysis are consistent with the conclusions in previous studies: we also found periodic signals in the range of $2-9$\,days. 
However, the most stable period in the K2 light curve is the one at almost exactly 3\,days. 
This period is consistent with the typical rotational period of T~Tauri stars \citep{broeg2006}, therefore, we will use this as the rotational period of DR Tau in this study. 
We did not attempt to carry out similar analysis for the ground-based data, as the cadence of those are much sparser than the space-borne data.
Furthermore, if the rotational period of DR~Tau is indeed around exactly 3\,days, it would be difficult to identify it from ground-based observations due to the windowing effect caused by the day-night variations.
We also note that due to the low inclination of the system, it is hard to measure the rotational period -- this is seen in other systems as well, such as TW~Hya \citep{herczeg2023} or RU~Lup \citep{siwak2016}.

\subsection{Structure function analysis}

When a light curve is not dominated by periodic signal, it becomes challenging to characterize its variability. 
The structure function is a powerful tool to overcome this difficulty as it is sensitive to both periodic and aperiodic variability patters, and it has been widely used in several fields of astrophysics \citep[e.g., AGN or quasar studies,][]{hughes1992} to analyze light curve variability. Recently, \cite{roelens2017} applied the variogram method --  which corresponds to a first-order structure function --  to study variability in Gaia data. 
The concept of the structure function lies in calculating the average amplitude of variability for all pairs of data points separated by a chosen time lag, and then repeating this procedure for all time lags that can possibly occur in our dataset between a minimum $\tau_{\textrm{min}}$ (which is the
minimum timescale of variability that can occur in the light curve) and a maximum $\tau_{\textrm{max}}$ (which is the maximum timescale of variability that can occur light curve). As the dataset we work with might not be evenly sampled, the time lags are defined in the following way in order to allow for a tolerance $\epsilon_{\tau}=0.03$ in grouping the pairs: $|t_i-t_j| = \tau \pm \epsilon_{\tau}$.

Here, we are following the methods by which the structure function was introduced to the field of star formation \citep[e.g,][]{sergison2020, venuti2021, lakeland2022}, and calculate and plot the square root of the structure function, as we aim to compare our results with previous works in star formation. Therefore, we use the following equation \citep{venuti2021} to calculate the structure function of the original high-cadence K2 and TESS data:
\begin{equation}
    \mathcal{SF}(\tau) = \sqrt{\frac{1}{N(\tau)}\sum_{j>i} (f_i - f_j)^2}
\end{equation}

\noindent where $f_i$ is the flux measured at time $t_i$, and  $N(\tau)$ is the number of pairs of light curve points $(i, j)$ separated in time by a lag of $\tau \pm \epsilon_{\tau}$.

We selected our minimum timescale to correspond to twice the light-curve cadence (i.e., $\tau_{\textrm{min}}\sim0.04$\,d), 
and our maximum timescale as half the total light-curve span (i.e., $\tau_{\textrm{max}}\sim40$\,d for the K2 data, and $\tau_{\textrm{max}}\sim25$\,d for the TESS data), 
and we sampled this interval in 100 bins in logarithmic space. In general, the resulting structure function can be divided to three main regions: (1) at very short time lags, the structure function is expected to be flat and probes the intrinsic noise of the observations; (2) as the time lag increases, the structure function can be characterized by a power-law; (3) at the longest time lags, the structure function flattens out or exhibits a slowly varying trend. The point where the structure function transitions from region (2) to region (3) is the characteristic timescale of the variability in the dataset. If the data is periodic, the structure function oscillates after this point. In case of DR~Tau, the transition occurs around $\tau\sim2-3$\,days -- which is close to the period that the wavelet analysis revealed as well -- but the structure function does not start clearly oscillating after this point, suggesting that the light curve is dominated by aperiodic or stochastic variations.

\cite{venuti2021} obtained the structure function for a large sample of K2 light curves of young stars, and divided them into different categories based on their light curve characteristics. The majority of their sample showed periodic, quasi-periodic or multiperiodic signal. When comparing the structure function of DR~Tau to their results, we found that it is most similar to the bursters or the stochastic stars with no noise dominated region.

   \begin{figure}[t!]
   \centering
   \includegraphics[width=\columnwidth]{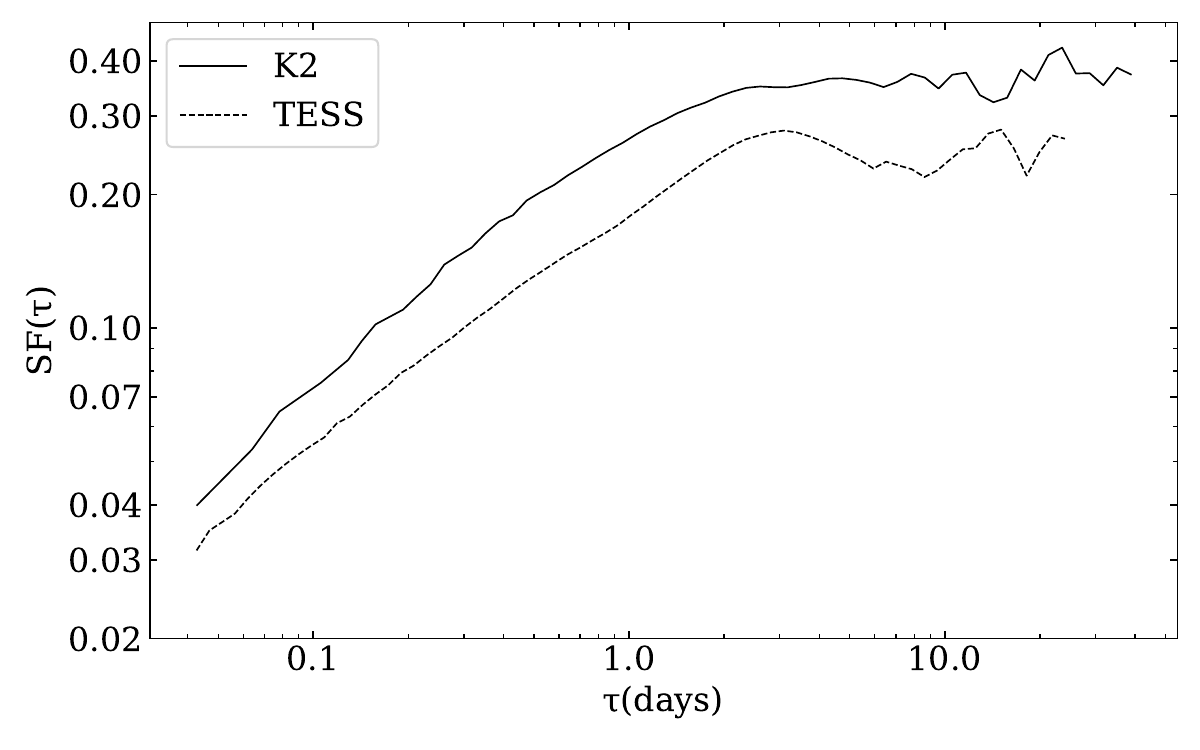}
      \caption{The structure functions of DR~Tau using the K2 and the TESS data.}
         \label{fig:drtau_variogram}
   \end{figure}

\subsection{Stellar properties and veiling}\label{sec:drtau_stellar_prop_veiling}

We derived the stellar parameters by fitting synthetic spectra to the observations using $\chi^2$ minimization.
Synthetic spectra were calculated using the \textsc{zeeman} spectrum synthesis code \citep{landstreet1988, wade2001}, which solves the polarized radiative transfer equations assuming local thermodynamic equilibrium (LTE).
The code uses a Levenberg–Marquardt $\chi^2$ minimization procedure \citep{folsom2012} to obtain the best-fitting parameters and determines the effective temperature, the radial velocity, $\log g$, $v \sin i$, microturbulent velocity, macroturbulent velocity and the veiling using line lists from the Vienna Atomic Line Database \citep[VALD\footnote{\url{http://vald.astro.uu.se/}},][]{ryabchikova1997} and MARCS model atmospheres \citep{gustafsson2008}.

\begin{figure}[t!]
   \centering
   \includegraphics[height=\columnwidth, angle=90]{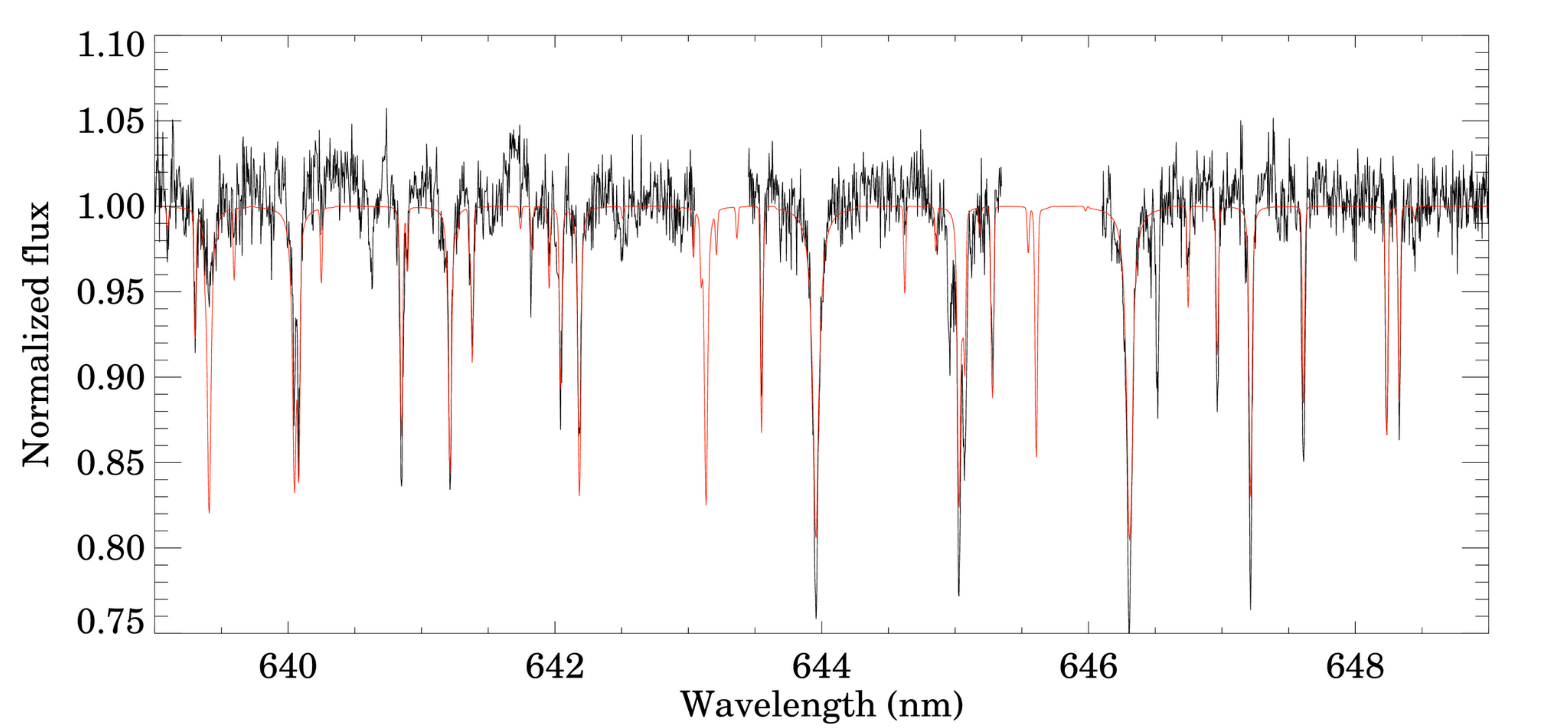}
      \caption{An example of the fit results. The black line shows the observed spectrum, and the red line indicates the model spectrum. The gaps in the observed spectrum are emission line regions, which were excluded during the fitting procedure.}
         \label{fig:drtau_fit_window}
   \end{figure}

As the absorption lines in the spectra of young stars can be affected by the veiling, we selected the spectrum with the lowest veiling ($\textrm{JD}=2457676.02$), and fit a synthetic spectrum using the \textsc{zeeman} code.
Since DR~Tau is a cool star, the density of the absorption lines becomes very high in the blue part of the spectrum, therefore, it is preferable to complete the fit on the red part.
Furthermore, we excluded those regions of the spectrum which were contaminated by emission lines.
First, we normalized the spectrum to the continuum level by fitting a low-order polynomial to each selected region. 
We performed the fit on 22 independent spectral windows between 547\,nm and 820\,nm covering 7 to 10\,nm wide wavelength ranges. 
In order to perform the fit, we fixed $\log g$ to 3.7 based on \cite{petrov2011}, assumed the macroturbulent velocity to be 2\,km\,s$^{-1}$ and we also assumed solar metallicity as it is expected in the case of a low mass pre-main sequence star in our Galaxy. 
We retrieved the atomic line data from the VALD data base by using the ‘extract stellar’ request, and we obtained a line list for stellar parameters approximately matching the estimated temperature of DR~Tau ($\teff=4100$\,K, $\log g=3.7$).
After obtaining the effective temperature, the radial velocity, the microturbulent velocity, $v \sin i$, and the veiling for each spectral window (see an example of a fit in Fig.~\ref{fig:drtau_fit_window}), we calculated the mean value for every parameter, and we excluded the 1-$\sigma$ outliers.
When obtaining the final parameters, we incorporated the mean values yielded by the independent windows as the final best-fit values, while the standard deviations of the individual fits were the estimators for the uncertainty.
The final values we obtained are the following: $\teff=4317 \pm 159$\,K, $v_r=22.97 \pm 0.1$\,km\,s$^{-1}$, veiling of $4.0 \pm 1.1$, $v \sin i=3.0 \pm 0.9$\,km\,s$^{-1}$, and $v_{\rm {mic}} = 2.7 \pm 0.5$\,km\,s$^{-1}$.
These values are in good agreement with the system parameters presented in the literature \citep[e.g.,][]{petrov2011}.

   \begin{figure}[t!]
   \centering
   \includegraphics[height=\columnwidth, angle=90]{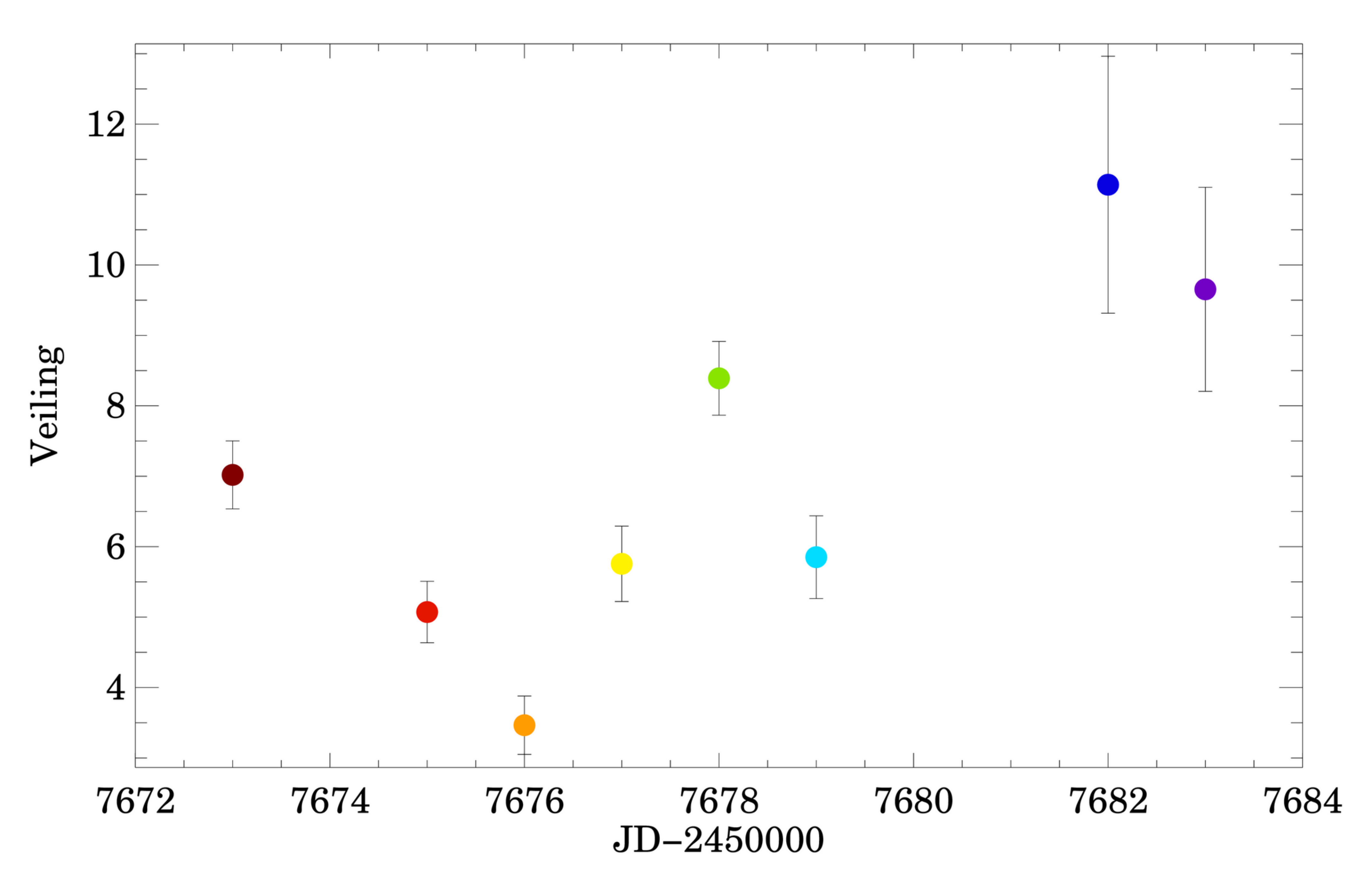}
      \caption{Veiling measurements of DR~Tau. The same colors were used to indicate each epoch as in Fig.~\ref{fig:drtau_var_profile} for easier comparability in later Sections.}
         \label{fig:drtau_veiling}
   \end{figure}

As the veiling is expected to vary on short timescales in case of highly accreting stars, we performed the fit for each observation obtained on the individual nights on the same spectral windows in order to determine the temporal variations of the veiling.
For this analysis, all fitted parameters were fixed to the values obtained previously, except for the veiling, as it is not expected to vary on daily timescale.
The veiling is expected to be wavelength dependent: higher values are expected at blue part of the spectrum and lower values are anticipated at the red part of the spectrum \citep[e.g.,][]{dodin2013} and the wavelength dependence is stronger at shorter wavelengths. Therefore, we used the spectral windows with $\lambda>700$\,nm in order to obtain the veiling.
For determining the best-fit value, we obtained the mean veiling for each night while rejecting the 1-$\sigma$ outliers. The final results are shown in Fig.~\ref{fig:drtau_veiling}.

\subsection{Overview of the spectra}

\begin{figure*}[t!]
   \centering
   \includegraphics[width=1.0\textwidth]{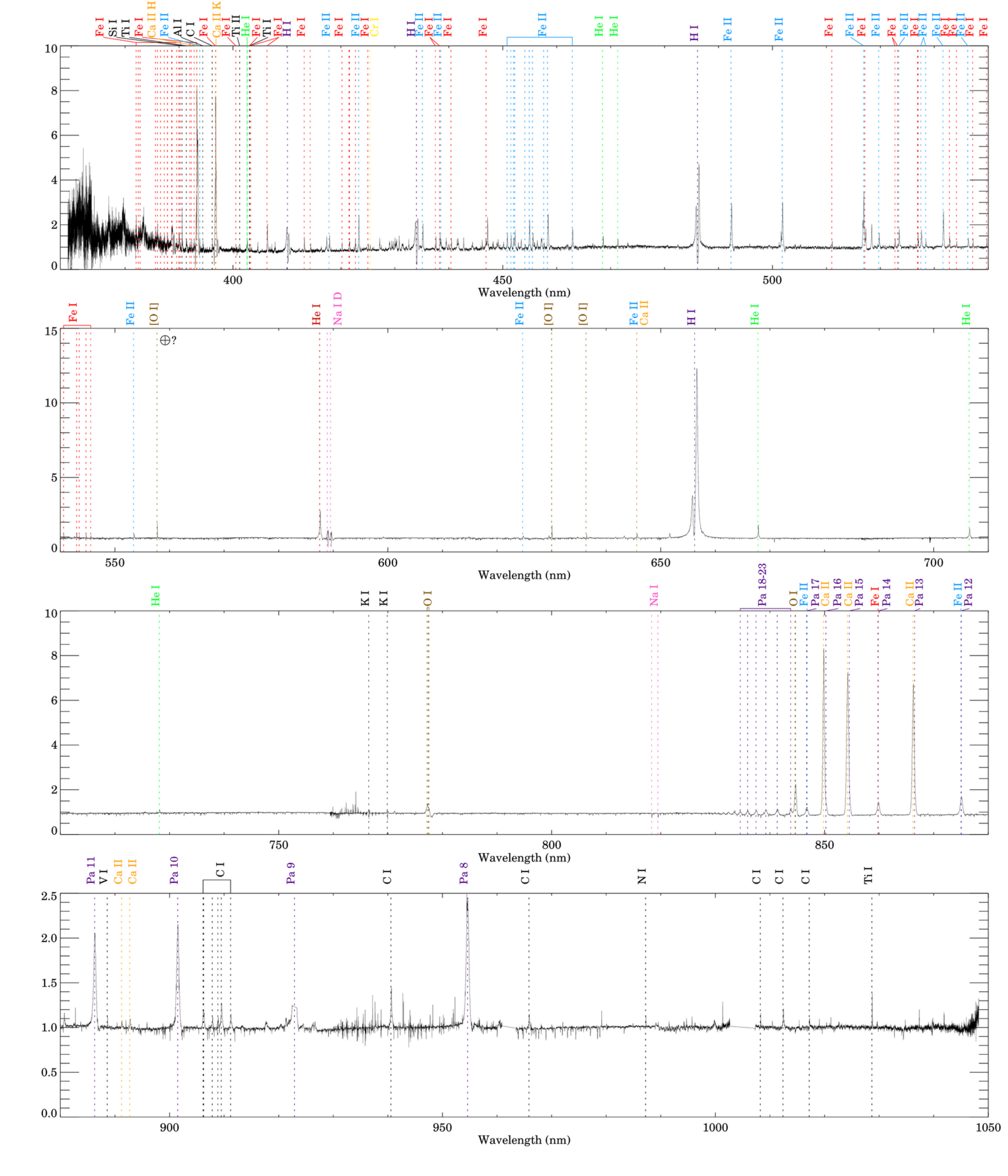}
      \caption{The average ESPaDOnS spectrum is shown with black lines. The spectral emission lines are identified with vertical dashed lines, and the same species are indicated with the same color.}
         \label{fig:drtau_line_inventory}
   \end{figure*}

The spectra of DR Tau exhibit several emission lines. 
Fig.~\ref{fig:drtau_line_inventory} shows the average of the eight normalized ESPaDOnS spectra and we identified emission lines with vertical dashed lines. 
The strongest detected emission lines are the H$\alpha$, H$\beta$, Ca\,II H and K lines and the Ca\,II infrared triplet. 
However, several other species were also found, such as He~I, He\,II, Fe\,I, Fe\,II, and Na\,I lines. 
We also identified numerous other members of the hydrogen Balmer and Paschen series, however, they exhibit a weaker profile.
The only detected forbidden emission lines are the [O\,I], the [N\,II], and the [S\,II] lines, which are also detected as weaker emission lines.
The strong emission lines, such as Fe\,I, Fe\,II, or He\,I, often have a narrow and a broad component. 
As the lines get stronger, these components become more and more distinguishable. 
The hydrogen Balmer lines often exhibit more components, with up to four components in case of the H$\alpha$ line.
These are further discussed in Sect.~\ref{subsect:drtau_line_variability}.

\subsection{Emission line morphology and their variability} \label{subsect:drtau_line_variability}

\begin{figure*}[h!]
   \centering
   \includegraphics[width=\textwidth]{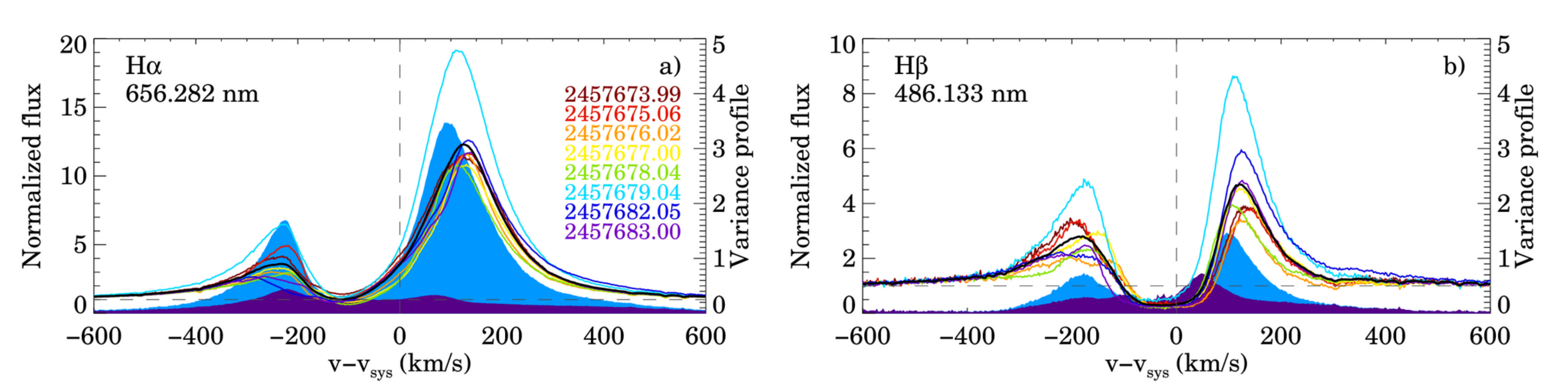}
      \caption{Variance profiles of emission lines. The profiles with different colors indicate observations made on the Julian days marked on panel a). The thick black curve shows the mean line profile, the light blue area is the variance profile, and the dark blue shaded area is the normalized variance profile. Additional panels of this Figure with further emission lines are shown in Appendix~\ref{sect:appendix_A}.}
         \label{fig:drtau_var_profile_main}
   \end{figure*}

The spectra of DR~Tau reveal diverse line profiles, and the emission lines show variations on daily timescale. 
We show a selection of emission lines in Figs.~\ref{fig:drtau_var_profile_main} and \ref{fig:drtau_var_profile}. 
The mean line profiles are indicated with thick black curves, and the individual observations are marked with the colored curves.
We calculated the variance profiles for each line, as defined by \cite{johns1995} and indicated them with the light blue shaded area.
Furthermore, we displayed the normalized variance profile with the dark blue shaded area, which is the variance profile divided by the average line profile, for better comparability of the results in Sect.~\ref{sect:drtau_shortterm_longterm} with other works.

The H$\alpha$ lines exhibit complex, asymmetric line profiles with multiple Gaussian components and with high variability (Fig.~\ref{fig:drtau_var_profile_main}\,$a$). 
A stronger redshifted peak appears around 150\,km/s and a weaker blueshifted peak is present around $-250$\,km/s. 
The blueshifted component changes its amplitude and shape during the observing season: in some cases, it shows a nearly Gaussian shape, whereas some observations reveal a more truncated profile (e.g., ${\rm JD}=2457682$, indicated with dark blue). 
On the contrary, only the amplitude of the redshifted peak changes conspicuously while the location of the peak fluctuates only slightly and no extreme variation was observed in the shape of this component.
The (normalized) variance profiles also suggest strong variability of the two emission peaks.
The H$\alpha$ line is expected to form not only in the accretion funnel, but also in the accretion spot, in the stellar winds and in the disk wind as well.
For this reason, the observed variability might arise from multiple physical mechanisms.

\begin{figure*}[h!]
   \centering
   \includegraphics[width=\textwidth]{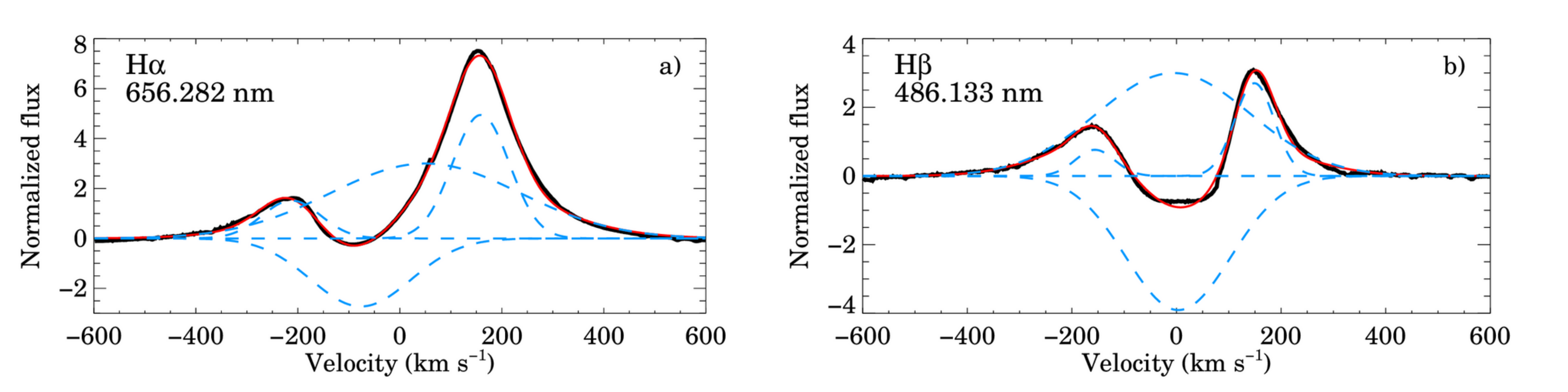}
      \caption{Spectral decomposition. The thick black lines are the observed spectra, the blue dashed curves show the individual components, and the red curve indicates the total fit. Additional panels of this Figure with further emission lines are shown in Appendix~\ref{sect:appendix_B}.}
         \label{fig:emission_fitteles_main}
   \end{figure*}

In order to examine their contribution to the overall line profile, we decomposed the H$\alpha$ line to multiple Gaussian components.
\cite{alencar2001} proposed a decomposition using three Gaussian components: (1) a strong redshifted emission peak originating from the accretion flow; (2) a blueshifted absorption component due to wind; and (3) a relatively low amplitude component with large FWHM centered at the rest velocity. They also reported that in their sample of 103 spectra, they found cases when a blue shoulder appeared in the Balmer lines which made it necessary to introduce an additional blue emission component, however, it appeared only in 15\% of their spectra. 
Our ESPaDOnS observations show profiles with two distinctive emission peaks and a slightly blueshifted absorption, therefore, we fitted the H$\alpha$ with four Gaussian components, including the low amplitude Gaussian with large FWHM to recover the broad line wings as well. 
We carried out this decomposition for each night, and we show the results for the average line profile in Fig.~\ref{fig:emission_fitteles_main}\,$a$.
It must be noted that other line decomposition is possible as well \citep{alencar2001}: a strong and broad emission component due to accretion centered around the systemic velocity can be applied along with redshifted and blueshifted absorption components.
However, this decomposition is less favorable, as it requires an emission component almost twice as strong as the peak of the observed profile.
The H$\beta$ lines (Fig.~\ref{fig:drtau_var_profile_main}\,$b$) resemble the behavior of the H$\alpha$ lines, and can be similarly decomposed into four Gaussians, as shown in Fig.~\ref{fig:emission_fitteles_main}\,$b$.

All three members of the Ca~II infrared triplet appear in strong emission in all the spectra (Fig.~\ref{fig:drtau_var_profile}\,$c$).
The high resolution spectra allow distinguishing a narrow component (NC) centered at the systemic velocity, and a broad component (BC) extending to almost $\pm$200\,km/s.
The narrow component shows less variability, and is considered to arise from the post-shock region. The amplitude and the center of the broad component fluctuate during the observing season, causing more apparent changes in the shape of the line. 
This component is associated with the accretion funnel.
It should also be noted that the red wings of these lines are blended with Paschen lines (the Pa 16 is blended with the 849.8\,nm CaI\,I line, Pa\,15 with the 854.2\,nm Ca\,II line and Pa\,14 with the 866.2\,nm Ca\,II line), appearing as a redshifted bump around 200\,km/s (see Fig.~\ref{fig:drtau_var_profile}\,$c$), which also contribute to the observed variability. When decomposing the Ca~II lines (Fig.~\ref{fig:emission_fitteles}\,$c$), we used three components on account of the blended Paschen line: a narrow component, a broad component and the Paschen line itself.

Broad and narrow components can be clearly distinguished in the case of the He~I lines as well, however, as opposed to the Ca~II triplet, the He~I lines exhibit more asymmetric profile (Fig.~\ref{fig:drtau_var_profile}\,$e$). 
Interestingly, the helium line shows the strongest peak in the emission at a different epoch than the Balmer or the Ca~II lines.
However, the broad component of the He~I line appears to vary in sync with the other accretion tracers, suggesting that the different components of this line originate from different regions of the accretion flow.
The narrow component is centered on the systemic velocity, shows notable amplitude variations, and is considered to form in the decelerating post-shock gas at the footpoint of the accretion column. 
On the other hand, the broad component has a flatter profile, shows a blueward asymmetry and exceeds $\pm$200\,km/s. 
This component is associated with the funnel flow, however, it was suggested that the BC itself is a composite including an origin in outflowing gas or hot wind likely powered by interactions between the
magnetosphere and the inner disc \citep{beristain2001}.
We decomposed the He~I line into two components and fitted them with Gaussian functions for each night.
The He~II lines also show an asymmetric profile (Fig.~\ref{fig:drtau_var_profile}\,$f$), however, the presence of a narrow and a broad component is not as evident as for the He~I line.

O~I lines were also detected in the spectra. 
Some of them display complex line profiles due to blend with neighbouring lines, however, the more distinct lines have one component centered at the systemic velocity and show small amplitude variations on daily timescale (Fig.~\ref{fig:drtau_var_profile}\,$g$). 
On the other hand, the forbidden [O~I] lines have a very narrow component and a small amplitude broad component (Fig.~\ref{fig:drtau_var_profile}\,$h$). This emission line mainly shows amplitude variations over the observing period, it does not exhibit as significant morphological changes as other emission lines.

Several Fe~II and Fe~I lines were found in emission in the spectra (Fig.~\ref{fig:drtau_var_profile}\,$i$ and $j$). 
They can be decomposed into a narrow and a broad components, and the Fe~I lines exhibit stronger narrow component. 
The numerous Paschen lines have one component centered at the systemic velocity, however, in some cases they show asymmetric profiles.

\subsection{Correlation matrices}

To examine how the variability of two sets of simultaneously observed spectral lines are related to each other, we employed the method of correlation matrices \citep{johns1995}. 
As some of the emission lines in the spectra of DR~Tau show complex profiles with multiple components, this analysis allows the examination of the relationship between different components, i.e. we can verify whether they were formed in the same region or not. 
Such analysis is therefore yielding hints about the underlying physical processes. 
We calculated autocorrelation matrices for the main accretion tracers, such as the H$\alpha$, H$\beta$, He~I, and Ca~II lines and we studied their cross-correlations as well (Figures~\ref{fig:corrmtx} and \ref{fig:corrmtx2}).
For comparison with long-term observations, we also recreated the correlation and autocorrelation matrices of \cite{alencar2001}.
We show the results from the current study in the left columns of Figures~\ref{fig:corrmtx} and \ref{fig:corrmtx2}, and the results from \cite{alencar2001} in the right columns.

The autocorrelation matrix for the H$\alpha$ line shows significant correlation along most of the line profile with the exception of the low velocity wing of the blueshifted emission component, which shows no correlation with the high velocity wing of the redshifted emission component. 
The lack of correlation supports the idea that these parts of the line profiles are shaped by different mechanisms: the former is expected to be truncated by absorption due to wind, whereas the latter is formed in the accretion funnel.
Similarly to \cite{alencar2001}, strong correlation is seen in the ESPaDOnS spectra between the redshifted emission component and the high velocity wing of the blueshifted emission component, suggesting similar region of origin. 
The H$\beta$ line shows correlation along the diagonal, similarly to \cite{alencar2001}, and also between the redshifted and the blueshifted emission peak. 
However, the low velocity wing of the blueshifted emission component shows slight anticorrelation with the rest of the line.
This resembles the behavior of the H$\alpha$ line.

The He\,I line indicates strong correlation along the broad component, in agreement with \cite{alencar2001}. 
Furthermore, we were also able to examine the behavior of the NC and the BC thanks to the high-resolution of the ESPaDOnS spectra.
We found no correlation between the narrow and the broad components, which supports the concept of them forming in different regions. 
The autocorrelation matrix of the Ca\,II line shows a slightly different structure than the one in \cite{alencar2001}, however, this might be due to the more clearly detected blend of the Paschen line with the red wing of the Ca\,II line in the ESPaDOnS spectra.
The region, which is affected by this blend in the red wing, shows slight to no correlation with the center and the blue wing of the Ca\,II line (Fig.~\ref{fig:corrmtx2} top left).
When examining the individual Ca\,II lines, a BC and a NC could be clearly distinguished.
They only show a weak correlation, and different regions of origin are not as evident as in the case of He\,I.
This might be due to the weaker NC of the Ca\,II, or the contribution of the Paschen line that smears out the differences.

We also calculated the cross-correlation matrices in order to compare the variations between different lines. 
The variations of the H$\alpha$ line mostly correlate with variations of the H$\beta$ line, with the exception of the low velocity region of the blueshifted component in the H$\beta$ line. 
The broad component of the He\,I line shows correlation with the Ca\,II line. 
The narrow component of the He\,I line indicates anticorrelation, which suggests that this component was formed in a different region. The comparison of the short-term and the long-term data is further discussed in Sect.~\ref{sect:drtau_shortterm_longterm}.

\section{Spectropolarimetric measurements}
\begin{figure}[t!]
   \centering
    \includegraphics[width=0.48\columnwidth]{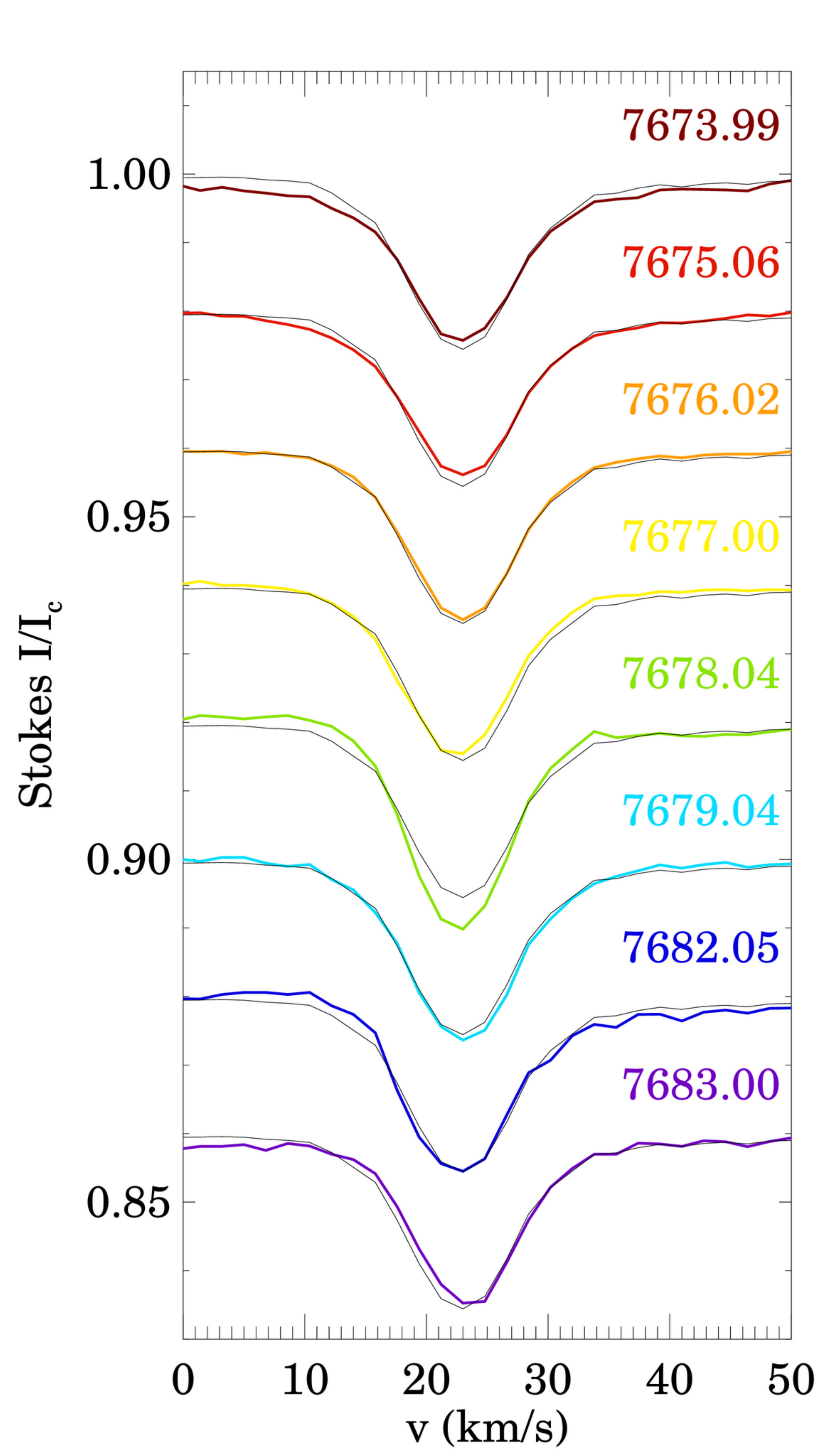}
    \includegraphics[width=0.48\columnwidth]{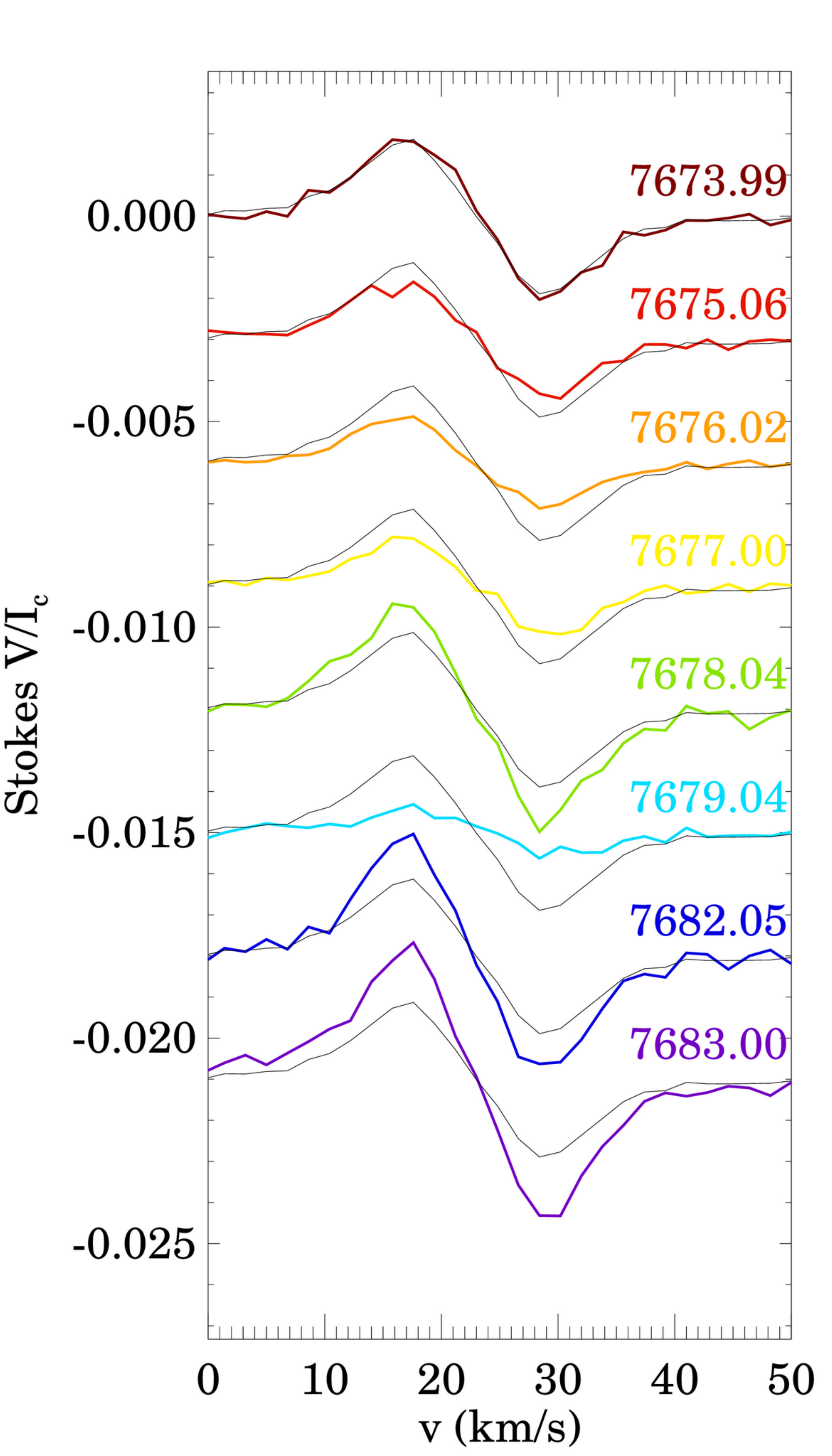}
    \includegraphics[height=\columnwidth, angle=90]{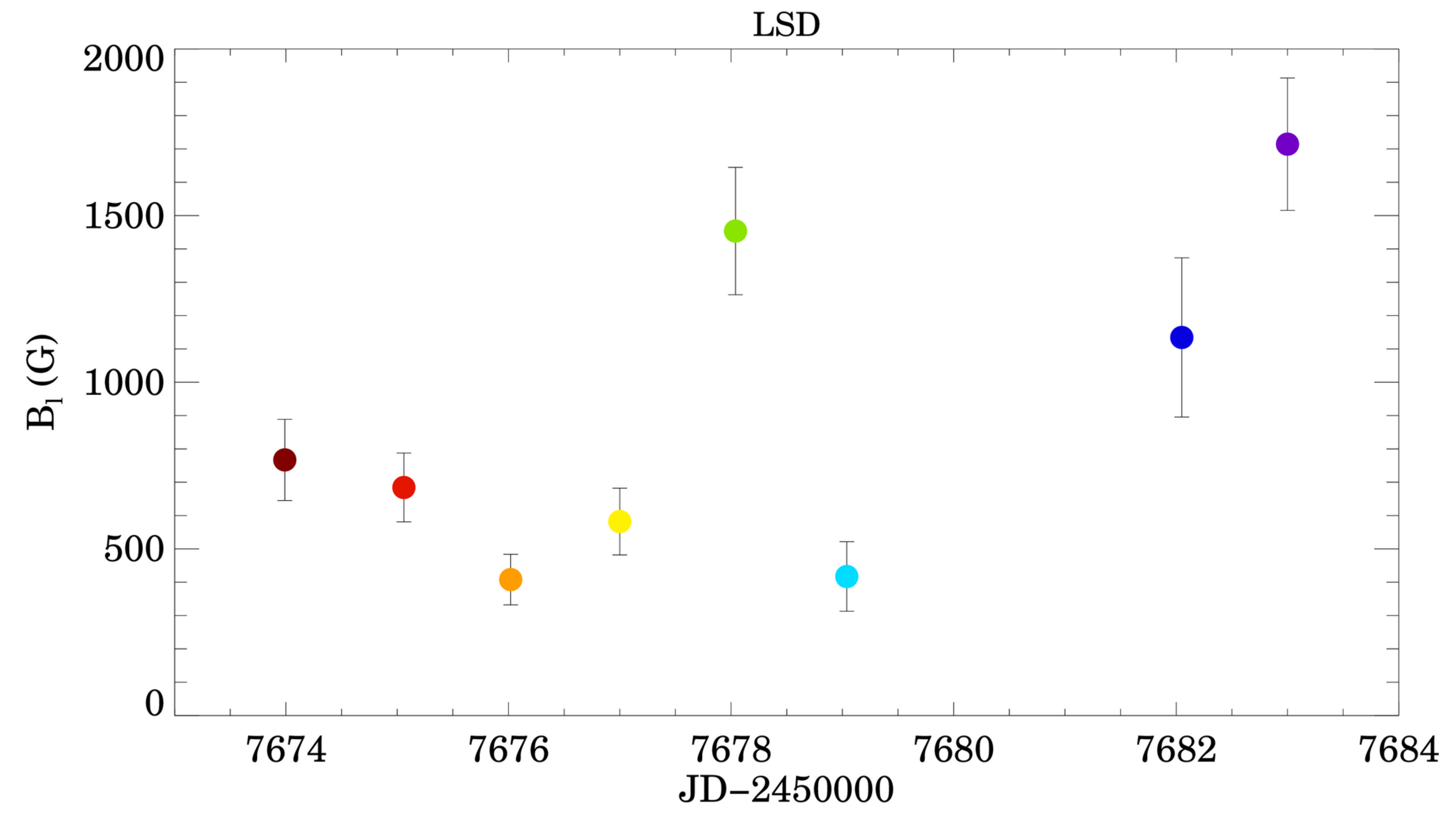}
    \caption{Spectropolarimetric measurements of DR~Tau. The top left panel shows the LSD Stokes \textit{I} profiles, the top right panel the Stokes \textit{V} profiles. The black lines in the top panes are the average line profiles. The lines were shifted along the $y$-axis by an arbitrary value for better display. The bottom panel the longitudinal magnetic field as calculated from the LSD profiles.}
         \label{fig:drtau_longitudinal}
   \end{figure}

We used the `least-squares deconvolution' (LSD) technique \citep{donati1997} in order to derive the mean Stokes $I$ (unpolarized) and $V$ (circularly polarized) photospheric absorption line profiles from the complete ESPaDOnS spectra (see top panels of Fig.~\ref{fig:drtau_longitudinal}). This technique assumes that all lines in the spectrum contain the same information, and the spectrum can be written as a convolution of a mean profile with a line mask. Therefore, the mean line profile can be determined by a deconvolution process. The detailed description of the method can be found in \cite{donati1997}.

The LSD technique is particularly well designed for measuring the Zeeman-signatures of the lines generated by magnetic fields at the surfaces of stars. Stokes $V$ signatures of line profiles are very small (with a typical amplitude of about 0.1\% peak-to-peak), however, LSD can reveal them, as it extracts line profile polarization information from thousands of spectral lines simultaneously. For DR~Tau, we used $\sim$7000 spectral lines from the whole spectrum covering the $370-1050$\,nm wavelength region.
The photospheric LSD profiles were normalized with a mean wavelength of 500\,nm, Doppler width of 1.8\,km/s, and Landé factor of 1.2.
As the veiling affects the spectral lines used this procedure, we scaled the obtained LSD $I$ and $V$ so that all profiles have the same equivalent width.
This allowed us to study the intrinsic variations of the shape of these lines, unaffected by the changes in the line depth due to changes in the continuum level.

The top panels of Fig.~\ref{fig:drtau_longitudinal} show the LSD~$I$ and $V$ profiles.
The LSD~$I$ profiles often show distortions due to starspots \citep[e.g.,][]{nowacki2023}, but the intensity profiles here show only slight changes in the line profiles.
The LSD~$V$ profiles, on the other hand, vary significantly over the observing period.

The Zeeman signatures of the large-scale surface magnetic field were detected in all circular polarization LSD line profiles. The strength of the longitudinal (i.e. the line-of-sight projected magnetic fields averaged over the visible hemisphere) magnetic field ($B_l$) can be obtained from the Stokes $V$ LSD profiles for each epoch \citep{donati1997, wade2000, folsom2016}. The bottom panel of Fig.~\ref{fig:drtau_longitudinal} shows that the longitudinal magnetic field of DR Tau varies between 400 G and 1800\,G.

Using the $B_l$ values, we can determine the strength of the dipole magnetic field using the following equations from \cite{preston1967}:

\begin{equation}\label{eq:b_dipole}
H_e = \frac{1}{20}\frac{15 + u}{3 - u} H_p (\cos \beta \cos i + \sin \beta \sin i \cos 2 \pi t / P),
\end{equation}

\noindent where $\beta$ is the angle between the rotation axis and the magnetic axis, $i$ is the angle between the rotation axis and the line of sight, $P$ is the rotational period, $u$ is the limb darkening coefficient, and $H_p$ is the polar field strength.
The ratio between the maximum and the minimum field strength, $r$ can be defined as follows:

\begin{equation}
r = \frac{H_e (\textrm{min})}{H_e (\textrm{max})} = \frac{\cos \beta \cos i - \sin \beta \sin i}{\cos \beta \cos i + \sin \beta \sin i},
\end{equation}

\noindent where
\begin{equation}
i = \arctan \left[ \left( \frac{1 - r}{1 + r}\right) \cot \beta  \right].
\end{equation}

\noindent 
By fitting a sinusoidal to the $B_l$ curve, we estimated $H_e (\textrm{min}) = 412$\,G, and $H_e (\textrm{max}) = 961$\,G. By using $i=26^{\circ}$ inclination, we determined $\beta = 0.33$. Using $\beta$, $H_e (\textrm{max})$, $i$, and $u=0.85$ \citep{claret2011}, Eq.~\ref{eq:b_dipole} resulted in 3287~G.

\section{Discussion}

\subsection{Short-term and long-term photometric changes}

The light curves of DR~Tau exhibit primarily stochastic or quasi-periodic behavior in all bands.
The ground-based multifilter observations, due to their nightly or sparser cadence, reveal variations on daily timescale.
The peak-to-peak amplitude of the changes differs for the three observing seasons, but the general trend of the decreasing amplitude towards the longer wavelength bands is always detected. 
This trend towards the infrared wavelengths  suggests that part of the disk may be optically thick and invariable, which causes the smaller infrared amplitudes compared to the optical ones.

\cite{grankin2007} presented a survey of the long-term photometric variability of 49 classical T~Tauri stars, including DR~Tau.
Their aim was to characterize the long-term variability from a set of statistical parameters using observations over up to 20\,years.
For DR~Tau, they found a mean brightness of $\overline{V_m} = 11.66$\,mag, and an average $V$-band photometric amplitude for all observing seasons $\overline{\Delta V} = 1.271$\,mag.
For their whole sample, they found a rather large distribution for the average $V$-band amplitude of variability ($0.1-2$\,mag): nearly half of their sample exhibited small variability ($0.1-0.4$\,mag), a similar number of sources showed larger changes ($0.6-1.5$\,mag), and only a few targets displayed brightness changes over 1.5\,mag.
The above cited parameters place DR~Tau among the sources with larger variability.
We also obtained these parameters for the data presented here, considering all three observing seasons, which lead to similar values: $\overline{V_m} = 12.02$\,mag, and $\overline{\Delta V} = 1.20$\,mag.
However, as \cite{grankin2007} report, the long-term variability merely reflect the shorter-term changes, it is instructive to make a comparison with the individual observing seasons as well to investigate how representative shorter observing runs are.
For the three observing seasons, both the average brightness level (11.67\,mag, 12.35\,mag, 12.00\,mag for 2009, 2017, 2021/22, respectively) and the amplitude of the variability shows differences (0.59\,mag, 1.46\,mag, 1.54\,mag for 2009, 2017, 2021/22, respectively): we found $\sim$2.6 times smaller amplitudes of variability for 2009 than for 2021/22.
This result might arise from the fact that the 2009 observing season was shorter than the 2021/22 observing run, but different levels of contribution from the occurring physical processes may also play a role.

The color-magnitude diagrams can help identifying the underlying physical mechanisms, which involve cold surface spots related to stellar activity, hot spots due to accretion, or variable circumstellar extinction.
Fig.~\ref{fig:drtau_cmd} indicates strong correlation between the observed color and the brightness: the system becomes bluer when brighter in the optical bands (top panels).
For DR~Tau, a variable extinction cause of variability is not likely, as the color slopes slightly differ from those expected from the interstellar reddening law indicated with black arrows in Fig.~\ref{fig:drtau_cmd}. Furthermore, the system is seen close to pole on, which would require the circumstellar matter to be lifted high above the disk plane to cause the observed effect.
Rotational modulation caused by cold spots is also not likely due to the low inclination.
This is supported by the period analysis presented in Sect.~\ref{sect:drtau_period}, which reveals no clear periodicity in the light curves. 
Forming and decaying active regions might contribute to the variability, however, it is more significant on longer timescales due to the longer lifetime of starspots.

\subsection{Short-term versus long-term changes of the H$\alpha$ line}
\label{sect:drtau_shortterm_longterm}

The spectra of DR~Tau, presented here, show several emission features that are related to the accretion process.
The permitted emission lines carry information on the complex region where the stellar magnetosphere interacts with the accretion disk. 
Within the 10~days of the ESPaDOnS observing campaign, these lines presented high variability, which suggests a very dynamic interaction between the star and the disk.
The H$\alpha$ line exhibits the most complex profiles, that can be decomposed with multiple Gaussian components.
They mainly vary in strength, and show morphological variation to smaller extent.

It is instructive to compare the results that we presented in the previous Sections with the long-term spectroscopic variability presented by \cite{alencar2001}. They examined 103 spectra of DR~Tau obtained over more than a decade, and found that the H$\alpha$ lines were generally strongly peaked in the red ($v\sim$100\,km/s).
In $\sim$75\% of the observed spectra, the H$\alpha$ lines display the type IVB P~Cygni profile according to the classification scheme of \cite{reipurth1996}, i.e., they have an absorption component which has sufficient velocity to be present beyond the wing of the emission line.
In these cases, their observed line profiles could be decomposed into a redshifted emission peak, a blueshifted absorption component, and a broad low amplitude component centered at the rest velocity.
However, the lines exhibit substantial morphological variations over the course of a decade.
On one hand, the significance and strength of the above mentioned line components vary, e.g., the blue absorption may disappear completely in some cases.
On the other hand, they found that in $\sim$15\% of their observations, and additional blue shoulder appears in the line profile, therefore, an additional blueshifted emission component was needed to fit the H$\alpha$ line.
This latter line profile is the only one that was detected in our ESPaDOnS spectra.
This suggests that in the case of a highly accreting and extremely variable T~Tauri star, such as DR~Tau, the changes observed over the course of $\sim$10~days might not be entirely representative of the overall long-term behavior of the system.

It is interesting to study what mechanisms are responsible for the morphological changes of the H$\alpha$ line in the long-term and in the short-term observations.
The strong redshifted peak, which is associated with the accretion flow, is present in all of the spectra.
The normalized variance profile indicates similar level of variability in this component in \cite{alencar2001} (see their Fig.~7) and in our spectra (Fig.~\ref{fig:drtau_var_profile_main} $a$).
This suggests that the typical accretion variability can be observed even during a short observing run.
The variability of the blueshifted side of the line might originate from the changes of the blueshifted absorption, the blueshifted emission, or their combination; and is much more significant in the long-term data.
In contrast, this is only slightly more variable than the redshifted peak in our spectra.
Furthermore, \cite{alencar2001} found no correlation between the redshifted and the blueshifted parts of the H$\alpha$ line, whereas our spectra suggest the presence of correlation between these components (Fig.~\ref{fig:corrmtx}).
It was suggested that the blue emission peak might be influenced by both the far side of the magnetospheric flow and a sporadic jet-like outflow launched relatively close to the star. The latter possibility of the outflow origin is further discussed in the next Section.

\subsection{Outflow} \label{sect:drtau_outflow}

A long-standing question in the formation of stars is how young stars rid themselves from the accreted angular momentum. One proposed explanation is that open magnetic field lines allow for stellar wind, which might be able to carry away enough angular momentum to counteract the accretion torque \citep{hartmann1989, tout1992}.
\cite{matt2005} suggested that protostellar winds can remove accreted angular momentum, provided that the stars have large mass-loss rates ($\dot{M_w} \sim 0.1\dot{M_a}$). They proposed that the energy driving the stellar wind derives from accretion power. 

Several previous studies found that accreting systems show wind signatures as well.
The complex structure of the H$\alpha$ line of DR~Tau suggests that the accretion process is not the only mechanism shaping the observed line profiles, but outflows also play a role. This can be further examined by searching for forbidden emission lines in the spectra, which are typical wind and outflow tracers.

The spectra of DR~Tau reveal three [O~I] lines (557.7\,nm, 630.0\,nm, 636.4\,nm), and the [S~II] 406.8\,nm  and 673.1\,nm lines.
The individual line profiles in Fig.~\ref{fig:drtau_var_profile} suggest that the variability of certain wind tracers might correlate with the variability of certain components of some of the accretion tracers. In order to further examine this, we constructed the cross-correlation diagrams of accretion and wind tracers (Fig.~\ref{fig:corrmtx_outflow}). 
The broad component of the [O\,I] 630.0\,nm line shows correlation with the the blueshifted parts of the Balmer lines (H$\alpha$ and H$\beta$), mainly where the absorption component occurs. Furthermore, there is correlation between the changes in the blueshifted side of the broad component of the Ca\,II line, and the variability of the [O\,I] line and the narrow component of the He\,I line. This latter correlation between the [O\,I] and the NC He\,I could suggest the presence of accretion powered wind.

However, we must note that  the spectra might be affected by telluric emission as well, which could contaminate the [O\,I] emission originating from the source. 
\cite{banzatti2019} and \cite{simon2016} studied the forbidden emission lines of a few T~Tauri stars, including DR~Tau, in the spectra obtained with the HIRES instrument of the Keck telescope. 
Their data reduction procedure eliminates the telluric contamination from the spectra. 
The corrected spectra of DR~Tau still shows significant [O\,I] emission, which suggests that emission originating from the source dominates over the telluric emission.
We directly compared the corrected [O\,I] 630.0\,nm line from \cite{banzatti2019} with our ESPaDOnS observations (see dotted line in panel $h$ of Fig.\ref{fig:drtau_var_profile}), and found that the corrected line is weaker than our ESPaDOnS observations. This suggests that the observed lines are contaminated by the telluric emission, however, we also note that some level of variability is expected in the [O\,I] lines on various timescales.
As the ESPaDOnS spectra presented here are not corrected for telluric emission, we examine its effect on our analysis in the following. 
As the telluric lines originate from the atmosphere of Earth, they should be shifted by the barycentric velocity.
Unfortunately, it is difficult to separate the contamination of the telluric line and the emission originating from the source, as the barycentric velocity for our observations ($21.5-23.5$km/s) is very close to the measured $v_{sys}$. 
In order to further disentangle the telluric contribution in the [O\,I] lines, we inspected the airmass changes throughout the ESPaDOnS observing period. 
We found that the changes in the airmass do not correlate with the variability in the strength of the [O\,I] lines (e.g. the airmass on the nights with the three strongest [O\,I] emission were 1.16, 1.007, and 1.011). 
This suggests that even though the [O I] lines could be contaminated by telluric emission, it is most probably a constant level of emission, and the observed line variability presumably originates from the source. 
Furthermore, this problem concerns only the narrow component of the [O I] lines, the broad component, which is clearly present in our data, is not affected.

   \begin{figure}[t!]
   \centering
   \includegraphics[height=\columnwidth, angle=90]{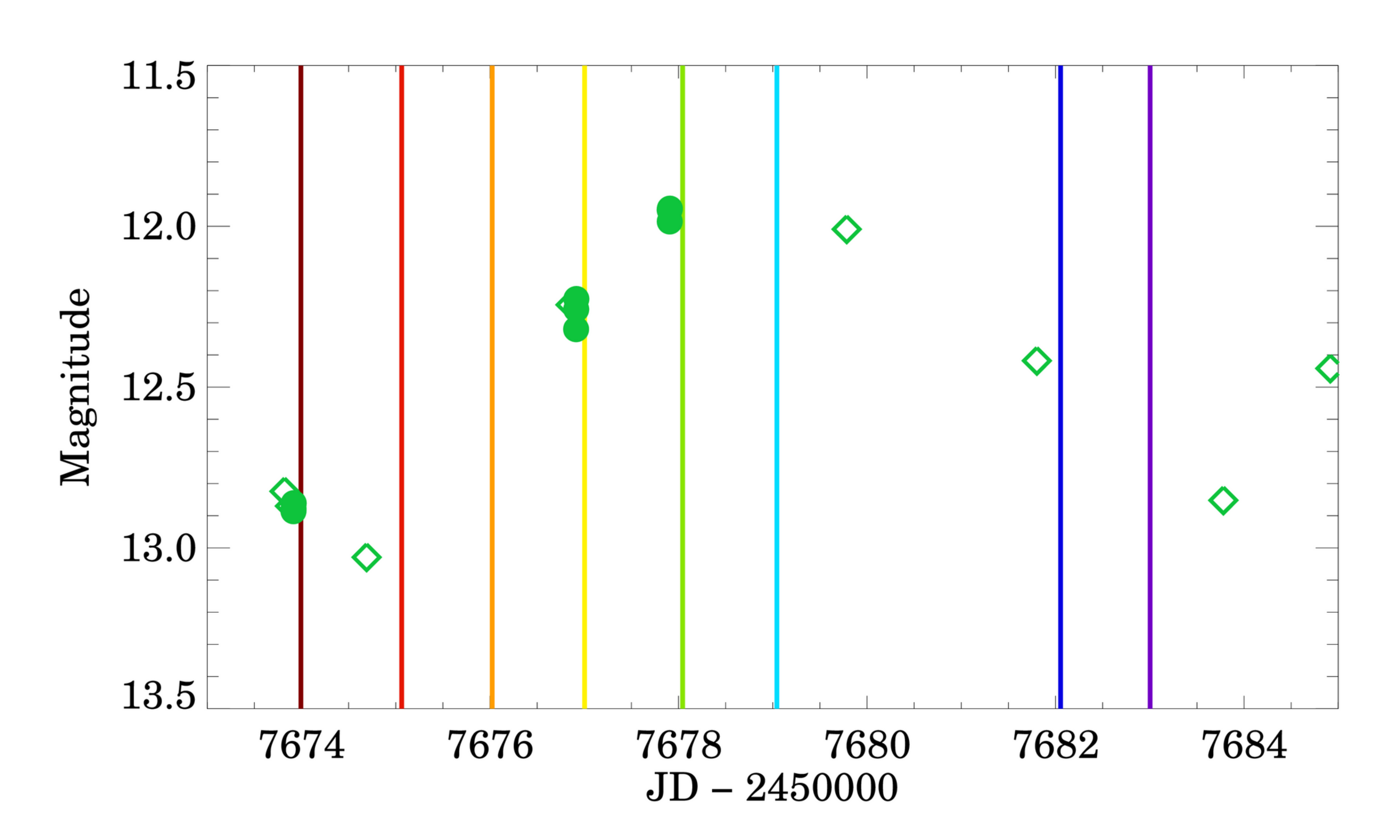}
   \caption{$V$ band light curve from the ASAS-SN (filled circles) and the AAVSO (open circles) databases for the ESPaDOnS observing period. The vertical lines indicate the epochs of the ESPaDOnS spectropolarimetric observations. The same colors were used to indicate each epoch as in Fig.~\ref{fig:drtau_var_profile_main} for easier comparability in earlier Sections. }
         \label{fig:drtau_asassn_esp}
   \end{figure}

\subsection{Veiling variations}
The photospheric absorption lines of classical T~Tauri stars are often weakened by excess emission, known as veiling, arising from the radiation of the heated gas at the footpoint of the accretion flows.
For this reason, changes of the veiling are expected to correlate with the accretion rate variations, and the stellar brightness is also expected to vary accordingly. This is often observed for T~Tauri stars \citep[e.g.,][]{dodin2012, fiorellino2022, zsidi2022_vw}, therefore the amount of veiling is often used to estimate the accretion rate.

The veiling variations of DR~Tau, however, do not resemble the amplitude variations of the accretion tracers, such as the Balmer lines, the Ca~II infrared triplet or the He~I lines (see Figs.~\ref{fig:drtau_var_profile} and \ref{fig:drtau_veiling}).
We calculated cross-correlation matrices between photospheric lines and H$\alpha$, the Ca~II, and the He~I lines, and the results confirm the lack of correlation between the veiling and the accretion tracers.
However, we must note that here, we are analyzing continuum normalized spectra, therefore the amplitude variations of the emission lines seen in Fig.~\ref{fig:drtau_var_profile} might not reveal directly the accretion rate changes. Additionally, our spectra are not corrected for the veiling effect, which might affect the correlations seen in Figs.~\ref{fig:corrmtx}, \ref{fig:corrmtx2}, and \ref{fig:corrmtx_outflow}.

As some evidence has been found that the veiling variations and the brightness changes do not always correlate \citep[e.g,][]{petrov2001, petrov2007, gahm2008}, it is instructive to further examine if this is the case for DR~Tau as well.
It was proposed that the photospheric lines can be filled in when the veiling becomes high, which results in large veiling factors, that are not related to the continuum emission from the shock region.
This phenomenon has mainly been observed in a few systems with strong and irregular light variations.
\cite{gahm2008} analyzed photometric and spectroscopic variations of four systems (RW~Aur~A, RU~Lup, S~CrA~NW, and S~CrA~SE) with unusually strong veiling. As the veiling is expected to originate from a variable excess emission, the stellar brightness should change accordingly. They did find some correlation for the low veiling states, but as the veiling increases, the correlation does not hold anymore. 
\cite{gahm2008} proposed a few mechanisms that can result in the observed effect. Firstly, variable extinction might balance the increase in brightness caused by increased accretion under those specific circumstances, when infalling dust grains survive an accretion event. However, this scenario is unlikely for a system seen nearly pole-on. Secondly, electron scattering can also decrease the strength of absorption lines. 
Electron scattering is a geometric effect when considering optically thick material; and the veiling depends on it, as the scattered particles can cover a fraction of the surface, and erase all underlying spectral information. 
However, producing the largest veiling values would require a large fraction of the stellar surface to be covered as a result of the electron scattering. Lastly, narrow emission in metallic lines could lead to the line cores being filled in. This could lead to an increase in the veiling without significantly increasing the brightness of the system.
\cite{gahm2008} also concluded that, because of the above listed effects, the degree of veiling could lead to and overestimation of the accretion rates in CTTS with large veiling values and rich emission spectra.

We did not obtain contemporaneous photometric data for the observing period of the ESPaDOs spectroscopic measurements. 
Therefore, we searched public databases, such as the ASAS-SN survey and the archive of the American Association of Variable Star Observers (AAVSO), which allowed us to examine the brightness changes of DR~Tau. 
We found a few $V$ band measurements from the desired observing period, which are close to the spectroscopic observations in time. 
These give a hint at the brightness state of the system (Fig.~\ref{fig:drtau_asassn_esp}) during our spectroscopic monitoring.
We find that for lower veiling values there is a hint of a correlation between the system brightness and the veiling, but the epoch for which the largest veiling was measured does not show the brightest photometric state.
\cite{petrov2011} proposed that the veiling measurements of DR~Tau might be affected by line veiling, and our observations also suggests that the continuum excess cannot be the only contributing factor to the observed veiling.

\subsection{Stellar magnetic field}

   \begin{figure}[t!]
   \centering
   \includegraphics[height=\columnwidth, angle=90]{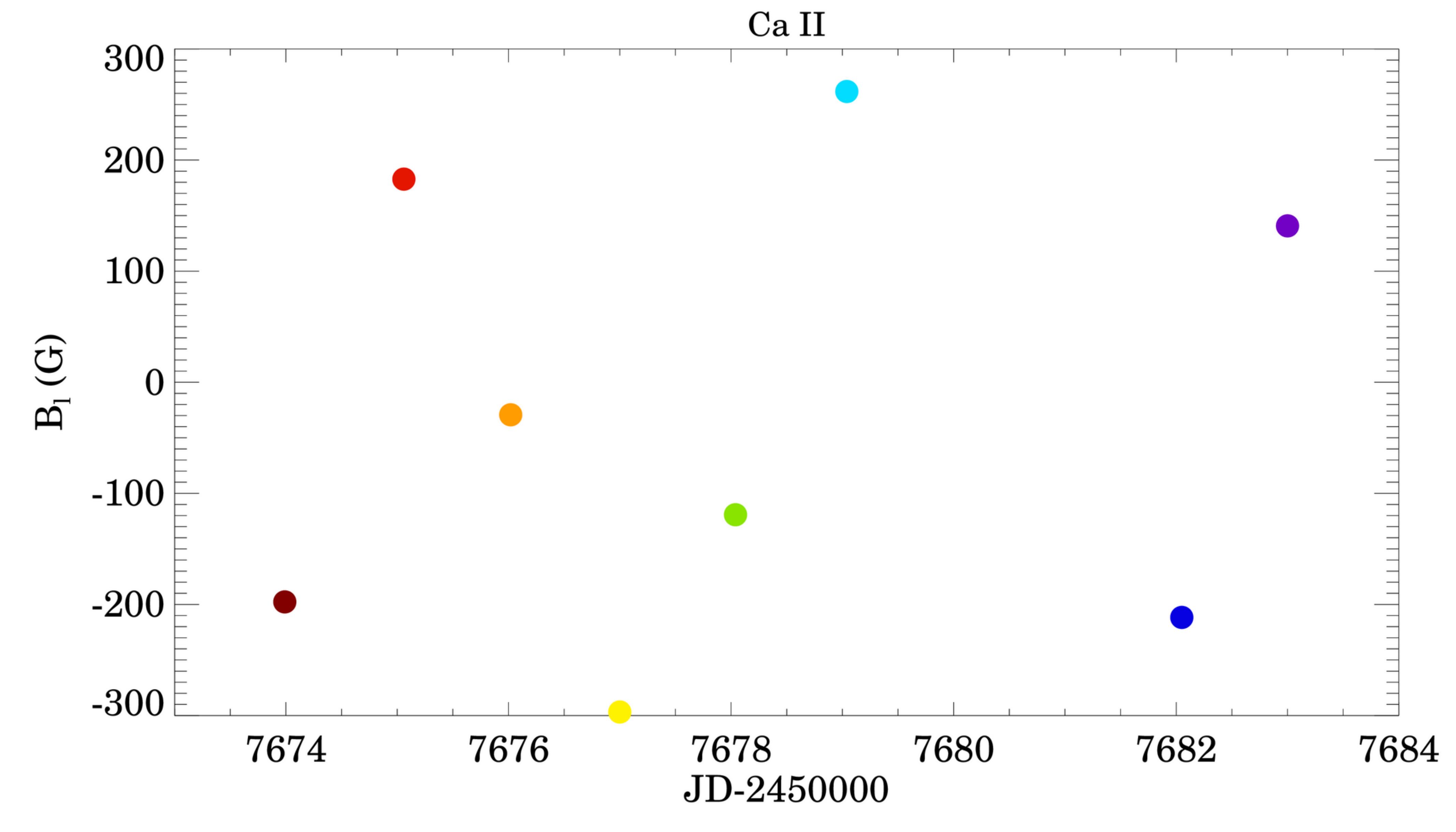}
   \includegraphics[height=\columnwidth, angle=90]{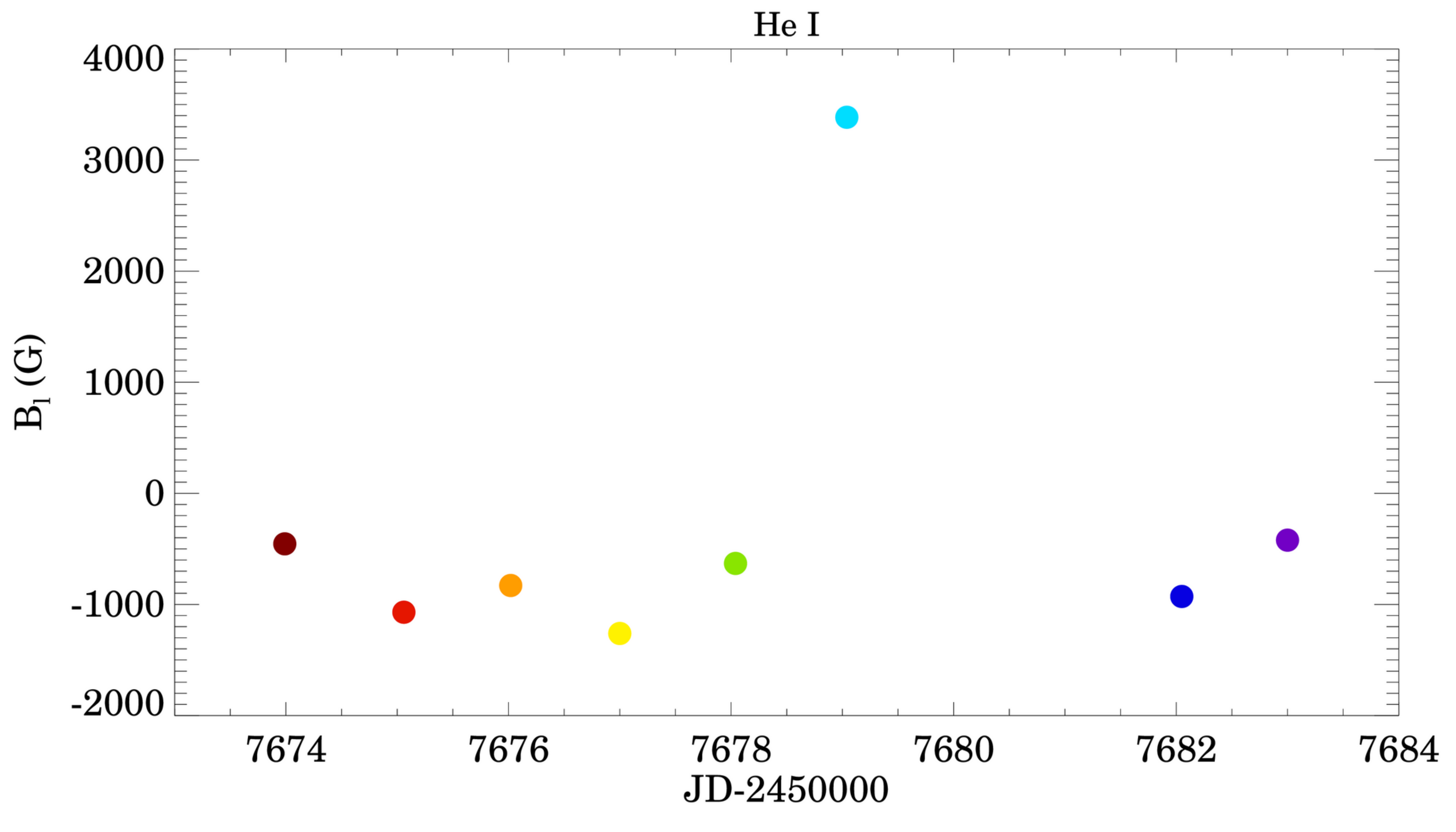}
      \caption{Longitudinal magnetic field of DR~Tau as calculated from the Ca~II and He~I lines}
         \label{fig:drtau_longitudinal_emission}
   \end{figure}

DR~Tau is a strongly accreting T~Tauri star, often so high accretion rates were measured that DR~Tau was suggested to be an EXor candidate \citep{giannini2022}. However, the light curves of DR~Tau do not display the month-long-outbursts, characteristic to EXors.
Magnetic field measurements of T~Tauri stars with such high accretion rates are rare: the magnetic field was measured only for one other extreme T~Tauri star, S~CrA~N, to our best knowledge.
\cite{nowacki2023} estimated the a mass accretion rate to be around around $10^{-7.3} - 10^{-6.2}\,M_{\odot} \rm{yr}^{-1}$ for S~CrA~N using the width of the H$\alpha$ at 10\% in their ESPaDONs spectra, and found a mean veiling value of 7. They obtained magnetic field measurements, which suggest strong and complex magnetic field, and found that the large-scale magnetic field measurements that best represent their Stokes~$V$ profiles is as strong as 5.4 kG. 
Using the the NC of their He\,I line, they found that the longitudinal field shows a complex modulation, which could be consistent with a complex magnetic field significantly differing from a dipole. 

We also looked for Zeeman-signatures in emission lines of DR~Tau.  We found strong signal in the Stokes $V$ spectra of the narrow components of several lines, such as He\,I, and Ca\,II IRT. This means that the magnetic field is strong in the accretion shock, and the shocks are expected to cover a region of the surface of the star that is the most magnetized. However, because of the higher temperature in the shock region, it contributes only little to the photospheric lines. This means that the magnetic field we measured in the photospheric LSD line profiles is underestimated. 

We found strong signatures in the the He~I 587.6\,nm line. The He~I line has a narrow and a broad component, and the Stokes $V$ profile has and asymmetric profile and extends from -50\,km/s to $\sim50$\,km/s -- which roughly aligns with the NC of the intensity profile.
This component is thought to trace the hot regions at the footpoint of the accretion funnel. 
We fitted both the broad and the narrow component with a Gaussian, and we subtracted the BC profile in order to study the longitudinal magnetic filed measurements in the NC. We measured the strength of the longitudinal magnetic field with the first moment method \citep{donati1997, wade2000}. We show the results in Fig.~\ref{fig:drtau_longitudinal_emission}.

Furthermore, we also detected clear Zeeman-signature in the core emission of the Ca\,II IRT (849.8\,nm, 854.2\,nm, and 866.2\,nm), which presumably probes both the chromosphere and the accretion regions. We estimated the broad component for each line with a Gaussian fit and subtracted it from the Ca\,II line, and calculated the longitudinal magnetic field. In the bottom panel of Fig.~\ref{fig:drtau_longitudinal_emission} show that the longitudinal magnetic filed in the Ca\,II formation zone varies in polarity and intensity.

\section{Conclusions}

We analyzed the short-term and long-term photometric and spectroscopic variations of DR~Tau using newly obtained observations, and combined them with yet unpublished archival data covering more than a decade in time. In addition, we compared our findings with already published results in order to extend the temporal coverage of our variability study.

We find that DR~Tau is highly variable on all timescales that are available in our dataset. The high-cadence K2 and TESS data revealed that the light curves are primarily dominated by stochastic changes, and the typical timescale of the variability is around 2-3\,days. The period analysis did not lead to a reliable rotational period, not even with the highest (1-day) cadence K2 data, but the wavelet analysis hints at a periodic signal around 3\,days.
The ground-based multifilter photometric observations show that the shape of the light curves is similar in all filters, but the amplitude of the variations decreases with increasing wavelength. This does not closely follow the extinction curve, and the trend at the longest (mid-IR) wavelengths suggest that the inner part of the disk might be optically thick and invariable. By comparing our results with previous findings from the literature, we found that DR~Tau is highly variable when compared to a large sample of CCTS sources. 
However, it must be noted that short observing seasons might be affected by unusually quiescent phases or larg outbursts, which leads to the measurement of lower or higher amplitude of variability, respectively.

Our spectroscopic monitoring observations cover a period of $\sim$10\,days. 
During this period, emission lines tracing accretion exhibited high variability in both their amplitudes and line profiles. The high-resolution spectra revealed multiple components in several accretion tracers. 
We decomposed the He~I lines and the Ca~II triplet into a broad and a narrow components, where the broad component is thought to originate from the accretion flow and its variability correlates with other accretion tracers, while the narrow component might arise from the post-shock region. 
The H$\alpha$ line exhibits the most complex and asymmetric line profile, and the variance profiles suggest that the two emission peaks show the highest variability. 
When comparing these shorter-term variations with long-term spectral changes, we found that the emission lines can show even more significant morphological variations when monitored over a decade, which suggest that the significance of the physical mechanisms forming the line profiles might change over time.
Furthermore, we found that the veiling variations do not always correlate with the amplitude changes of accretion tracers and the brightness changes of the system, which suggests that veiling might not solely be related to an excess continuum emission in DR~Tau.

\begin{acknowledgements}
This project has received funding from the European Research Council (ERC) under the European Union's Horizon 2020 research and innovation programme under grant agreement No~716155 (SACCRED) and grant agreement No~742095 (SPIDI). 
We thank M. R\'acz for obtaining photometric data in 2009.
GZs acknowledges support from the Leverhulme Trust (grant number RPG-2021-380).
CFM is Funded by the European Union (ERC, WANDA, 101039452). Views and opinions expressed are however those of the author(s) only and do not necessarily reflect those of the European Union or the European Research Council Executive Agency. Neither the European Union nor the granting authority can be held responsible for them. 
This work was supported by the Hungarian National Research, Development and Innovation Office grants OTKA K131508 and KH-130526, and the \'Elvonal grant KKP-143986. KV and LK are supported by the Bolyai J\'anos Research Scholarship of the Hungarian Academy of Sciences, KV is supported by the Bolyai+ grant \'UNKP-22-5-ELTE-1093. 
Zs.M.Sz. acknowledges funding from a St Leonards scholarship from the University of St Andrews.  
For the purpose of open access, the author has applied a Creative Commons Attribution (CC BY) license to any Author Accepted Manuscript version arising.
Zs.M.Sz. is a member of the International Max Planck Research School (IMPRS) for Astronomy and Astrophysics at the Universities of Bonn and Cologne.
LK acknowledges the Hungarian National Research, Development and
Innovation Office
grant OTKA PD-134784.
This work received funding from the OTKA K-138962 grant.
This work was also supported by the NKFIH NKKP grant ADVANCED 149943 and the NKFIH excellence grant TKP2021-NKTA-64. Project no.149943 has been implemented with the support provided by the Ministry of Culture and Innovation of Hungary from the National Research, Development and Innovation Fund, financed under the NKKP ADVANCED funding scheme.
\end{acknowledgements}

%
%

\bibliographystyle{aa}
\bibliography{drtau_ref.bib}

\begin{thebibliography}{71}
\expandafter\ifx\csname natexlab\endcsname\relax\def\natexlab#1{#1}\fi

\bibitem[{{Alencar} {et~al.}(2001){Alencar}, {Johns-Krull}, \&
  {Basri}}]{alencar2001}
{Alencar}, S. H.~P., {Johns-Krull}, C.~M., \& {Basri}, G. 2001, \aj, 122, 3335

\bibitem[{{Banzatti} {et~al.}(2014){Banzatti}, {Meyer}, {Manara},
  {Pontoppidan}, \& {Testi}}]{banzatti2014}
{Banzatti}, A., {Meyer}, M.~R., {Manara}, C.~F., {Pontoppidan}, K.~M., \&
  {Testi}, L. 2014, \apj, 780, 26

\bibitem[{{Banzatti} {et~al.}(2019){Banzatti}, {Pascucci}, {Edwards}, {Fang},
  {Gorti}, \& {Flock}}]{banzatti2019}
{Banzatti}, A., {Pascucci}, I., {Edwards}, S., {et~al.} 2019, \apj, 870, 76

\bibitem[{{Beristain} {et~al.}(2001){Beristain}, {Edwards}, \&
  {Kwan}}]{beristain2001}
{Beristain}, G., {Edwards}, S., \& {Kwan}, J. 2001, \apj, 551, 1037

\bibitem[{{Bouvier} {et~al.}(1993){Bouvier}, {Cabrit}, {Fernandez}, {Martin},
  \& {Matthews}}]{bouvier1993}
{Bouvier}, J., {Cabrit}, S., {Fernandez}, M., {Martin}, E.~L., \& {Matthews},
  J.~M. 1993, \aap, 272, 176

\bibitem[{{Bouvier} {et~al.}(1995){Bouvier}, {Covino}, {Kovo}, {Martin},
  {Matthews}, {Terranegra}, \& {Beck}}]{bouvier1995}
{Bouvier}, J., {Covino}, E., {Kovo}, O., {et~al.} 1995, \aap, 299, 89

\bibitem[{{Broeg} {et~al.}(2006){Broeg}, {Joergens}, {Fern{\'a}ndez}, {Husar},
  {Hearty}, {Ammler}, \& {Neuh{\"a}user}}]{broeg2006}
{Broeg}, C., {Joergens}, V., {Fern{\'a}ndez}, M., {et~al.} 2006, \aap, 450,
  1135

\bibitem[{{Cardelli} {et~al.}(1989){Cardelli}, {Clayton}, \&
  {Mathis}}]{cardelli1989}
{Cardelli}, J.~A., {Clayton}, G.~C., \& {Mathis}, J.~S. 1989, \apj, 345, 245

\bibitem[{{Chavarria-K.}(1979)}]{chavarria1979}
{Chavarria-K.}, C. 1979, \aap, 79, L18

\bibitem[{{Claret} \& {Bloemen}(2011)}]{claret2011}
{Claret}, A. \& {Bloemen}, S. 2011, \aap, 529, A75

\bibitem[{{Cutri} {et~al.}(2003){Cutri}, {Skrutskie}, {van Dyk}, {Beichman},
  {Carpenter}, {Chester}, {Cambresy}, {Evans}, {Fowler}, {Gizis}, {Howard},
  {Huchra}, {Jarrett}, {Kopan}, {Kirkpatrick}, {Light}, {Marsh}, {McCallon},
  {Schneider}, {Stiening}, {Sykes}, {Weinberg}, {Wheaton}, {Wheelock}, \&
  {Zacarias}}]{cutri2003}
{Cutri}, R.~M., {Skrutskie}, M.~F., {van Dyk}, S., {et~al.} 2003, {2MASS All
  Sky Catalog of point sources.}

\bibitem[{{Dodin} \& {Lamzin}(2012)}]{dodin2012}
{Dodin}, A.~V. \& {Lamzin}, S.~A. 2012, Astronomy Letters, 38, 649

\bibitem[{{Dodin} \& {Lamzin}(2013)}]{dodin2013}
{Dodin}, A.~V. \& {Lamzin}, S.~A. 2013, Astronomy Letters, 39, 389

\bibitem[{{Donati} {et~al.}(2019){Donati}, {Bouvier}, {Alencar}, {Hill},
  {Carmona}, {Folsom}, {M{\'e}nard}, {Gregory}, {Hussain}, {Grankin}, {Moutou},
  {Malo}, {Takami}, {Herczeg}, \& {MaTYSSE Collaboration}}]{donati2019}
{Donati}, J.~F., {Bouvier}, J., {Alencar}, S.~H., {et~al.} 2019, \mnras, 483,
  L1

\bibitem[{{Donati} {et~al.}(2011){Donati}, {Gregory}, {Alencar}, {Bouvier},
  {Hussain}, {Skelly}, {Dougados}, {Jardine}, {M{\'e}nard}, {Romanova}, \&
  {Unruh}}]{donati2011}
{Donati}, J.~F., {Gregory}, S.~G., {Alencar}, S.~H.~P., {et~al.} 2011, \mnras,
  417, 472

\bibitem[{{Donati} {et~al.}(1997){Donati}, {Semel}, {Carter}, {Rees}, \&
  {Collier Cameron}}]{donati1997}
{Donati}, J.~F., {Semel}, M., {Carter}, B.~D., {Rees}, D.~E., \& {Collier
  Cameron}, A. 1997, \mnras, 291, 658

\bibitem[{{Dworetsky}(1983)}]{dworetsky1983}
{Dworetsky}, M.~M. 1983, \mnras, 203, 917

\bibitem[{{Fiorellino} {et~al.}(2022){Fiorellino}, {Park}, {K{\'o}sp{\'a}l}, \&
  {{\'A}brah{\'a}m}}]{fiorellino2022}
{Fiorellino}, E., {Park}, S., {K{\'o}sp{\'a}l}, {\'A}., \& {{\'A}brah{\'a}m},
  P. 2022, \apj, 928, 81

\bibitem[{{Fischer} {et~al.}(2023){Fischer}, {Hillenbrand}, {Herczeg},
  {Johnstone}, {Kospal}, \& {Dunham}}]{fischer2023}
{Fischer}, W.~J., {Hillenbrand}, L.~A., {Herczeg}, G.~J., {et~al.} 2023, in
  Astronomical Society of the Pacific Conference Series, Vol. 534, Protostars
  and Planets VII, ed. S.~{Inutsuka}, Y.~{Aikawa}, T.~{Muto}, K.~{Tomida}, \&
  M.~{Tamura}, 355

\bibitem[{{Folsom} {et~al.}(2012){Folsom}, {Bagnulo}, {Wade}, {Alecian},
  {Landstreet}, {Marsden}, \& {Waite}}]{folsom2012}
{Folsom}, C.~P., {Bagnulo}, S., {Wade}, G.~A., {et~al.} 2012, \mnras, 422, 2072

\bibitem[{{Folsom} {et~al.}(2016){Folsom}, {Petit}, {Bouvier}, {L{\`e}bre},
  {Amard}, {Palacios}, {Morin}, {Donati}, {Jeffers}, {Marsden}, \&
  {Vidotto}}]{folsom2016}
{Folsom}, C.~P., {Petit}, P., {Bouvier}, J., {et~al.} 2016, \mnras, 457, 580

\bibitem[{{Gahm} {et~al.}(2008){Gahm}, {Walter}, {Stempels}, {Petrov}, \&
  {Herczeg}}]{gahm2008}
{Gahm}, G.~F., {Walter}, F.~M., {Stempels}, H.~C., {Petrov}, P.~P., \&
  {Herczeg}, G.~J. 2008, \aap, 482, L35

\bibitem[{{Gaia Collaboration} {et~al.}(2023){Gaia Collaboration}, {Vallenari},
  {Brown}, {Prusti}, {de Bruijne}, {Arenou}, {Babusiaux}, {Biermann},
  {Creevey}, {Ducourant}, {Evans}, {Eyer}, {Guerra}, {Hutton}, {Jordi},
  {Klioner}, {Lammers}, {Lindegren}, {Luri}, {Mignard}, {Panem}, {Pourbaix},
  {Randich}, {Sartoretti}, {Soubiran}, {Tanga}, {Walton}, {Bailer-Jones},
  {Bastian}, {Drimmel}, {Jansen}, {Katz}, {Lattanzi}, {van Leeuwen}, {Bakker},
  {Cacciari}, {Casta{\~n}eda}, {De Angeli}, {Fabricius}, {Fouesneau},
  {Fr{\'e}mat}, {Galluccio}, {Guerrier}, {Heiter}, {Masana}, {Messineo},
  {Mowlavi}, {Nicolas}, {Nienartowicz}, {Pailler}, {Panuzzo}, {Riclet}, {Roux},
  {Seabroke}, {Sordo}, {Th{\'e}venin}, {Gracia-Abril}, {Portell}, {Teyssier},
  {Altmann}, {Andrae}, {Audard}, {Bellas-Velidis}, {Benson}, {Berthier},
  {Blomme}, {Burgess}, {Busonero}, {Busso}, {C{\'a}novas}, {Carry}, {Cellino},
  {Cheek}, {Clementini}, {Damerdji}, {Davidson}, {de Teodoro}, {Nu{\~n}ez
  Campos}, {Delchambre}, {Dell'Oro}, {Esquej}, {Fern{\'a}ndez-Hern{\'a}ndez},
  {Fraile}, {Garabato}, {Garc{\'\i}a-Lario}, {Gosset}, {Haigron}, {Halbwachs},
  {Hambly}, {Harrison}, {Hern{\'a}ndez}, {Hestroffer}, {Hodgkin}, {Holl},
  {Jan{\ss}en}, {Jevardat de Fombelle}, {Jordan}, {Krone-Martins}, {Lanzafame},
  {L{\"o}ffler}, {Marchal}, {Marrese}, {Moitinho}, {Muinonen}, {Osborne},
  {Pancino}, {Pauwels}, {Recio-Blanco}, {Reyl{\'e}}, {Riello}, {Rimoldini},
  {Roegiers}, {Rybizki}, {Sarro}, {Siopis}, {Smith}, {Sozzetti}, {Utrilla},
  {van Leeuwen}, {Abbas}, {{\'A}brah{\'a}m}, {Abreu Aramburu}, {Aerts},
  {Aguado}, {Ajaj}, {Aldea-Montero}, {Altavilla}, {{\'A}lvarez}, {Alves},
  {Anders}, {Anderson}, {Anglada Varela}, {Antoja}, {Baines}, {Baker},
  {Balaguer-N{\'u}{\~n}ez}, {Balbinot}, {Balog}, {Barache}, {Barbato},
  {Barros}, {Barstow}, {Bartolom{\'e}}, {Bassilana}, {Bauchet}, {Becciani},
  {Bellazzini}, {Berihuete}, {Bernet}, {Bertone}, {Bianchi}, {Binnenfeld},
  {Blanco-Cuaresma}, {Blazere}, {Boch}, {Bombrun}, {Bossini}, {Bouquillon},
  {Bragaglia}, {Bramante}, {Breedt}, {Bressan}, {Brouillet}, {Brugaletta},
  {Bucciarelli}, {Burlacu}, {Butkevich}, {Buzzi}, {Caffau}, {Cancelliere},
  {Cantat-Gaudin}, {Carballo}, {Carlucci}, {Carnerero}, {Carrasco},
  {Casamiquela}, {Castellani}, {Castro-Ginard}, {Chaoul}, {Charlot}, {Chemin},
  {Chiaramida}, {Chiavassa}, {Chornay}, {Comoretto}, {Contursi}, {Cooper},
  {Cornez}, {Cowell}, {Crifo}, {Cropper}, {Crosta}, {Crowley}, {Dafonte},
  {Dapergolas}, {David}, {David}, {de Laverny}, {De Luise}, {De March}, {De
  Ridder}, {de Souza}, {de Torres}, {del Peloso}, {del Pozo}, {Delbo},
  {Delgado}, {Delisle}, {Demouchy}, {Dharmawardena}, {Di Matteo}, {Diakite},
  {Diener}, {Distefano}, {Dolding}, {Edvardsson}, {Enke}, {Fabre}, {Fabrizio},
  {Faigler}, {Fedorets}, {Fernique}, {Fienga}, {Figueras}, {Fournier},
  {Fouron}, {Fragkoudi}, {Gai}, {Garcia-Gutierrez}, {Garcia-Reinaldos},
  {Garc{\'\i}a-Torres}, {Garofalo}, {Gavel}, {Gavras}, {Gerlach}, {Geyer},
  {Giacobbe}, {Gilmore}, {Girona}, {Giuffrida}, {Gomel}, {Gomez},
  {Gonz{\'a}lez-N{\'u}{\~n}ez}, {Gonz{\'a}lez-Santamar{\'\i}a},
  {Gonz{\'a}lez-Vidal}, {Granvik}, {Guillout}, {Guiraud},
  {Guti{\'e}rrez-S{\'a}nchez}, {Guy}, {Hatzidimitriou}, {Hauser}, {Haywood},
  {Helmer}, {Helmi}, {Sarmiento}, {Hidalgo}, {Hilger}, {H{\l}adczuk}, {Hobbs},
  {Holland}, {Huckle}, {Jardine}, {Jasniewicz}, {Jean-Antoine Piccolo},
  {Jim{\'e}nez-Arranz}, {Jorissen}, {Juaristi Campillo}, {Julbe}, {Karbevska},
  {Kervella}, {Khanna}, {Kontizas}, {Kordopatis}, {Korn}, {K{\'o}sp{\'a}l},
  {Kostrzewa-Rutkowska}, {Kruszy{\'n}ska}, {Kun}, {Laizeau}, {Lambert},
  {Lanza}, {Lasne}, {Le Campion}, {Lebreton}, {Lebzelter}, {Leccia}, {Leclerc},
  {Lecoeur-Taibi}, {Liao}, {Licata}, {Lindstr{\o}m}, {Lister}, {Livanou},
  {Lobel}, {Lorca}, {Loup}, {Madrero Pardo}, {Magdaleno Romeo}, {Managau},
  {Mann}, {Manteiga}, {Marchant}, {Marconi}, {Marcos}, {Marcos Santos},
  {Mar{\'\i}n Pina}, {Marinoni}, {Marocco}, {Marshall}, {Martin Polo},
  {Mart{\'\i}n-Fleitas}, {Marton}, {Mary}, {Masip}, {Massari},
  {Mastrobuono-Battisti}, {Mazeh}, {McMillan}, {Messina}, {Michalik}, {Millar},
  {Mints}, {Molina}, {Molinaro}, {Moln{\'a}r}, {Monari}, {Mongui{\'o}},
  {Montegriffo}, {Montero}, {Mor}, {Mora}, {Morbidelli}, {Morel}, {Morris},
  {Muraveva}, {Murphy}, {Musella}, {Nagy}, {Noval}, {Oca{\~n}a}, {Ogden},
  {Ordenovic}, {Osinde}, {Pagani}, {Pagano}, {Palaversa}, {Palicio},
  {Pallas-Quintela}, {Panahi}, {Payne-Wardenaar}, {Pe{\~n}alosa Esteller},
  {Penttil{\"a}}, {Pichon}, {Piersimoni}, {Pineau}, {Plachy}, {Plum}, {Poggio},
  {Pr{\v{s}}a}, {Pulone}, {Racero}, {Ragaini}, {Rainer}, {Raiteri}, {Rambaux},
  {Ramos}, {Ramos-Lerate}, {Re Fiorentin}, {Regibo}, {Richards}, {Rios Diaz},
  {Ripepi}, {Riva}, {Rix}, {Rixon}, {Robichon}, {Robin}, {Robin}, {Roelens},
  {Rogues}, {Rohrbasser}, {Romero-G{\'o}mez}, {Rowell}, {Royer}, {Ruz Mieres},
  {Rybicki}, {Sadowski}, {S{\'a}ez N{\'u}{\~n}ez}, {Sagrist{\`a} Sell{\'e}s},
  {Sahlmann}, {Salguero}, {Samaras}, {Sanchez Gimenez}, {Sanna},
  {Santove{\~n}a}, {Sarasso}, {Schultheis}, {Sciacca}, {Segol}, {Segovia},
  {S{\'e}gransan}, {Semeux}, {Shahaf}, {Siddiqui}, {Siebert}, {Siltala},
  {Silvelo}, {Slezak}, {Slezak}, {Smart}, {Snaith}, {Solano}, {Solitro},
  {Souami}, {Souchay}, {Spagna}, {Spina}, {Spoto}, {Steele},
  {Steidelm{\"u}ller}, {Stephenson}, {S{\"u}veges}, {Surdej}, {Szabados},
  {Szegedi-Elek}, {Taris}, {Taylor}, {Teixeira}, {Tolomei}, {Tonello}, {Torra},
  {Torra}, {Torralba Elipe}, {Trabucchi}, {Tsounis}, {Turon}, {Ulla}, {Unger},
  {Vaillant}, {van Dillen}, {van Reeven}, {Vanel}, {Vecchiato}, {Viala},
  {Vicente}, {Voutsinas}, {Weiler}, {Wevers}, {Wyrzykowski}, {Yoldas}, {Yvard},
  {Zhao}, {Zorec}, {Zucker}, \& {Zwitter}}]{gaia2023}
{Gaia Collaboration}, {Vallenari}, A., {Brown}, A.~G.~A., {et~al.} 2023, \aap,
  674, A1

\bibitem[{{Giannini} {et~al.}(2022){Giannini}, {Giunta}, {Gangi}, {Carini},
  {Lorenzetti}, {Antoniucci}, {Caratti o Garatti}, {Cassar{\'a}}, {Nisini},
  {Rossi}, {Testa}, \& {Vitali}}]{giannini2022}
{Giannini}, T., {Giunta}, A., {Gangi}, M., {et~al.} 2022, \apj, 929, 129

\bibitem[{{Grankin} {et~al.}(2007){Grankin}, {Melnikov}, {Bouvier}, {Herbst},
  \& {Shevchenko}}]{grankin2007}
{Grankin}, K.~N., {Melnikov}, S.~Y., {Bouvier}, J., {Herbst}, W., \&
  {Shevchenko}, V.~S. 2007, \aap, 461, 183

\bibitem[{{Gustafsson} {et~al.}(2008){Gustafsson}, {Edvardsson}, {Eriksson},
  {J{\o}rgensen}, {Nordlund}, \& {Plez}}]{gustafsson2008}
{Gustafsson}, B., {Edvardsson}, B., {Eriksson}, K., {et~al.} 2008, \aap, 486,
  951

\bibitem[{{Hartmann} {et~al.}(2016){Hartmann}, {Herczeg}, \&
  {Calvet}}]{hartmann2016}
{Hartmann}, L., {Herczeg}, G., \& {Calvet}, N. 2016, \araa, 54, 135

\bibitem[{{Hartmann} \& {Stauffer}(1989)}]{hartmann1989}
{Hartmann}, L. \& {Stauffer}, J.~R. 1989, \aj, 97, 873

\bibitem[{{Herczeg} {et~al.}(2023){Herczeg}, {Chen}, {Donati}, {Dupree},
  {Walter}, {Hillenbrand}, {Johns-Krull}, {Manara}, {G{\"u}nther}, {Fang},
  {Schneider}, {Valenti}, {Alencar}, {Venuti}, {Alcal{\'a}}, {Frasca},
  {Arulanantham}, {Linsky}, {Bouvier}, {Brickhouse}, {Calvet}, {Espaillat},
  {Campbell-White}, {Carpenter}, {Chang}, {Cruz}, {Dahm}, {Eisl{\"o}ffel},
  {Edwards}, {Fischer}, {Guo}, {Henning}, {Ji}, {Jose}, {Kastner}, {Launhardt},
  {Principe}, {Robinson}, {Serna}, {Siwak}, {Sterzik}, \&
  {Takasao}}]{herczeg2023}
{Herczeg}, G.~J., {Chen}, Y., {Donati}, J.-F., {et~al.} 2023, \apj, 956, 102

\bibitem[{{Herczeg} \& {Hillenbrand}(2014)}]{herczeg2014}
{Herczeg}, G.~J. \& {Hillenbrand}, L.~A. 2014, \apj, 786, 97

\bibitem[{{Hessman} \& {Guenther}(1997)}]{hessman1997}
{Hessman}, F.~V. \& {Guenther}, E.~W. 1997, \aap, 321, 497

\bibitem[{{Horne} \& {Baliunas}(1986)}]{horne1986}
{Horne}, J.~H. \& {Baliunas}, S.~L. 1986, \apj, 302, 757

\bibitem[{{Howell} {et~al.}(2014){Howell}, {Sobeck}, {Haas}, {Still},
  {Barclay}, {Mullally}, {Troeltzsch}, {Aigrain}, {Bryson}, {Caldwell},
  {Chaplin}, {Cochran}, {Huber}, {Marcy}, {Miglio}, {Najita}, {Smith},
  {Twicken}, \& {Fortney}}]{howell2014}
{Howell}, S.~B., {Sobeck}, C., {Haas}, M., {et~al.} 2014, \pasp, 126, 398

\bibitem[{{Hughes} {et~al.}(1992){Hughes}, {Aller}, \& {Aller}}]{hughes1992}
{Hughes}, P.~A., {Aller}, H.~D., \& {Aller}, M.~F. 1992, \apj, 396, 469

\bibitem[{{Johns} \& {Basri}(1995)}]{johns1995}
{Johns}, C.~M. \& {Basri}, G. 1995, \aj, 109, 2800

\bibitem[{{Johns-Krull}(2007)}]{johns-krull2007}
{Johns-Krull}, C.~M. 2007, \apj, 664, 975

\bibitem[{{Johnstone} {et~al.}(2014){Johnstone}, {Jardine}, {Gregory},
  {Donati}, \& {Hussain}}]{johnstone2014}
{Johnstone}, C.~P., {Jardine}, M., {Gregory}, S.~G., {Donati}, J.~F., \&
  {Hussain}, G. 2014, \mnras, 437, 3202

\bibitem[{{Jordi} {et~al.}(2006){Jordi}, {Grebel}, \& {Ammon}}]{jordi2006}
{Jordi}, K., {Grebel}, E.~K., \& {Ammon}, K. 2006, \aap, 460, 339

\bibitem[{{Joy}(1945)}]{joy1945}
{Joy}, A.~H. 1945, \apj, 102, 168

\bibitem[{{Kausch} {et~al.}(2015){Kausch}, {Noll}, {Smette}, {Kimeswenger},
  {Barden}, {Szyszka}, {Jones}, {Sana}, {Horst}, \& {Kerber}}]{kausch2015}
{Kausch}, W., {Noll}, S., {Smette}, A., {et~al.} 2015, \aap, 576, A78

\bibitem[{{Kenyon} {et~al.}(1994){Kenyon}, {Hartmann}, {Hewett}, {Carrasco},
  {Cruz-Gonzalez}, {Recillas}, {Salas}, {Serrano}, {Strom}, {Strom}, \&
  {Newton}}]{kenyon1994}
{Kenyon}, S.~J., {Hartmann}, L., {Hewett}, R., {et~al.} 1994, \aj, 107, 2153

\bibitem[{{K{\'o}sp{\'a}l} {et~al.}(2018){K{\'o}sp{\'a}l}, {{\'A}brah{\'a}m},
  {Zsidi}, {Vida}, {Szab{\'o}}, {Mo{\'o}r}, \& {P{\'a}l}}]{kospal2018}
{K{\'o}sp{\'a}l}, {\'A}., {{\'A}brah{\'a}m}, P., {Zsidi}, G., {et~al.} 2018,
  \apj, 862, 44

\bibitem[{{Kun} {et~al.}(2011){Kun}, {Szegedi-Elek}, {Mo{\'o}r},
  {K{\'o}sp{\'a}l}, {{\'A}brah{\'a}m}, {Apai}, {Kiss}, {Klagyivik}, {Magakian},
  {Mez{\H{o}}}, {Movsessian}, {P{\'a}l}, {R{\'a}cz}, \& {Rogers}}]{kun2011}
{Kun}, M., {Szegedi-Elek}, E., {Mo{\'o}r}, A., {et~al.} 2011, \mnras, 413, 2689

\bibitem[{{Kurosawa} {et~al.}(2011){Kurosawa}, {Romanova}, \&
  {Harries}}]{kurosawa2011}
{Kurosawa}, R., {Romanova}, M.~M., \& {Harries}, T.~J. 2011, \mnras, 416, 2623

\bibitem[{{Lakeland} \& {Naylor}(2022)}]{lakeland2022}
{Lakeland}, B.~S. \& {Naylor}, T. 2022, \mnras, 514, 2736

\bibitem[{{Landstreet}(1988)}]{landstreet1988}
{Landstreet}, J.~D. 1988, \apj, 326, 967

\bibitem[{{Long} {et~al.}(2019){Long}, {Herczeg}, {Harsono}, {Pinilla},
  {Tazzari}, {Manara}, {Pascucci}, {Cabrit}, {Nisini}, {Johnstone}, {Edwards},
  {Salyk}, {Menard}, {Lodato}, {Boehler}, {Mace}, {Liu}, {Mulders}, {Hendler},
  {Ragusa}, {Fischer}, {Banzatti}, {Rigliaco}, {van de Plas}, {Dipierro},
  {Gully-Santiago}, \& {Lopez-Valdivia}}]{long2019}
{Long}, F., {Herczeg}, G.~J., {Harsono}, D., {et~al.} 2019, \apj, 882, 49

\bibitem[{{Manara} {et~al.}(2013){Manara}, {Testi}, {Rigliaco}, {Alcal{\'a}},
  {Natta}, {Stelzer}, {Biazzo}, {Covino}, {Covino}, {Cupani}, {D'Elia}, \&
  {Randich}}]{manara2013}
{Manara}, C.~F., {Testi}, L., {Rigliaco}, E., {et~al.} 2013, \aap, 551, A107

\bibitem[{{Matt} \& {Pudritz}(2005)}]{matt2005}
{Matt}, S. \& {Pudritz}, R.~E. 2005, \apjl, 632, L135

\bibitem[{{Nowacki} {et~al.}(2023){Nowacki}, {Alecian}, {Perraut}, {Zaire},
  {Folsom}, {Pouilly}, {Bouvier}, {Manick}, {Pantolmos}, {Sousa}, {Dougados},
  {Hussain}, {Alencar}, \& {Le Bouquin}}]{nowacki2023}
{Nowacki}, H., {Alecian}, E., {Perraut}, K., {et~al.} 2023, \aap, 678, A86

\bibitem[{{P{\'a}l} {et~al.}(2020){P{\'a}l}, {Szak{\'a}ts}, {Kiss}, {B{\'o}di},
  {Bogn{\'a}r}, {Kalup}, {Kiss}, {Marton}, {Moln{\'a}r}, {Plachy},
  {S{\'a}rneczky}, {Szab{\'o}}, \& {Szab{\'o}}}]{pal2020}
{P{\'a}l}, A., {Szak{\'a}ts}, R., {Kiss}, C., {et~al.} 2020, \apjs, 247, 26

\bibitem[{{Petrov} {et~al.}(2001){Petrov}, {Gahm}, {Gameiro}, {Duemmler},
  {Ilyin}, {Laakkonen}, {Lago}, \& {Tuominen}}]{petrov2001}
{Petrov}, P.~P., {Gahm}, G.~F., {Gameiro}, J.~F., {et~al.} 2001, \aap, 369, 993

\bibitem[{{Petrov} {et~al.}(2011){Petrov}, {Gahm}, {Stempels}, {Walter}, \&
  {Artemenko}}]{petrov2011}
{Petrov}, P.~P., {Gahm}, G.~F., {Stempels}, H.~C., {Walter}, F.~M., \&
  {Artemenko}, S.~A. 2011, \aap, 535, A6

\bibitem[{{Petrov} \& {Kozack}(2007)}]{petrov2007}
{Petrov}, P.~P. \& {Kozack}, B.~S. 2007, Astronomy Reports, 51, 500

\bibitem[{{Plachy} {et~al.}(2021){Plachy}, {P{\'a}l}, {B{\'o}di}, {Szab{\'o}},
  {Moln{\'a}r}, {Szabados}, {Benk{\H{o}}}, {Anderson}, {Bellinger}, {Bhardwaj},
  {Ebadi}, {Gazeas}, {Hambsch}, {Hasanzadeh}, {Jurkovic}, {Kalaee}, {Kervella},
  {Kolenberg}, {Miko{\l}ajczyk}, {Nardetto}, {Nemec}, {Netzel}, {Ngeow},
  {Ozuyar}, {Pascual-Granado}, {Pilecki}, {Ripepi}, {Skarka}, {Smolec},
  {S{\'o}dor}, {Szab{\'o}}, {Christensen-Dalsgaard}, {Jenkins}, {Kjeldsen},
  {Ricker}, \& {Vanderspek}}]{plachy2021}
{Plachy}, E., {P{\'a}l}, A., {B{\'o}di}, A., {et~al.} 2021, \apjs, 253, 11

\bibitem[{{Preston}(1967)}]{preston1967}
{Preston}, G.~W. 1967, \apj, 150, 547

\bibitem[{{Rei} {et~al.}(2018){Rei}, {Petrov}, \& {Gameiro}}]{rei2018}
{Rei}, A.~C.~S., {Petrov}, P.~P., \& {Gameiro}, J.~F. 2018, \aap, 610, A40

\bibitem[{{Reipurth} {et~al.}(1996){Reipurth}, {Pedrosa}, \&
  {Lago}}]{reipurth1996}
{Reipurth}, B., {Pedrosa}, A., \& {Lago}, M.~T.~V.~T. 1996, \aaps, 120, 229

\bibitem[{{Roelens} {et~al.}(2017){Roelens}, {Eyer}, {Mowlavi},
  {Lecoeur-Ta{\"\i}bi}, {Rimoldini}, {Blanco-Cuaresma}, {Palaversa},
  {S{\"u}veges}, {Charnas}, \& {Wevers}}]{roelens2017}
{Roelens}, M., {Eyer}, L., {Mowlavi}, N., {et~al.} 2017, \mnras, 472, 3230

\bibitem[{{Ryabchikova} {et~al.}(1997){Ryabchikova}, {Piskunov}, {Kupka}, \&
  {Weiss}}]{ryabchikova1997}
{Ryabchikova}, T.~A., {Piskunov}, N.~E., {Kupka}, F., \& {Weiss}, W.~W. 1997,
  Baltic Astronomy, 6, 244

\bibitem[{{Sergison} {et~al.}(2020){Sergison}, {Naylor}, {Littlefair}, {Bell},
  \& {Williams}}]{sergison2020}
{Sergison}, D.~J., {Naylor}, T., {Littlefair}, S.~P., {Bell}, C. P.~M., \&
  {Williams}, C.~D.~H. 2020, \mnras, 491, 5035

\bibitem[{{Simon} {et~al.}(2016){Simon}, {Pascucci}, {Edwards}, {Feng},
  {Gorti}, {Hollenbach}, {Rigliaco}, \& {Keane}}]{simon2016}
{Simon}, M.~N., {Pascucci}, I., {Edwards}, S., {et~al.} 2016, \apj, 831, 169

\bibitem[{{Siwak} {et~al.}(2016){Siwak}, {Ogloza}, {Rucinski}, {Moffat},
  {Matthews}, {Cameron}, {Guenther}, {Kuschnig}, {Rowe}, {Sasselov}, \&
  {Weiss}}]{siwak2016}
{Siwak}, M., {Ogloza}, W., {Rucinski}, S.~M., {et~al.} 2016, \mnras, 456, 3972

\bibitem[{{Smette} {et~al.}(2015){Smette}, {Sana}, {Noll}, {Horst}, {Kausch},
  {Kimeswenger}, {Barden}, {Szyszka}, {Jones}, {Gallenne}, {Vinther},
  {Ballester}, \& {Taylor}}]{smette2015}
{Smette}, A., {Sana}, H., {Noll}, S., {et~al.} 2015, \aap, 576, A77

\bibitem[{{Stempels} \& {Piskunov}(2003)}]{stempels2003}
{Stempels}, H.~C. \& {Piskunov}, N. 2003, \aap, 408, 693

\bibitem[{{Tout} \& {Pringle}(1992)}]{tout1992}
{Tout}, C.~A. \& {Pringle}, J.~E. 1992, \mnras, 256, 269

\bibitem[{{Venuti} {et~al.}(2021){Venuti}, {Cody}, {Rebull}, {Beccari},
  {Irwin}, {Thanvantri}, {Howell}, \& {Barentsen}}]{venuti2021}
{Venuti}, L., {Cody}, A.~M., {Rebull}, L.~M., {et~al.} 2021, \aj, 162, 101

\bibitem[{{Wade} {et~al.}(2001){Wade}, {Bagnulo}, {Kochukhov}, {Landstreet},
  {Piskunov}, \& {Stift}}]{wade2001}
{Wade}, G.~A., {Bagnulo}, S., {Kochukhov}, O., {et~al.} 2001, \aap, 374, 265

\bibitem[{{Wade} {et~al.}(2000){Wade}, {Donati}, {Landstreet}, \&
  {Shorlin}}]{wade2000}
{Wade}, G.~A., {Donati}, J.~F., {Landstreet}, J.~D., \& {Shorlin}, S.~L.~S.
  2000, \mnras, 313, 851

\bibitem[{{Zsidi} {et~al.}(2022{\natexlab{a}}){Zsidi}, {Fiorellino},
  {K{\'o}sp{\'a}l}, {{\'A}brah{\'a}m}, {B{\'o}di}, {Hussain}, {Manara}, \&
  {P{\'a}l}}]{zsidi2022_vw}
{Zsidi}, G., {Fiorellino}, E., {K{\'o}sp{\'a}l}, {\'A}., {et~al.}
  2022{\natexlab{a}}, \apj, 941, 177

\bibitem[{{Zsidi} {et~al.}(2022{\natexlab{b}}){Zsidi}, {Manara},
  {K{\'o}sp{\'a}l}, {Hussain}, {{\'A}brah{\'a}m}, {Alecian}, {B{\'o}di},
  {P{\'a}l}, \& {Sarkis}}]{zsidi2022_cr}
{Zsidi}, G., {Manara}, C.~F., {K{\'o}sp{\'a}l}, {\'A}., {et~al.}
  2022{\natexlab{b}}, \aap, 660, A108

\end{thebibliography}

\begin{appendix}

\onecolumn

\section{Line profile variability}\label{sect:appendix_A}
   \begin{figure*}[h!]
   \centering
   \includegraphics[width=\textwidth]{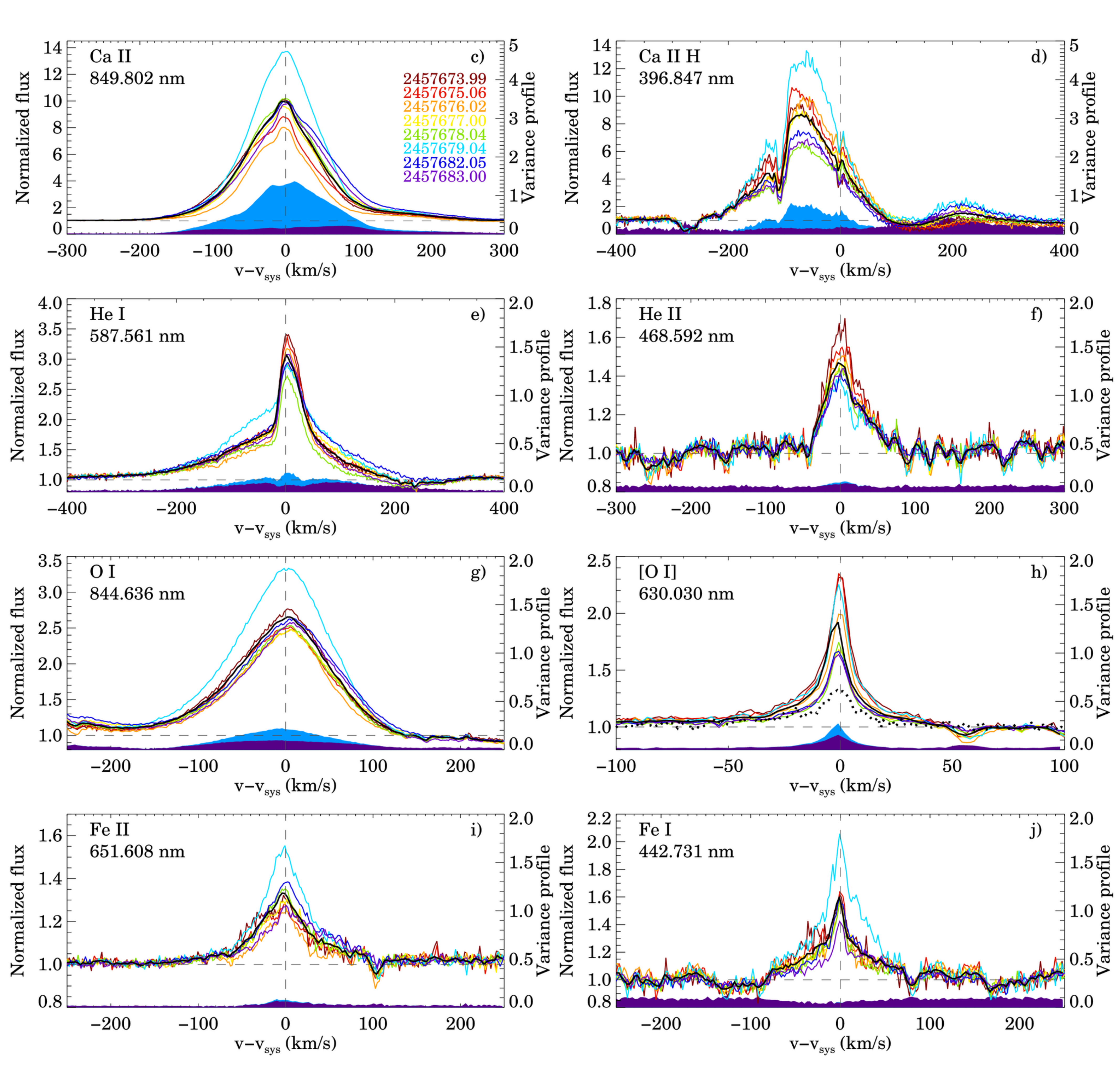}
      \caption{Variance profiles of emission lines. The profiles with different colors indicate observations made on the Julian days marked on the first panel. The thick black curve shows the mean line profile, the light blue area is the variance profile, and the dark blue shaded area is the normalized variance profile. The dotted line in panel $h$ indicate data from \cite{banzatti2019}.}
         \label{fig:drtau_var_profile}
   \end{figure*}

\newpage
\section{Decomposition of spectral lines}\label{sect:appendix_B}
   \begin{figure*}[h!]
   \centering
   \includegraphics[width=\textwidth]{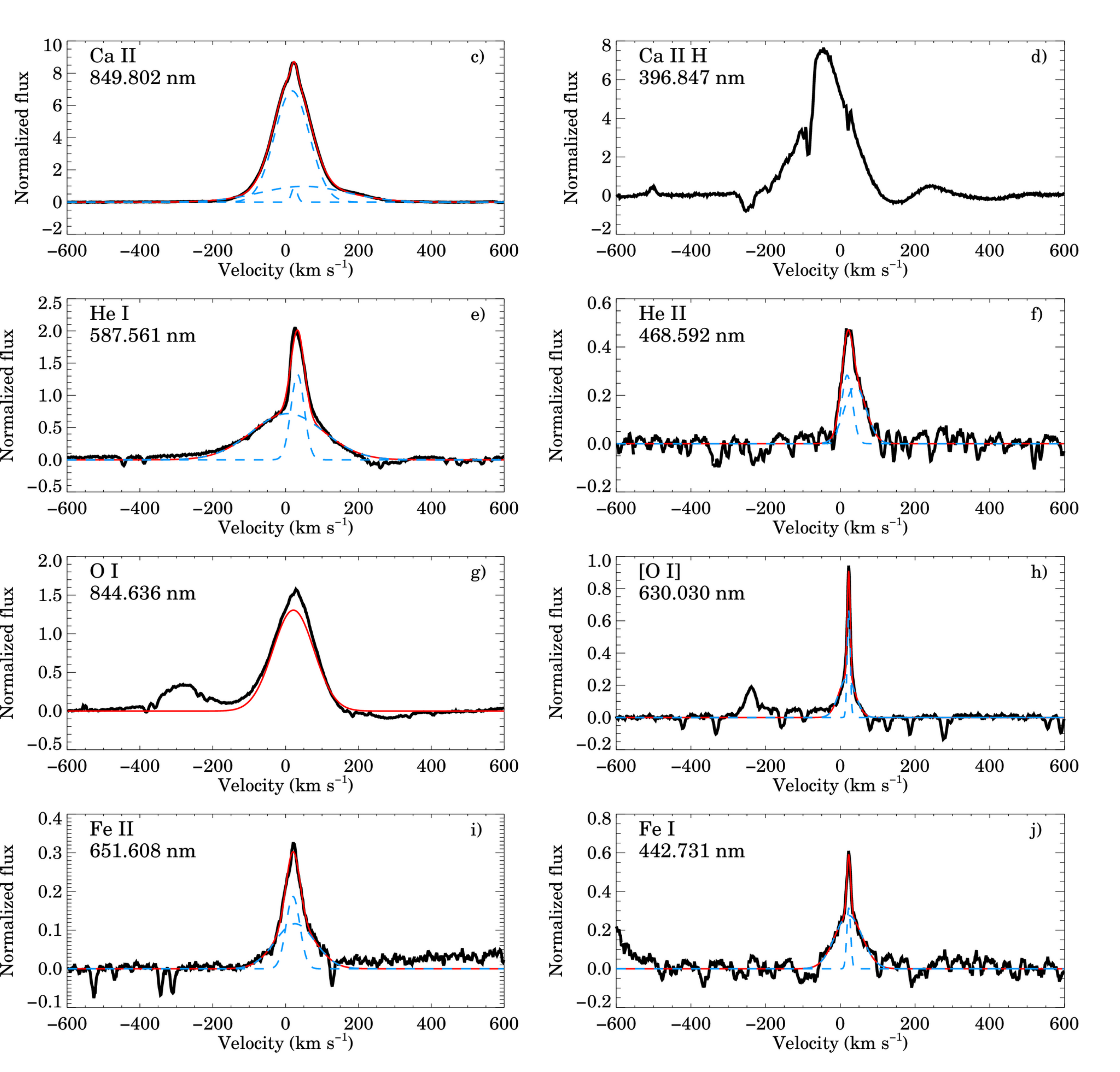}
      \caption{Spectral decomposition. The thick black lines are the observed spectra, the blue dashed curves show the individual components, and the red curve indicates the total fit.}
         \label{fig:emission_fitteles}
   \end{figure*}

\onecolumn
\section{Correlation and cross-correlation matrices for emission lines}
   \begin{figure*}[h!]
   \centering
    \includegraphics[width=0.45\textwidth]{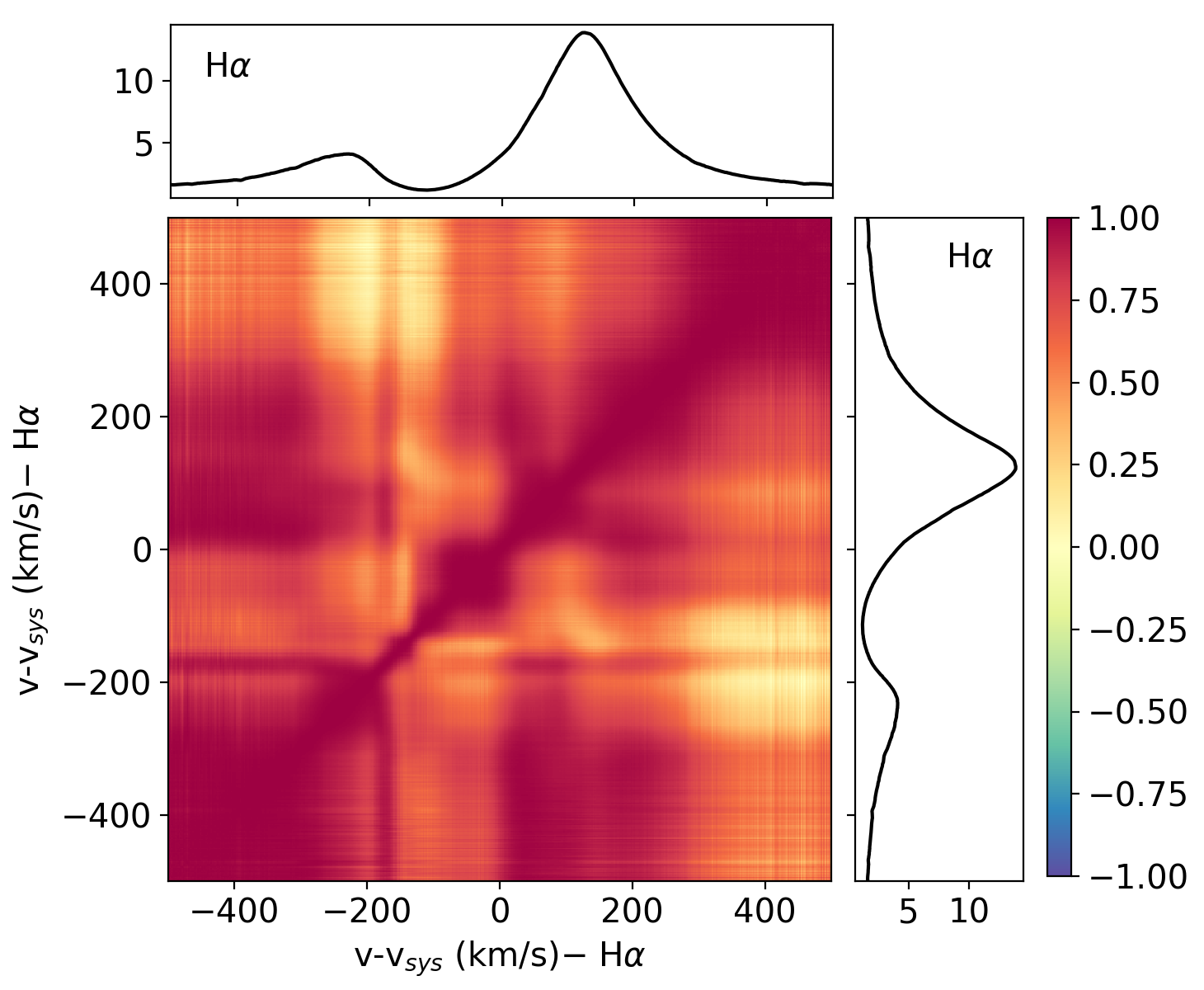}
    \includegraphics[width=0.45\textwidth]{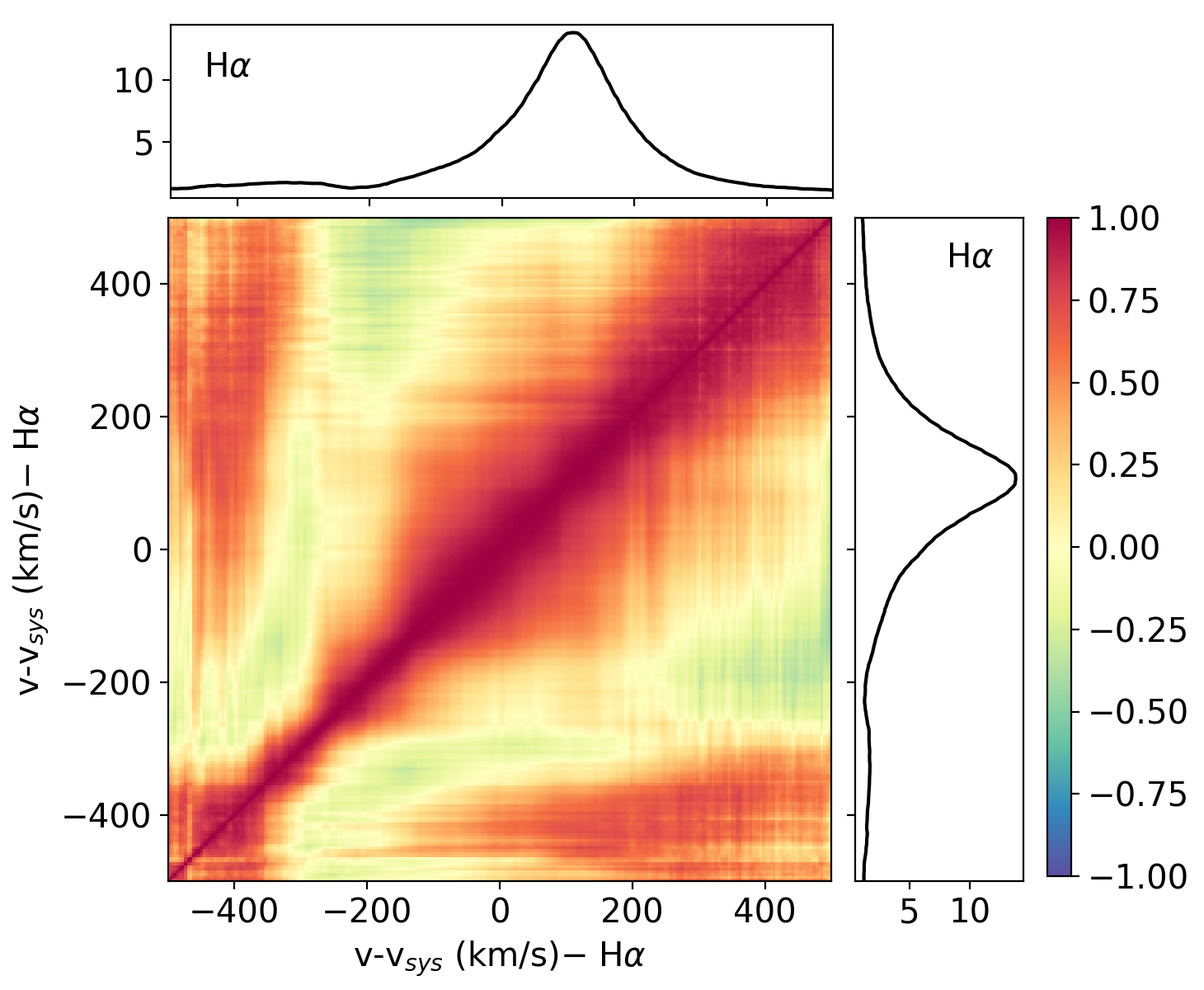}

    \includegraphics[width=0.45\textwidth]{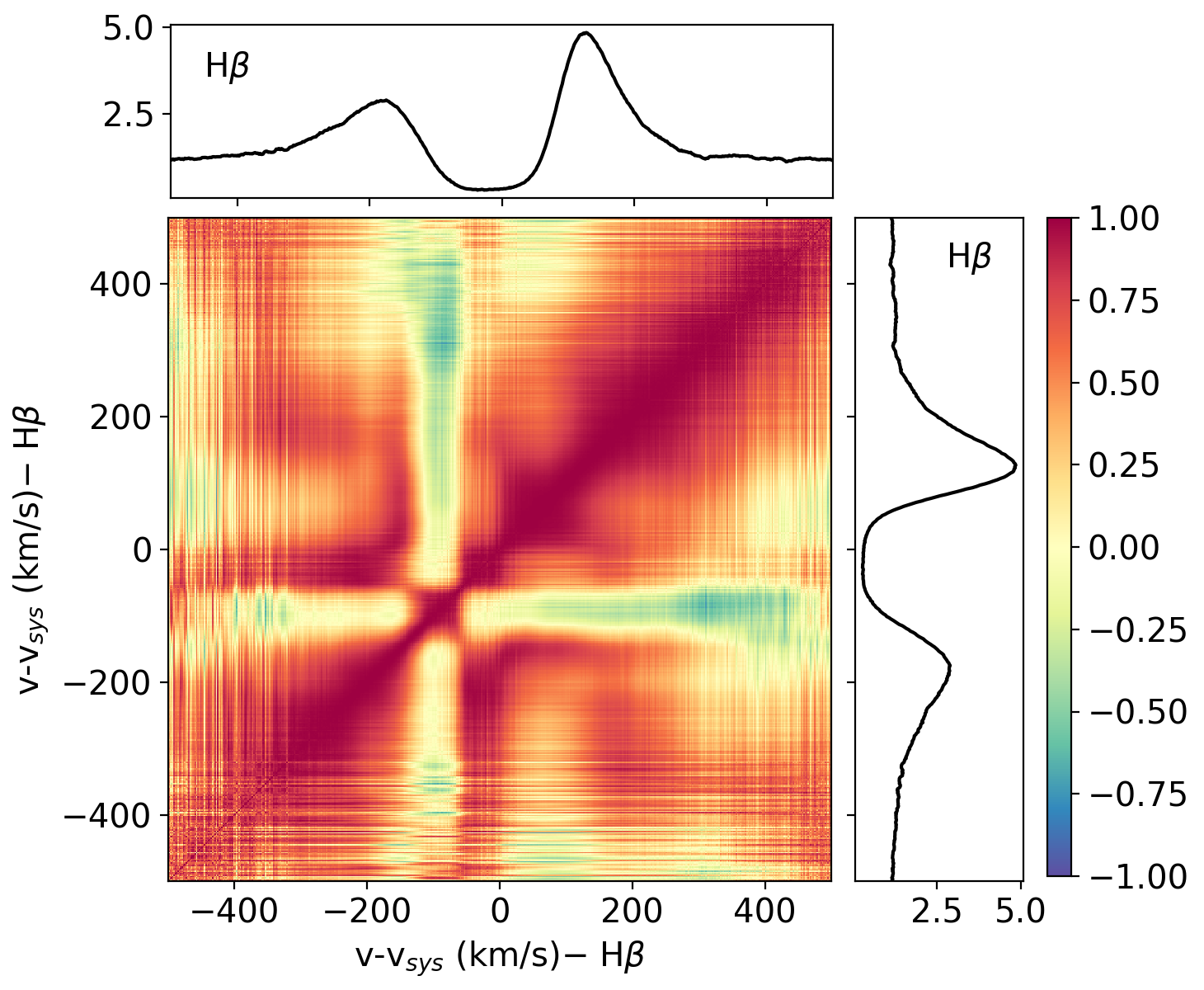}
    \includegraphics[width=0.45\textwidth]{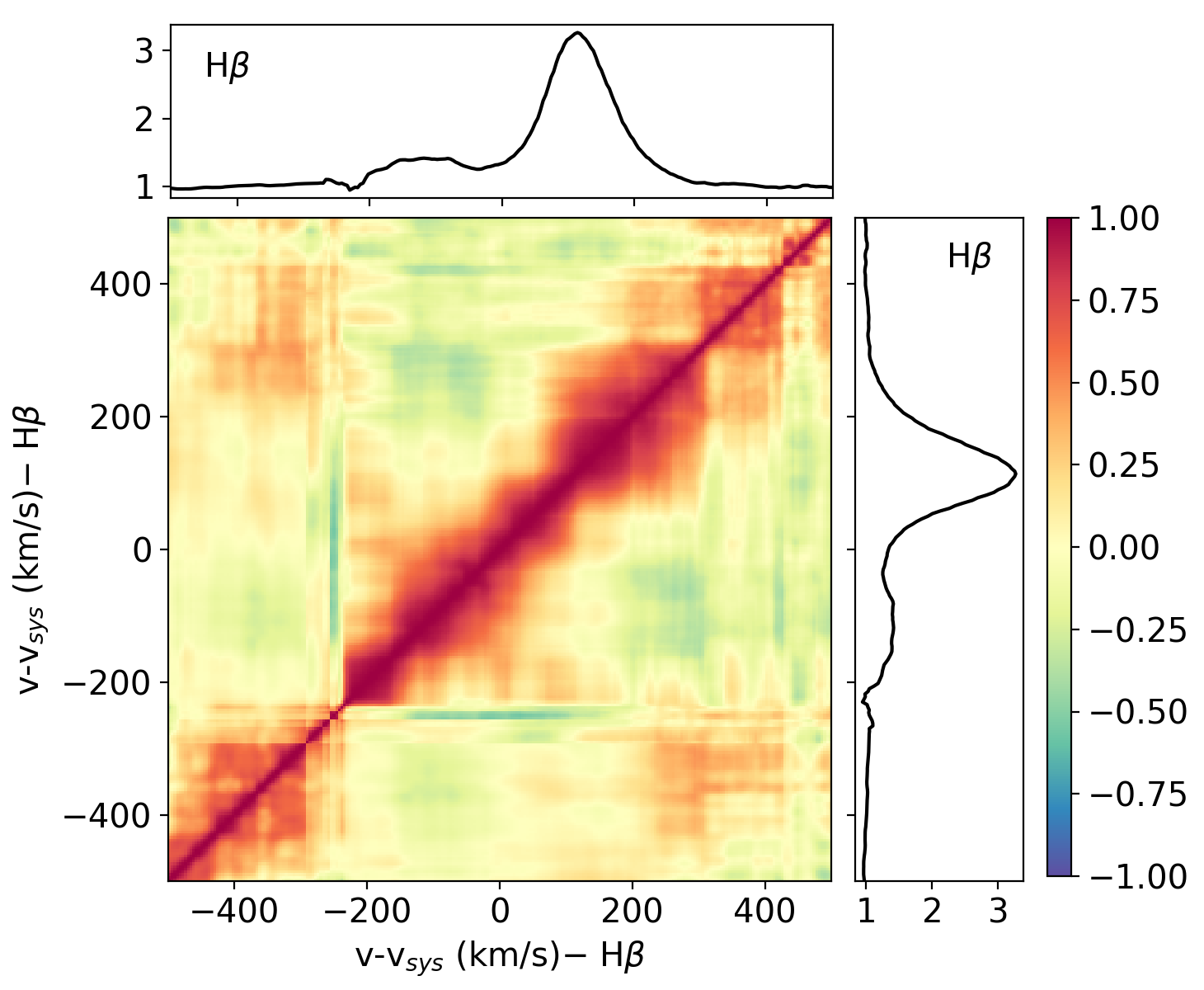}
    
    \includegraphics[width=0.45\textwidth]{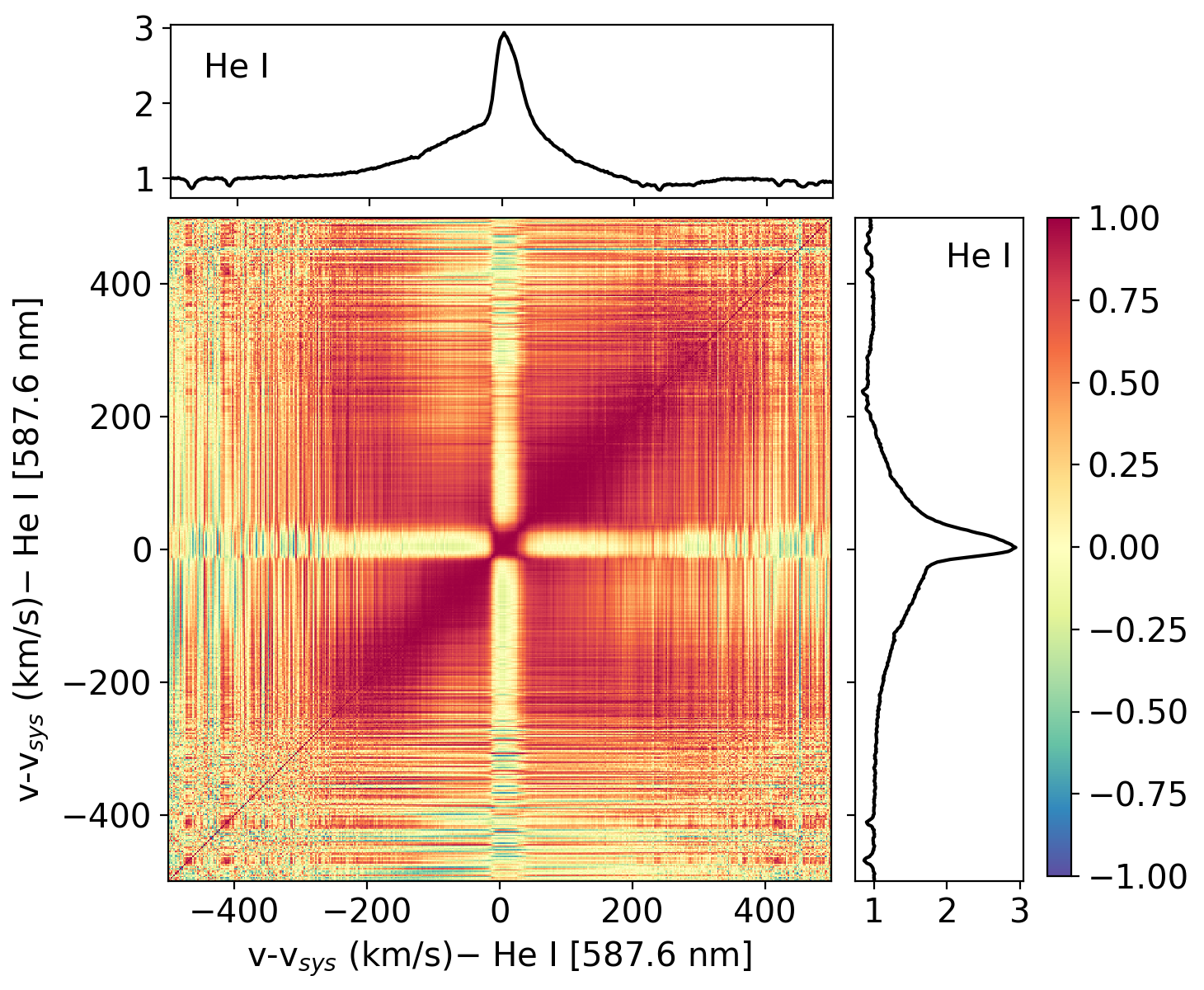}
    \includegraphics[width=0.45\textwidth]{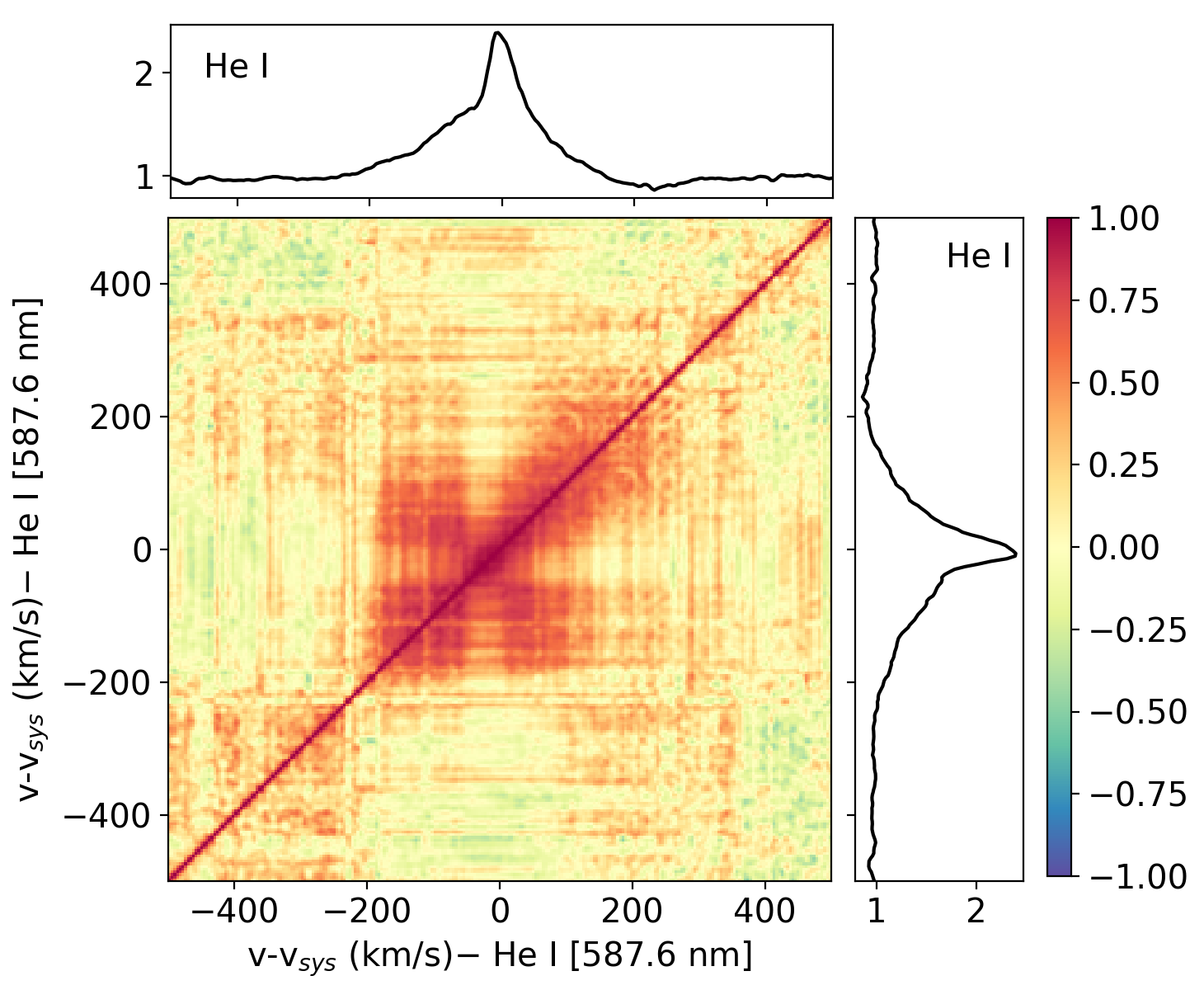}

     \caption{Correlation and cross-correlation matrices for H$\alpha$, H$\beta$, He I and Ca II lines. The average line profiles are indicated with black curves. The left column shows the correlation matrices based on the data published here, and the right column shows the correlation matrices based on the data published in \cite{alencar2001}}
   \label{fig:corrmtx}

   \end{figure*}

   \begin{figure*}[t!]
   \centering
   
     \includegraphics[width=0.45\textwidth]{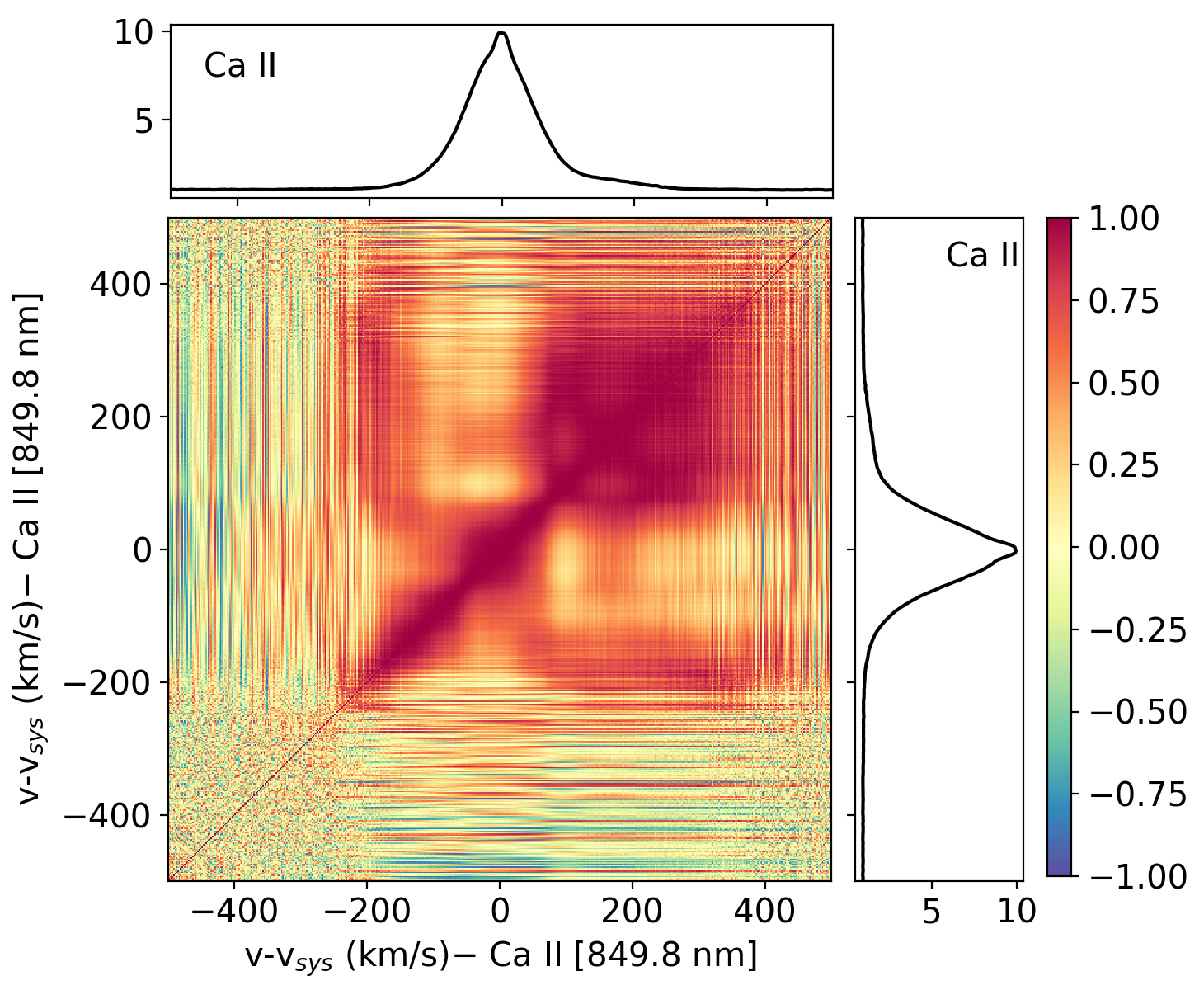}
    \includegraphics[width=0.45\textwidth]{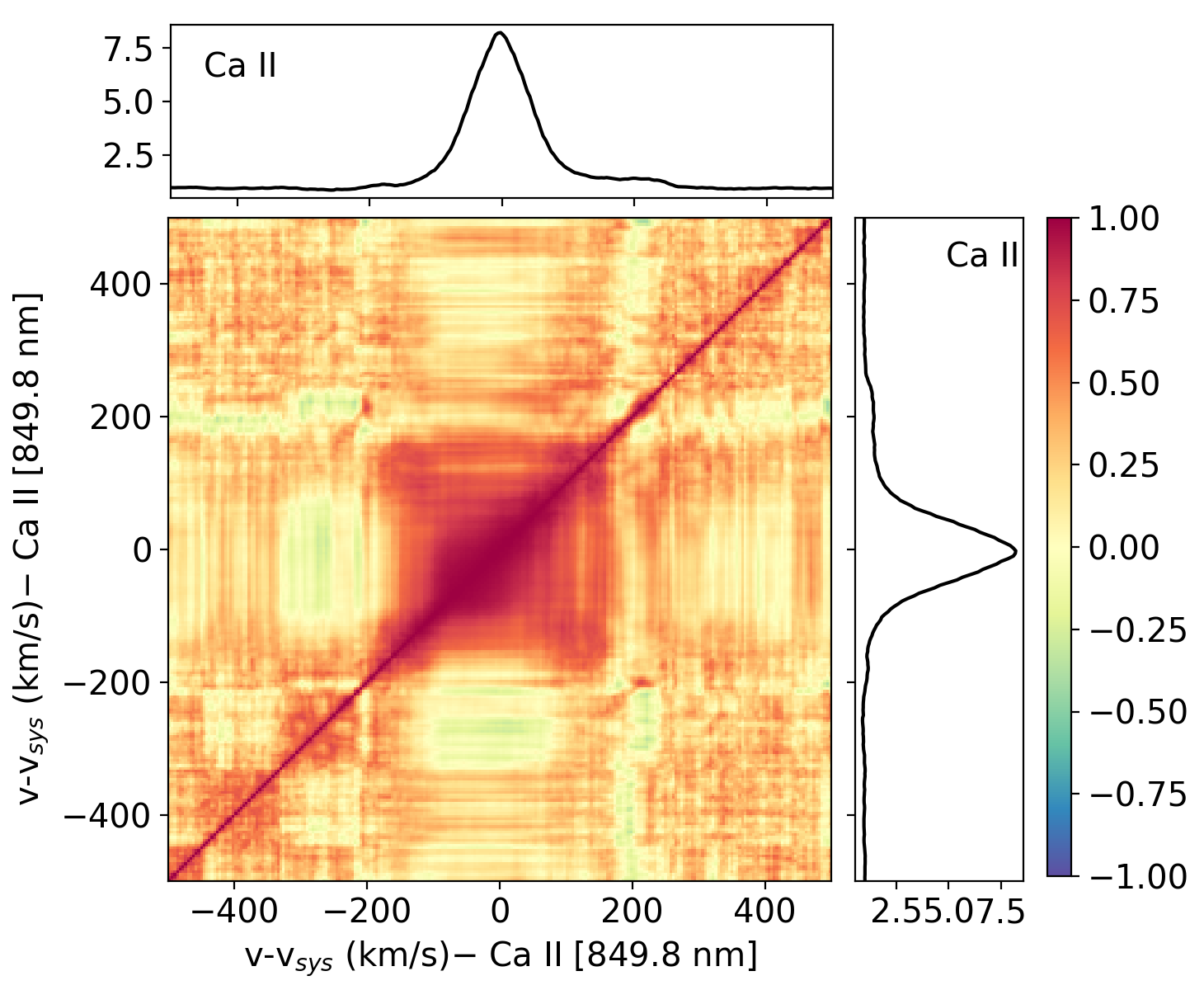}

    \includegraphics[width=0.45\textwidth]{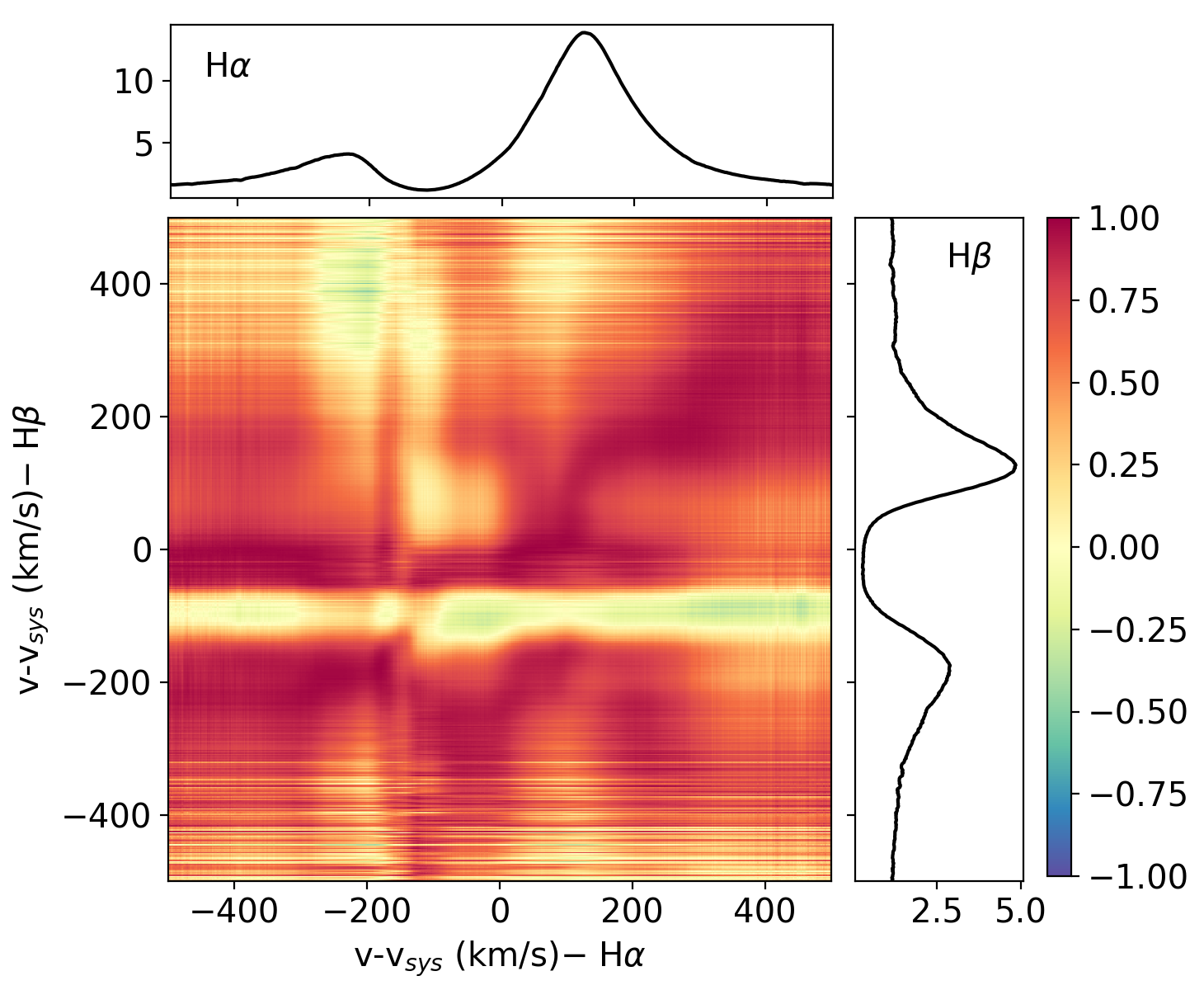}
    \includegraphics[width=0.45\textwidth]{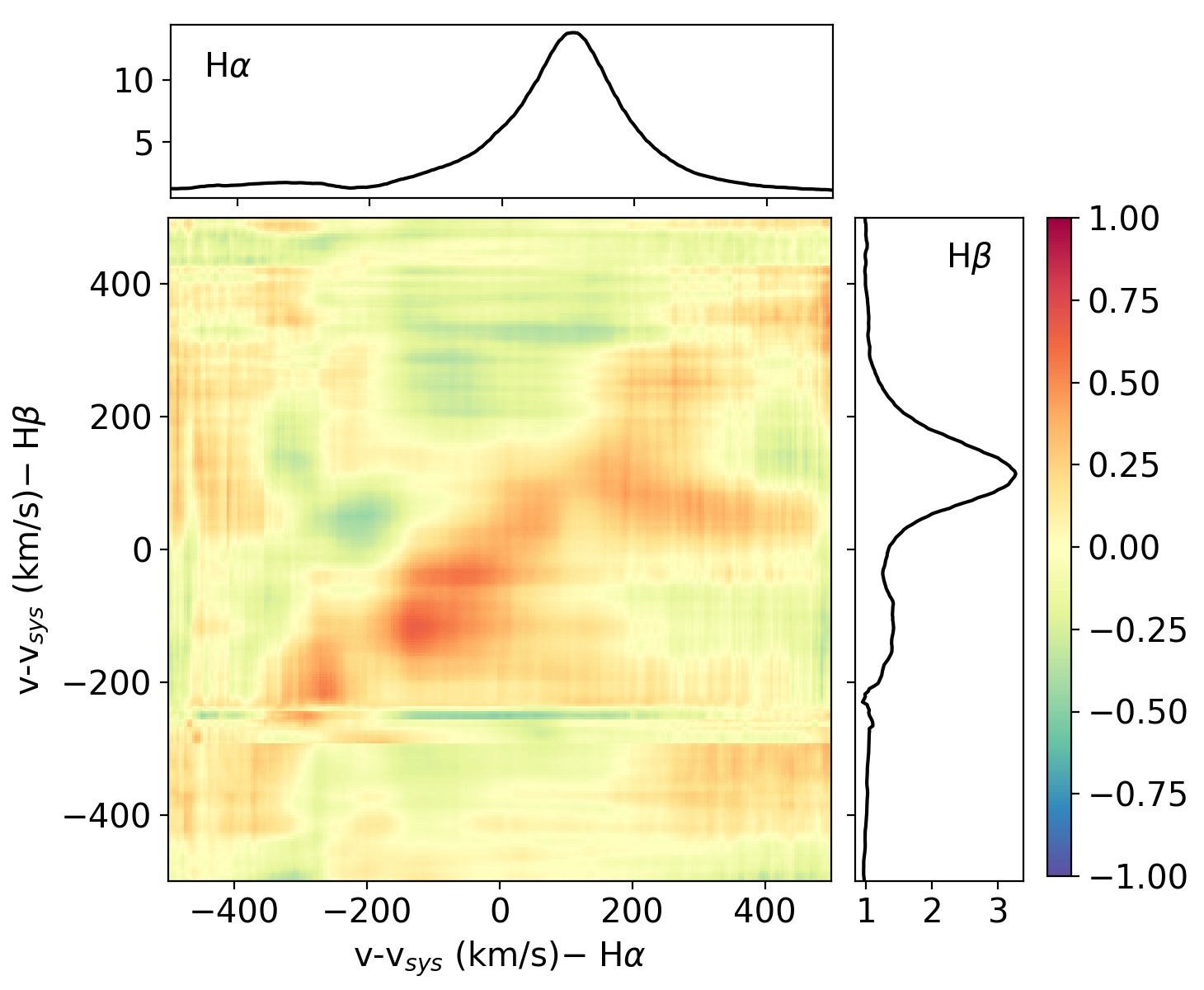}

    \includegraphics[width=0.45\textwidth]{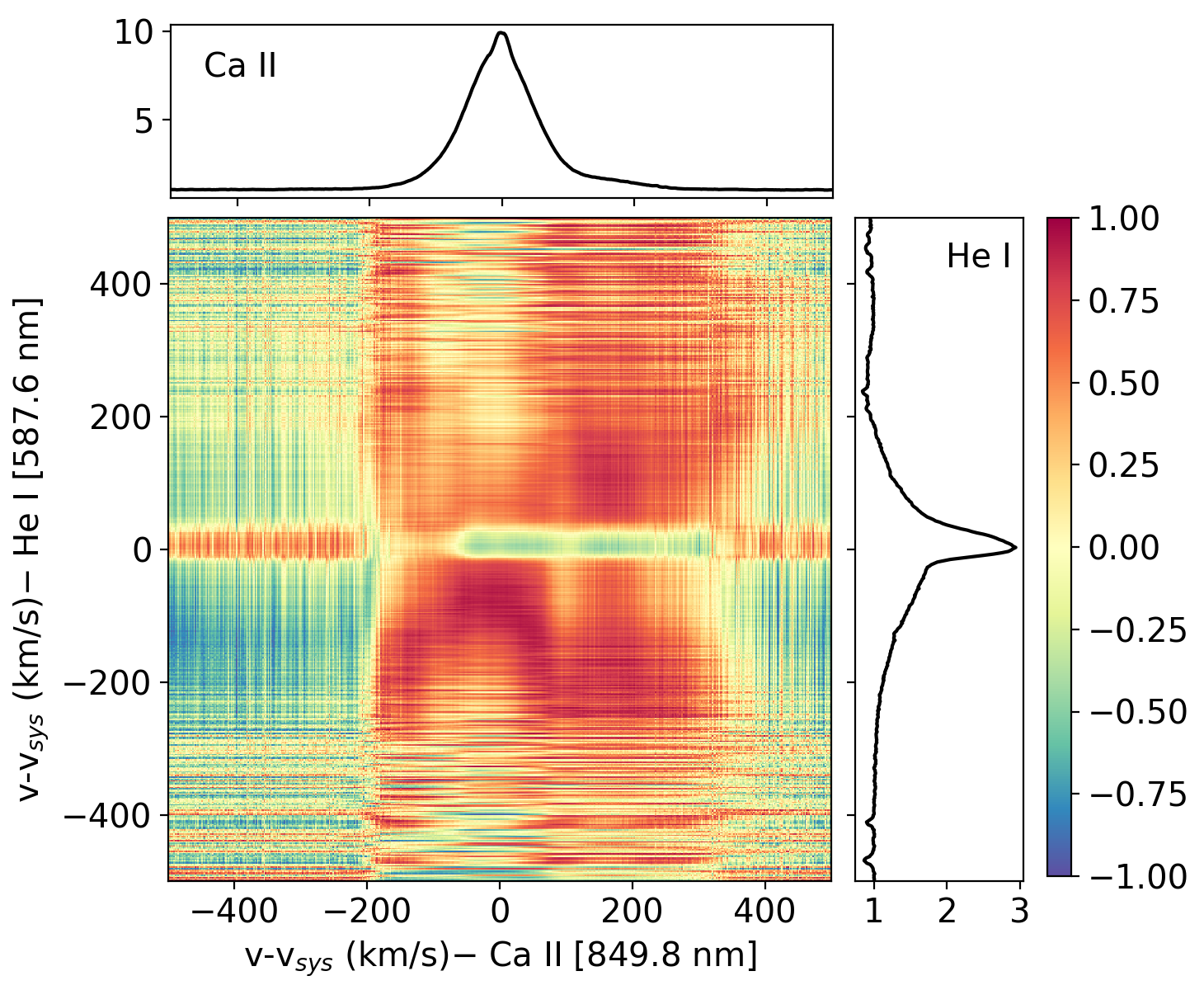}
    \includegraphics[width=0.45\textwidth]{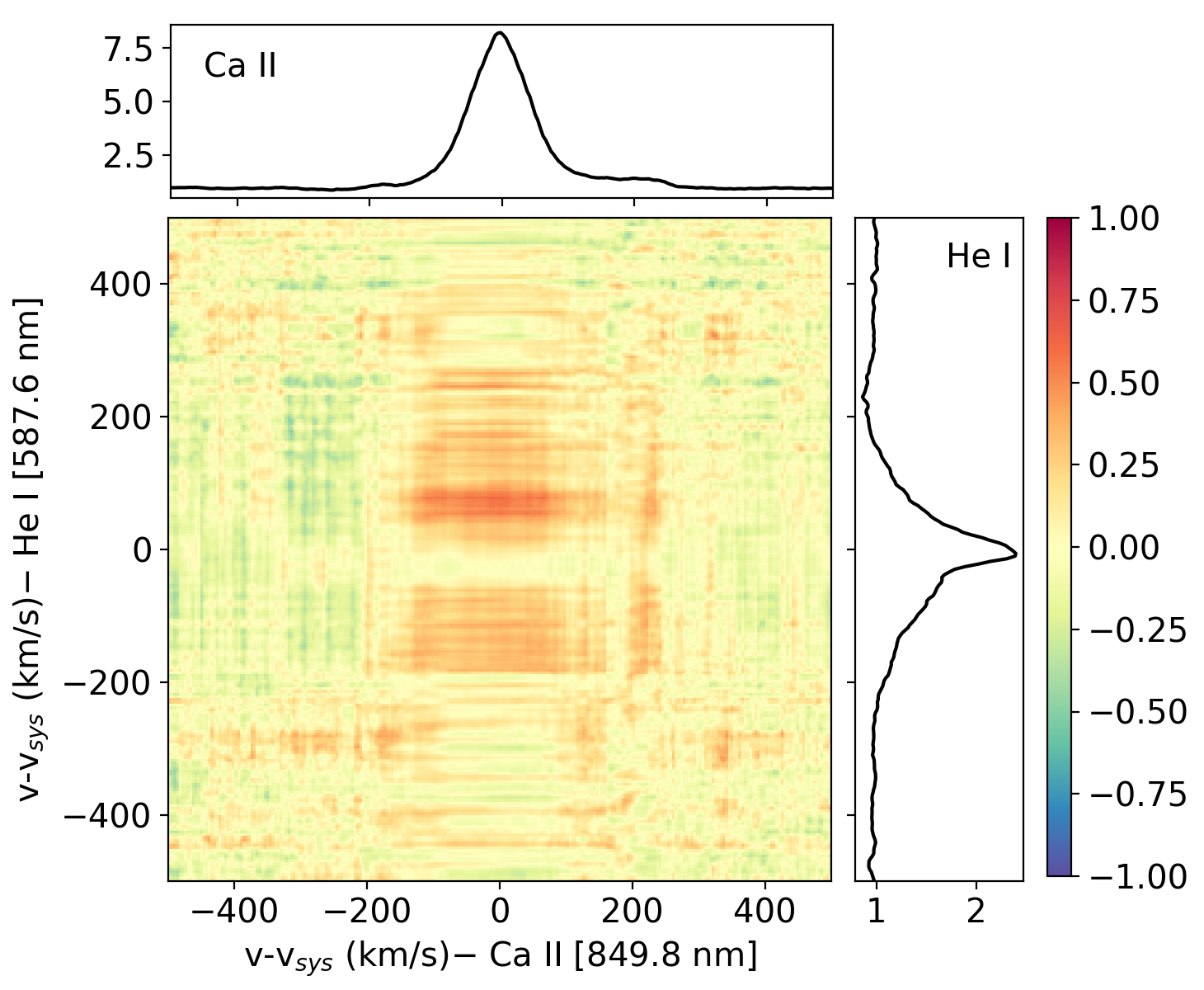}
 
     \caption{Correlation and cross-correlation matrices for H$\alpha$, H$\beta$, He I and Ca II lines. The average line profiles are indicated with black curves. The left column shows the correlation matrices based on the data published here, and the right column shows the correlation matrices based on the data published in \cite{alencar2001}}
   \label{fig:corrmtx2}
   \end{figure*}

   \begin{figure*}[h!]
   \centering
    \includegraphics[width=0.45\textwidth]{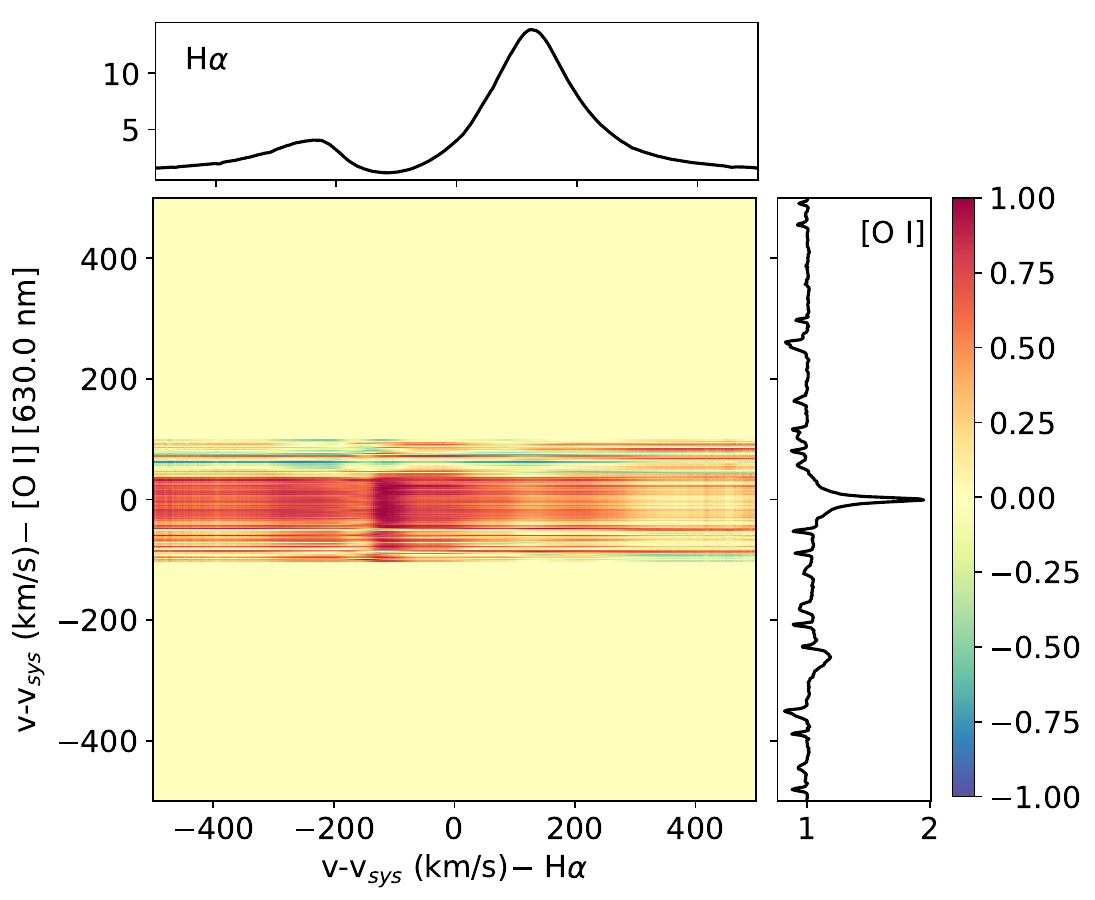}
    \includegraphics[width=0.45\textwidth]{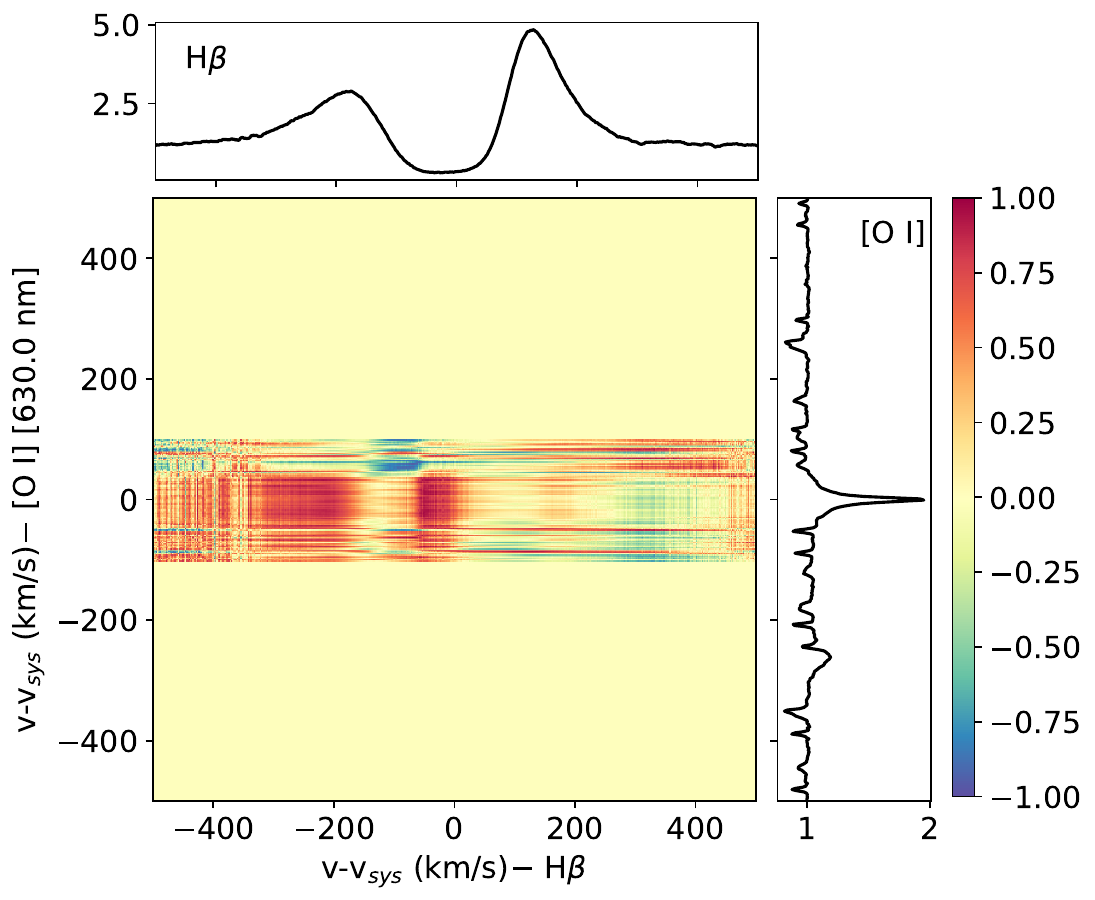}

    \includegraphics[width=0.45\textwidth]{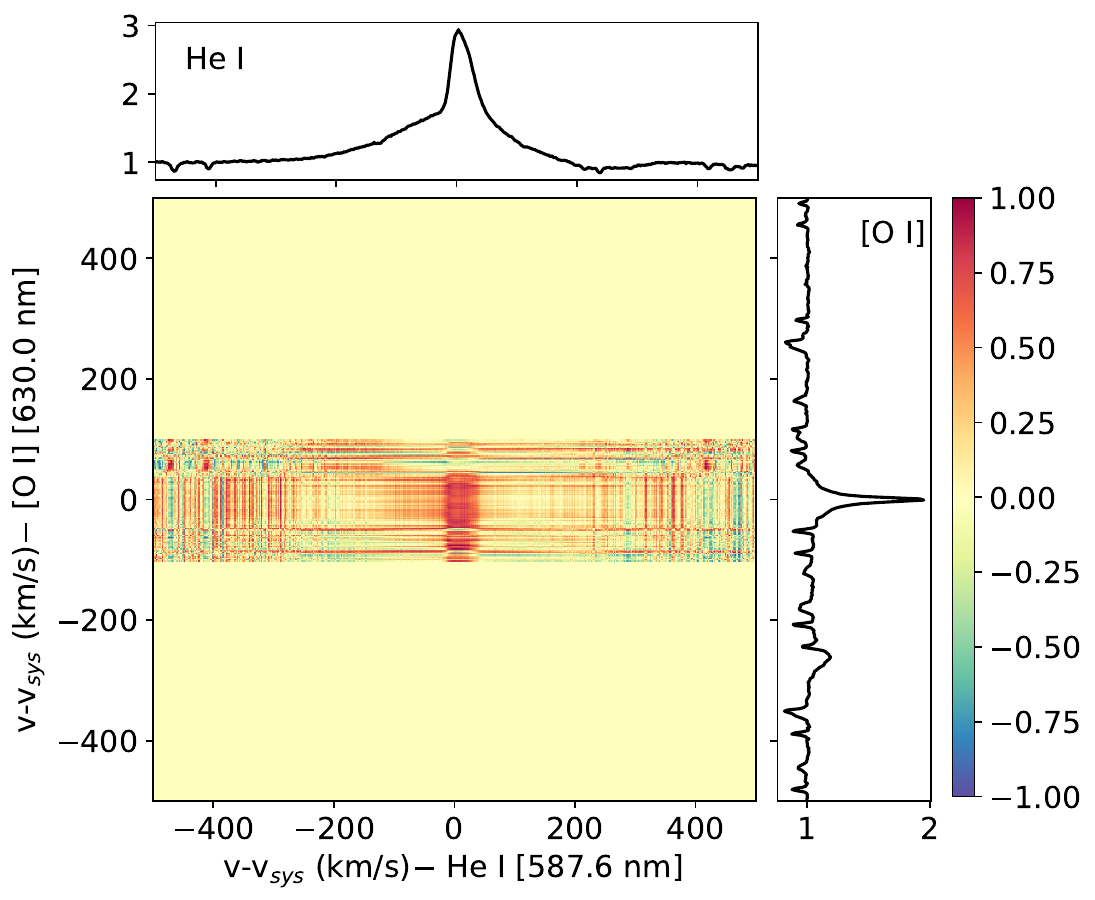}
    \includegraphics[width=0.45\textwidth]{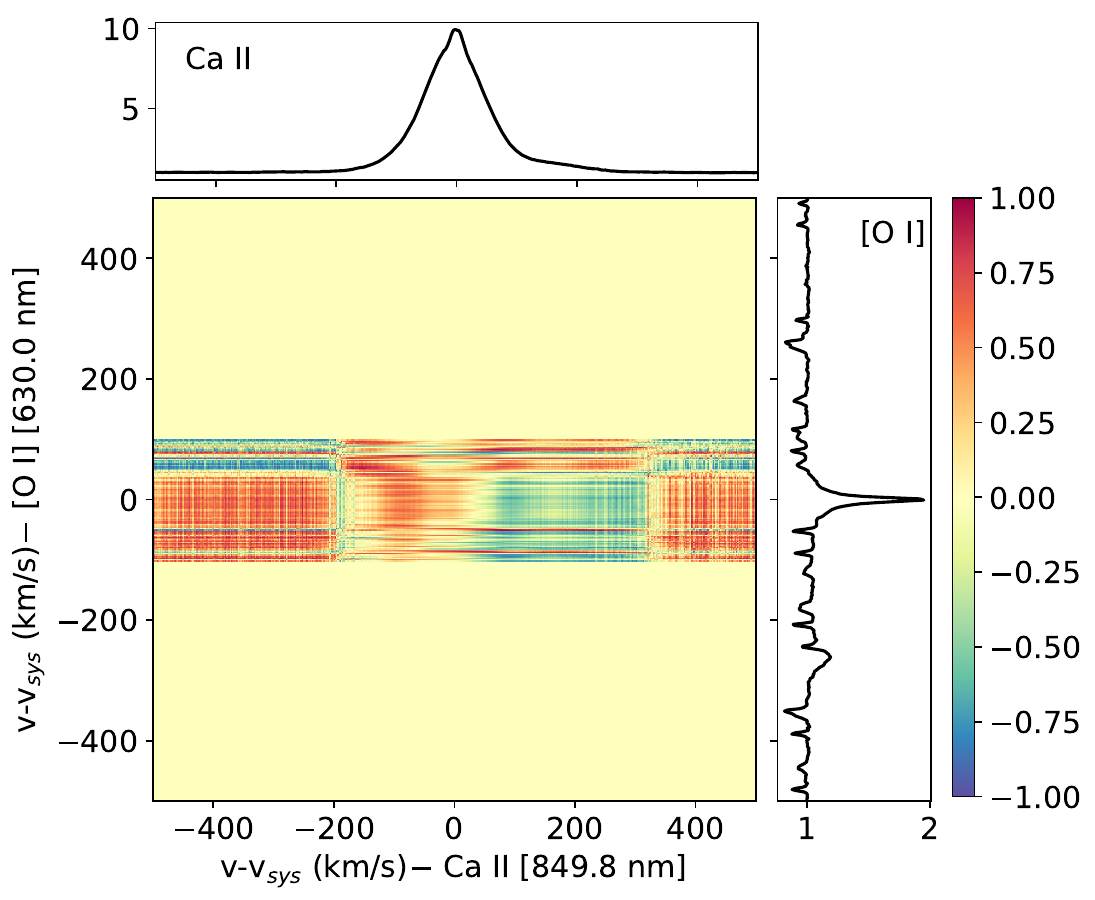}

     \caption{Cross-correlation matrices for the [O\,I] and the H$\alpha$, H$\beta$, He\,I, and Ca\,II lines. The average line profiles are indicated with black curves.}
   \label{fig:corrmtx_outflow}

   \end{figure*}

\end{appendix}

\end{document}